\documentclass[acmsmall,screen]{acmart}

\usepackage{mathtools}
\usepackage{amsmath, mathtools}
\usepackage[bb=stix]{mathalpha}
\usepackage{ebproof}
\usepackage{fontawesome5}
\usepackage{stmaryrd}
\usepackage{lipsum}
\usepackage{xspace}
\usepackage{amsthm}
\usepackage{float}
\usepackage[svgnames]{xcolor}
\usepackage{listings}
\usepackage[thicklines]{cancel}
\usepackage{longtable}
\usepackage{multirow}
\usepackage[titleref, user]{zref}[2025-06-08]
\usepackage{zref-clever}[2024-11-28]
\usepackage{mathpartir}
\usepackage[inline]{enumitem}
\usepackage{tikz}
\usetikzlibrary{tikzmark,calc}
\usetikzlibrary{arrows.meta,positioning,decorations.pathmorphing,matrix}
\usepackage{bm}

\allowdisplaybreaks

\usetikzlibrary{
  arrows,
  cd,
  decorations.pathmorphing,
}
\tikzset{
  >=stealth',
  line around/.style={decoration={pre length=#1, post length=#1}},
  stage/.style={
      decorate,
      decoration={snake, segment length=0.8em, amplitude=0.4ex},
      line around=0.3333em,
    },
  stage/.value forbidden,
  Transition/.pic={\draw[<-, #1] (0, 0) to +(2em, 0);},
  Scope/.pic={\pic{Transition};},
  Stage/.pic={\pic{Transition={stage, decoration={post length=0.2em}}};},
}
\NewDocumentCommand\Scope{}{\tikz[baseline=-axis_height] \pic{Scope};\xspace}
\NewDocumentCommand\Stage{}{\tikz[baseline=-axis_height] \pic{Stage};\xspace}

\newtheorem{definition}{Definition}
\newtheorem{lemma}{Lemma}
\newtheorem{theorem}{Theorem}
\newtheorem{corollary}{Corollary}

\lstdefinelanguage{MetaML}{
morekeywords= [1]{
    let,in,fun,rec,if,then,else,run,ref,deref,true,false
  },
sensitive=true,
morecomment=[s]{(*}{*)},
morestring=[b]",
literate=
  {->}{$\to$}2
{c!}{!}1
{cg}{$\gamma$}1
{cg1}{$\gamma_1$}1
{cg2}{$\gamma_2$}1
{cg3}{$\gamma_3$}1
{cg4}{$\gamma_4$}1
{cg5}{$\gamma_5$}1
{cg6}{$\gamma_6$}1
{ch0}{$\delta_0$}1
{ch1}{$\delta_1$}1
{ch2}{$\delta_2$}1
{/!}{@!}2
{/g}{@$\gamma$}2
{/g1}{@$\gamma_1$}2
{/g2}{@$\gamma_2$}2
{/g3}{@$\gamma_3$}2
{/g4}{@$\gamma_4$}2
{/g5}{@$\gamma_5$}2
{/g6}{@$\gamma_6$}2
{/h0}{@$\delta_0$}2
{/h1}{@$\delta_1$}2
{/h2}{@$\delta_2$}2
{@@!}{{\,\color{DimGray}\textsuperscript{\textbf{!}}}}2
{@@g}{{\,\color{DimGray}\textsuperscript{$\bm{\gamma}$}}}2
{@@g1}{{\,\color{DimGray}\textsuperscript{$\bm{\gamma_1}$}}}2
{@@g2}{{\,\color{DimGray}\textsuperscript{$\bm{\gamma_2}$}}}2
{@@g3}{{\,\color{DimGray}\textsuperscript{$\bm{\gamma_3}$}}}2
{@@g4}{{\,\color{DimGray}\textsuperscript{$\bm{\gamma_4}$}}}2
{@@g5}{{\,\color{DimGray}\textsuperscript{$\bm{\gamma_5}$}}}2
{@@g6}{{\,\color{DimGray}\textsuperscript{$\bm{\gamma_6}$}}}2
{@@h0}{{\,\color{DimGray}\textsuperscript{$\bm{\delta_0}$}}}2
{@@h1}{{\,\color{DimGray}\textsuperscript{$\bm{\delta_1}$}}}2
{@@h2}{{\,\color{DimGray}\textsuperscript{$\bm{\delta_2}$}}}2
{@g1}{{\,\color{ClassifierNew}\textsuperscript{$\bm{\gamma_1}$}}}1
{@g2}{{\,\color{ClassifierNew}\textsuperscript{$\bm{\gamma_2}$}}}1
{@g3}{{\,\color{ClassifierNew}\textsuperscript{$\bm{\gamma_3}$}}}1
{@g4}{{\,\color{ClassifierNew}\textsuperscript{$\bm{\gamma_4}$}}}1
{@g5}{{\,\color{ClassifierNew}\textsuperscript{$\bm{\gamma_5}$}}}1
{@g6}{{\,\color{ClassifierNew}\textsuperscript{$\bm{\gamma_6}$}}}1
{@g7}{{\,\color{ClassifierNew}\textsuperscript{$\bm{\gamma_7}$}}}1
{@g8}{{\,\color{ClassifierNew}\textsuperscript{$\bm{\gamma_8}$}}}1
{@g9}{{\,\color{ClassifierNew}\textsuperscript{$\bm{\gamma_9}$}}}1
{@g10}{{\,\color{ClassifierNew}\textsuperscript{$\bm{\gamma_{10}}$}}}1
{@h0}{{\,\color{ClassifierNew}\textsuperscript{$\bm{\delta_0}$}}}1
{@h1}{{\,\color{ClassifierNew}\textsuperscript{$\bm{\delta_1}$}}}1
{@h2}{{\,\color{ClassifierNew}\textsuperscript{$\bm{\delta_2}$}}}1
{dg1}{{\color{ClassifierNew}$\bm{\gamma_1}$}}1
{dg2}{{\color{ClassifierNew}$\bm{\gamma_2}$}}1
{dg3}{{\color{ClassifierNew}$\bm{\gamma_3}$}}1
{dg4}{{\color{ClassifierNew}$\bm{\gamma_4}$}}1
{dh1}{{\color{ClassifierNew}$\bm{\delta_1}$}}1
{dh2}{{\color{ClassifierNew}$\bm{\delta_2}$}}1
{.~}{{{\bfseries\$}}}1
{>.}{{{\bfseries>}}}1
{.< }{{{\bfseries<}}}1
{*~}{{{\bfseries\$}}}1
{>*}{{{\bfseries>}}}1
{*< }{{{\bfseries<}}}1
{reft}{{ref}}3
{runt}{{run}}3
{:>}{{$\mathrel{:\succeq}$}}2
}

\lstdefinestyle{metaml}{
  language=MetaML,
  basicstyle=\ttfamily\small,
  keywordstyle=[1]{\color{MidnightBlue}\bfseries},
  commentstyle=\color{SeaGreen},
  stringstyle=\color{SaddleBrown},
  columns=fullflexible,
  keepspaces=true,
  showstringspaces=false,
  mathescape=true,
  frame=single,
  framerule=0.3pt,
  rulecolor=\color{LightGray},
  xleftmargin=0.6em,
  aboveskip=0.6\baselineskip,
  belowskip=0.6\baselineskip,
  breaklines=true,
  breakatwhitespace=false
}

\ebproofset{
  center=false,
  label separation=\labelsep,
  rule margin=\fboxsep,
  template={%
      \normalfont
      \setbox0\hbox{$\inserttext$}%
      \raisebox{\dp0-\dp\strutbox}{$\UseName{m@th}\inserttext$}%
    },
}
\AddToHook{env/prooftree/begin}{\let\&=&}

\newcounter{rule}

\letcs\NewEbproofStatement{__ebproof_new_statement:nnn}
\newdimen\treewidth

\ebproofnewstyle{caption}{
  proof style=downwards,
  template={\makebox[\treewidth][l]{\normalfont\small\inserttext}\mathstrut},
  rule margin=\smallskipamount,
  rule style=no rule,
}
\NewEbproofStatement{caption}{>{\TrimSpaces}m e\label}{
\AddToHookNext{env/prooftree/end}{
\GetTitleStringExpand{#1}
\global\cslet{@currentlabel}\GetTitleStringResult
\global\cslet{@currentlabelname}\GetTitleStringResult
\rewrite{\global\treewidth\wd\treebox\copy\treebox}
\infer[caption]1{\IfValueT{#2}{\stepcounter{rule}\UseName{ltx@label}{#2}}#1}
}
}

\UseName{zref@setdefault}{\textcolor{red}{\textbf{??}}}
\ztitlerefsetup{
  expand=true,
  cleanup={\let\ref\zref},
}
\zcsetup{
  hyperref,
  nameinlink=false,
}
\zcRefTypeSetup{item}{
  Name-sg=,
  name-sg=,
  Name-pl=,
  name-pl=,
  refbounds={, \lparen, \rparen, },
}
\zcRefTypeSetup{rule}{
  Name-sg=Rule,
  name-sg=rule,
  Name-pl=Rules,
  name-pl=rules,
  refbounds={, \lparen, \rparen, },
}
\NewDocumentCommand\Zcref{}{\zcref[cap]}

\DeclareExpandableDocumentCommand\MathparLineskip{}
{\deflength{\lineskiplimit}{1\abovedisplayskip+\jot}\relax
  \deflength{\lineskip}{\abovedisplayskip+\jot}}
\ExpandArgs{c}
\apptocmd\MathparBindings{\UseName{@restorepar}}{}{}

\AtBeginDocument{%
  }

\begin{document}
\newcommand{\lamcirc}{\ensuremath{\lambda^{\bigcirc}}\xspace}
\newcommand{\lambox}{\ensuremath{\lambda^{\Box}}\xspace}
\newcommand{\lamfb}{\ensuremath{\lambda_{\forall[]}}}
\newcommand{\njcalc}{\ensuremath{\langle NJ\rangle}\xspace}
\newcommand{\lambra}{\ensuremath{\lambda^{[]} }}

\newcommand{\bmtt}{\text{BMTT}\xspace}
\newcommand{\lamgamma}{\ensuremath{\lambda^{\forall\gamma}}\xspace}
\newcommand{\lamgammacs}{\ensuremath{\lambda^{\forall\gamma *}}\xspace}
\newcommand{\minimlbmtt}{\ensuremath{\text{MiniML}^{\forall\gamma *}}\xspace}

\newcommand{\pluseq}{\mathrel{+}=}

\newcommand{\todo}[1]{\textcolor{red}{[TODO: #1]}}

\NewDocumentCommand{\ctext}{m}{%
  \raisebox{0.2ex}{\textcircled{\footnotesize #1}}%
}

\colorlet{ClassifierNew}{FireBrick}

\newcounter{inferencerule}
\providecommand*{\theHinferencerule}{\arabic{inferencerule}}

\makeatletter
\NewDocumentCommand{\myinferrule}{o m m}{%
  \IfNoValueTF{#1}{%
    \inferrule{#2}{#3}%
  }{%
    \refstepcounter{inferencerule}%
    \protected@edef\@currentlabel{#1}%
    \inferrule[#1]{#2}{#3}%
  }%
}
\makeatother

\newcommand{\refenvdrule}[4][]{{\displaystyle\frac{\begin{array}{l}#2\end{array}}{#3}\quad\refenvdrulename{#4}}}
\newcommand{\refenvusedrule}[1]{\[#1\]}
\newcommand{\refenvpremise}[1]{ #1 \\}
\newenvironment{refenvdefnblock}[3][]{ \framebox{\mbox{#2}} \quad #3 \\[0pt]}{}
\newenvironment{refenvfundefnblock}[3][]{ \framebox{\mbox{#2}} \quad #3 \\[0pt]\begin{displaymath}\begin{array}{l}}{\end{array}\end{displaymath}}
\newcommand{\refenvfunclause}[2]{ #1 \equiv #2 \\}
\newcommand{\refenvnt}[1]{\mathit{#1}}
\newcommand{\refenvmv}[1]{\mathit{#1}}
\newcommand{\refenvkw}[1]{\mathbf{#1}}
\newcommand{\refenvsym}[1]{#1}
\newcommand{\refenvcom}[1]{\text{#1}}
\newcommand{\refenvdrulename}[1]{\textsc{#1}}
\newcommand{\refenvcomplu}[5]{\overline{#1}^{\,#2\in #3 #4 #5}}
\newcommand{\refenvcompu}[3]{\overline{#1}^{\,#2<#3}}
\newcommand{\refenvcomp}[2]{\overline{#1}^{\,#2}}
\newcommand{\refenvgrammartabular}[1]{\begin{supertabular}{llcllllll}#1\end{supertabular}}
\newcommand{\refenvmetavartabular}[1]{\begin{supertabular}{ll}#1\end{supertabular}}
\newcommand{\refenvrulehead}[3]{$#1$ & & $#2$ & & & \multicolumn{2}{l}{#3}}
\newcommand{\refenvprodline}[6]{& & $#1$ & $#2$ & $#3 #4$ & $#5$ & $#6$}
\newcommand{\refenvfirstprodline}[6]{\refenvprodline{#1}{#2}{#3}{#4}{#5}{#6}}
\newcommand{\refenvlongprodline}[2]{& & $#1$ & \multicolumn{4}{l}{$#2$}}
\newcommand{\refenvfirstlongprodline}[2]{\refenvlongprodline{#1}{#2}}
\newcommand{\refenvbindspecprodline}[6]{\refenvprodline{#1}{#2}{#3}{#4}{#5}{#6}}
\newcommand{\refenvprodnewline}{\\}
\newcommand{\refenvinterrule}{\\[5.0mm]}
\newcommand{\refenvafterlastrule}{\\}
\newcommand\token[1]{%
  \mathbf{#1}%
}
\newcommand{\nominalbinder}{
  \mathord{\reflectbox{\sf N} }
}
\newcommand{\binder}[1]{
  \textcolor{ClassifierNew}{#1}
}
\newcommand{\pos}[1]{
  \textcolor{DodgerBlue}{#1}
}
\NewDocumentCommand\exclam{}{\mathord{\boldsymbol{!} } }

\newcommand{\refenvmetavars}{
\refenvmetavartabular{
 $ \refenvmv{x} ,\, \refenvmv{y} ,\, \refenvmv{z} ,\, \refenvmv{f} ,\, g $ & \refenvcom{variables} \\
 $ \refenvmv{l} $ & \refenvcom{locations} \\
 $ \refenvmv{p} ,\, \refenvmv{q} $ & \refenvcom{savepoints} \\
 $ \refenvmv{natliteral} $ & \refenvcom{meta variables for natural numbers} \\
 $ \gamma ,\, \delta $ & \refenvcom{classifiers} \\
}}

\newcommand{\refenvi}{
\refenvrulehead{\refenvnt{i}  ,\ \refenvnt{j}  ,\ \refenvnt{k}}{::=}{}\refenvprodnewline
\refenvfirstprodline{|}{\refenvmv{natliteral}}{}{}{}{}\refenvprodnewline
\refenvprodline{|}{\refenvnt{k_{{\mathrm{1}}}}  \refenvsym{+}  \refenvnt{k_{{\mathrm{2}}}}}{}{}{}{}\refenvprodnewline
\refenvprodline{|}{\refenvnt{k_{{\mathrm{1}}}}  \refenvsym{-}  \refenvnt{k_{{\mathrm{2}}}}}{}{}{}{}\refenvprodnewline
\refenvprodline{|}{ \refenvnt{k} ^{+} }{}{}{}{}\refenvprodnewline
\refenvprodline{|}{ \refenvnt{k} }{}{}{}{}\refenvprodnewline
\refenvprodline{|}{ \token{len} ( \refenvnt{clsseq} ) }{}{}{}{}\refenvprodnewline
\refenvprodline{|}{ \token{lv}( \Gamma ) }{}{}{}{}\refenvprodnewline
\refenvprodline{|}{ \token{len} ( \Psi , \Gamma ) }{}{}{}{}\refenvprodnewline
\refenvprodline{|}{ \token{deg} ( \Psi , \Gamma , \refenvnt{cls} ) }{}{}{}{}}

\newcommand{\refenvn}{
\refenvrulehead{\refenvnt{n}}{::=}{}\refenvprodnewline
\refenvfirstprodline{|}{\refenvmv{natliteral}}{}{}{}{}}

\newcommand{\refenvM}{
\refenvrulehead{t  ,\ \refenvnt{e}  ,\ \refenvnt{v}  ,\ {e_{I} }}{::=}{\refenvcom{terms, which can be values or expressions}}\refenvprodnewline
\refenvfirstprodline{|}{ t ^{ \refenvnt{k}  } }{}{}{}{}\refenvprodnewline
\refenvprodline{|}{\refenvnt{n}}{}{}{}{}\refenvprodnewline
\refenvprodline{|}{\refenvmv{x}}{}{}{}{}\refenvprodnewline
\refenvprodline{|}{ \lambda \binder{ \refenvmv{x} }.\  t }{}{}{}{}\refenvprodnewline
\refenvprodline{|}{ t_{{\mathrm{1}}} t_{{\mathrm{2}}} }{}{}{}{}\refenvprodnewline
\refenvprodline{|}{ \langle  t  \rangle }{}{}{}{}\refenvprodnewline
\refenvprodline{|}{ { \mathdollar }_{ \refenvnt{k} }\lbrace t \rbrace }{}{}{}{}\refenvprodnewline
\refenvprodline{|}{ \token{rec}\ \binder{ \refenvmv{f} }(\binder{ \refenvmv{x} }).\  t }{}{}{}{}\refenvprodnewline
\refenvprodline{|}{\token{ref} \, \refenvsym{(}  t  \refenvsym{)}}{}{}{}{}\refenvprodnewline
\refenvprodline{|}{ t_{{\mathrm{1}}} \coloneqq t_{{\mathrm{2}}} }{}{}{}{}\refenvprodnewline
\refenvprodline{|}{ \token{deref}( t ) }{}{}{}{}\refenvprodnewline
\refenvprodline{|}{\token{clos} \, \refenvsym{(}  E  \refenvsym{,}  R  \refenvsym{,}  t  \refenvsym{)}}{}{}{}{}\refenvprodnewline
\refenvprodline{|}{\refenvmv{l}}{}{}{}{}\refenvprodnewline
\refenvprodline{|}{\token{rename} \, \refenvsym{(}  R  \refenvsym{,}  \refenvmv{x}  \refenvsym{)}}{}{}{}{}\refenvprodnewline
\refenvprodline{|}{\refenvsym{(}  t  \refenvsym{)}}{}{}{}{}\refenvprodnewline
\refenvprodline{|}{ t }{}{}{}{}\refenvprodnewline
\refenvprodline{|}{ { t _{ \refenvnt{k} } } }{}{}{}{}\refenvprodnewline
\refenvprodline{|}{ \ldots }{}{}{}{}}

\newcommand{\refenvVSet}{
\refenvrulehead{\refenvnt{VSet}}{::=}{\refenvcom{variable sets}}\refenvprodnewline
\refenvfirstprodline{|}{\emptyset}{}{}{}{}\refenvprodnewline
\refenvprodline{|}{\refenvmv{x}}{}{}{}{}\refenvprodnewline
\refenvprodline{|}{ \overrightarrow{ \refenvmv{x} } }{}{}{}{}\refenvprodnewline
\refenvprodline{|}{\refenvnt{VSet_{{\mathrm{1}}}}  \refenvsym{,}  \refenvnt{VSet_{{\mathrm{2}}}}}{}{}{}{}\refenvprodnewline
\refenvprodline{|}{ \token{FV}( t ) }{}{}{}{}\refenvprodnewline
\refenvprodline{|}{\refenvnt{VSet_{{\mathrm{1}}}} \, \cap \, \refenvnt{VSet_{{\mathrm{2}}}}}{}{}{}{}\refenvprodnewline
\refenvprodline{|}{\refenvnt{VSet_{{\mathrm{1}}}} \, \cup \, \refenvnt{VSet_{{\mathrm{2}}}}}{}{}{}{}\refenvprodnewline
\refenvprodline{|}{ \token{Dom}_{V}( \Gamma ) }{}{}{}{}\refenvprodnewline
\refenvprodline{|}{ \token{Dom}_{V}( \Psi ) }{}{}{}{}\refenvprodnewline
\refenvprodline{|}{\refenvsym{(}  \refenvnt{VSet}  \refenvsym{)}}{}{}{}{}\refenvprodnewline
\refenvprodline{|}{ \refenvnt{VSet} }{}{}{}{}\refenvprodnewline
\refenvprodline{|}{ \token{BV}( E ) }{}{}{}{}\refenvprodnewline
\refenvprodline{|}{ \token{BV}( \refenvkw{et} ) }{}{}{}{}\refenvprodnewline
\refenvprodline{|}{N}{}{}{}{}\refenvprodnewline
\refenvprodline{|}{\refenvsym{\{}  \refenvmv{x}  \refenvsym{\}}}{}{}{}{}}

\newcommand{\refenvvenvElm}{
\refenvrulehead{\refenvnt{venvElm}}{::=}{}\refenvprodnewline
\refenvfirstprodline{|}{\refenvmv{x}  \coloneqq  t}{}{}{}{}}

\newcommand{\refenvvenv}{
\refenvrulehead{E}{::=}{}\refenvprodnewline
\refenvfirstprodline{|}{\varepsilon}{}{}{}{}\refenvprodnewline
\refenvprodline{|}{\refenvnt{venvElm}}{}{}{}{}\refenvprodnewline
\refenvprodline{|}{E_{{\mathrm{1}}}  \refenvsym{,}  E_{{\mathrm{2}}}}{}{}{}{}\refenvprodnewline
\refenvprodline{|}{\refenvsym{(}  E  \refenvsym{)}}{}{}{}{}}

\newcommand{\refenvrenv}{
\refenvrulehead{R}{::=}{}\refenvprodnewline
\refenvfirstprodline{|}{ \token{id} }{}{}{}{}\refenvprodnewline
\refenvprodline{|}{\refenvmv{x}  \coloneqq  \refenvmv{y}}{}{}{}{}\refenvprodnewline
\refenvprodline{|}{R_{{\mathrm{1}}}  \refenvsym{,}  R_{{\mathrm{2}}}}{}{}{}{}\refenvprodnewline
\refenvprodline{|}{\refenvsym{(}  R  \refenvsym{)}}{}{}{}{}}

\newcommand{\refenvnh}{
\refenvrulehead{N}{::=}{}\refenvprodnewline
\refenvfirstprodline{|}{\varepsilon}{}{}{}{}\refenvprodnewline
\refenvprodline{|}{\refenvmv{x}}{}{}{}{}\refenvprodnewline
\refenvprodline{|}{N_{{\mathrm{1}}}  \refenvsym{,}  N_{{\mathrm{2}}}}{}{}{}{}\refenvprodnewline
\refenvprodline{|}{\refenvsym{(}  N  \refenvsym{)}}{}{}{}{}\refenvprodnewline
\refenvprodline{|}{ \token{Dom}_{V}( \Psi ) }{}{}{}{}}

\newcommand{\refenvloc}{
\refenvrulehead{\refenvnt{loc}}{::=}{}\refenvprodnewline
\refenvfirstprodline{|}{\refenvmv{l}}{}{}{}{}\refenvprodnewline
\refenvprodline{|}{ { \refenvnt{loc} _{ \refenvnt{k} } } }{}{}{}{}}

\newcommand{\refenvstoreElm}{
\refenvrulehead{\refenvnt{storeElm}}{::=}{}\refenvprodnewline
\refenvfirstprodline{|}{\refenvmv{l}  \coloneqq  t}{}{}{}{}}

\newcommand{\refenvstore}{
\refenvrulehead{S}{::=}{}\refenvprodnewline
\refenvfirstprodline{|}{\varepsilon}{}{}{}{}\refenvprodnewline
\refenvprodline{|}{\refenvnt{storeElm}}{}{}{}{}\refenvprodnewline
\refenvprodline{|}{S_{{\mathrm{1}}}  \refenvsym{,}  S_{{\mathrm{2}}}}{}{}{}{}\refenvprodnewline
\refenvprodline{|}{\refenvsym{(}  S  \refenvsym{)}}{}{}{}{}\refenvprodnewline
\refenvprodline{|}{ \refenvnt{loc_{{\mathrm{1}}}} \coloneqq t_{{\mathrm{1}}} ,\dotsc, \refenvnt{loc_{{\mathrm{2}}}} \coloneqq t_{{\mathrm{2}}} }{}{}{}{}}

\newcommand{\refenvcls}{
\refenvrulehead{\refenvnt{cls}}{::=}{}\refenvprodnewline
\refenvfirstprodline{|}{ \exclam }{}{}{}{}\refenvprodnewline
\refenvprodline{|}{\gamma}{}{}{}{}\refenvprodnewline
\refenvprodline{|}{ \refenvnt{cls} _{ \refenvnt{k} } }{}{}{}{}\refenvprodnewline
\refenvprodline{|}{\refenvnt{cls}  \refenvsym{[}  \sigma  \refenvsym{]}}{}{}{}{}\refenvprodnewline
\refenvprodline{|}{ \refenvnt{cls} }{}{}{}{}}

\newcommand{\refenvclsseq}{
\refenvrulehead{\refenvnt{clsseq}}{::=}{}\refenvprodnewline
\refenvfirstprodline{|}{ \overrightarrow{ \gamma } }{}{}{}{}\refenvprodnewline
\refenvprodline{|}{ \overrightarrow{ \gamma }^{+} }{}{}{}{}\refenvprodnewline
\refenvprodline{|}{\varepsilon}{}{}{}{}\refenvprodnewline
\refenvprodline{|}{\refenvnt{cls}}{}{}{}{}\refenvprodnewline
\refenvprodline{|}{\refenvnt{clsseq_{{\mathrm{1}}}}  \refenvsym{,}  \refenvnt{clsseq_{{\mathrm{2}}}}}{}{}{}{}\refenvprodnewline
\refenvprodline{|}{ \refenvnt{cls_{{\mathrm{1}}}} ,\dotsc, \refenvnt{cls_{{\mathrm{2}}}} }{}{}{}{}\refenvprodnewline
\refenvprodline{|}{ \dotsc }{}{}{}{}\refenvprodnewline
\refenvprodline{|}{\refenvnt{clsseq}  \refenvsym{[}  \sigma  \refenvsym{]}}{}{}{}{}\refenvprodnewline
\refenvprodline{|}{\refenvsym{(}  \refenvnt{clsseq}  \refenvsym{)}}{}{}{}{}\refenvprodnewline
\refenvprodline{|}{ \refenvnt{clsseq} }{}{}{}{}}

\newcommand{\refenvbasetype}{
\refenvrulehead{\iota}{::=}{}\refenvprodnewline
\refenvfirstprodline{|}{ \texttt{int} }{}{}{}{}\refenvprodnewline
\refenvprodline{|}{ \texttt{str} }{}{}{}{}}

\newcommand{\refenvpolyclsdeclElm}{
\refenvrulehead{\refenvnt{polyclsdeclElm}}{::=}{}\refenvprodnewline
\refenvfirstprodline{|}{ \binder{ \refenvnt{cls_{{\mathrm{1}}}} }  \mathbin{:\succeq}   \refenvnt{cls_{{\mathrm{2}}}} }{}{}{}{}}

\newcommand{\refenvpolyclsdecl}{
\refenvrulehead{\varrho}{::=}{}\refenvprodnewline
\refenvfirstprodline{|}{\varepsilon}{}{}{}{}\refenvprodnewline
\refenvprodline{|}{\refenvnt{polyclsdeclElm}}{}{}{}{}\refenvprodnewline
\refenvprodline{|}{\varrho_{{\mathrm{1}}}  \refenvsym{,}  \varrho_{{\mathrm{2}}}}{}{}{}{}\refenvprodnewline
\refenvprodline{|}{ \overrightarrow{\binder{ \refenvnt{cls_{{\mathrm{1}}}} }  \mathbin{:\succeq}   \refenvnt{cls_{{\mathrm{2}}}} } }{}{}{}{}\refenvprodnewline
\refenvprodline{|}{ \binder{ \refenvnt{cls_{{\mathrm{1}}}} }  \mathbin{:\succeq}   \refenvnt{cls_{{\mathrm{2}}}} ,\dotsc,\binder{ \refenvnt{cls_{{\mathrm{3}}}} }  \mathbin{:\succeq}   \refenvnt{cls_{{\mathrm{4}}}} }{}{}{}{}\refenvprodnewline
\refenvprodline{|}{ \varrho _{ \refenvnt{k} } }{}{}{}{}\refenvprodnewline
\refenvprodline{|}{\varrho  \refenvsym{[}  \sigma  \refenvsym{]}}{}{}{}{}\refenvprodnewline
\refenvprodline{|}{\refenvsym{(}  \varrho  \refenvsym{)}}{}{}{}{}}

\newcommand{\refenvA}{
\refenvrulehead{\refenvnt{A}  ,\ \refenvnt{B}}{::=}{}\refenvprodnewline
\refenvfirstprodline{|}{\iota}{}{}{}{}\refenvprodnewline
\refenvprodline{|}{\refenvnt{A}  \rightarrow  \refenvnt{B}}{}{}{}{}\refenvprodnewline
\refenvprodline{|}{ \langle \refenvnt{A} \rangle^{ \refenvnt{cls} } }{}{}{}{}\refenvprodnewline
\refenvprodline{|}{ \refenvnt{A} \/\ \token{ref} }{}{}{}{}\refenvprodnewline
\refenvprodline{|}{\refenvsym{(}  \refenvnt{A}  \refenvsym{)}}{}{}{}{}\refenvprodnewline
\refenvprodline{|}{\refenvnt{A}  \refenvsym{[}  \sigma  \refenvsym{]}}{}{}{}{}\refenvprodnewline
\refenvprodline{|}{ { \refenvnt{A} _{ \refenvnt{k} } } }{}{}{}{}\refenvprodnewline
\refenvprodline{|}{ \forall \binder{ \refenvnt{cls_{{\mathrm{1}}}} }_{ \mathbin{:\succeq} \refenvnt{cls_{{\mathrm{2}}}} }. \refenvnt{A} }{}{}{}{}\refenvprodnewline
\refenvprodline{|}{ \forall  \varrho . \refenvnt{A} }{}{}{}{}\refenvprodnewline
\refenvprodline{|}{ \forall  \varrho_{{\mathrm{1}}} \dotsc\forall \varrho_{{\mathrm{2}}} . \refenvnt{A} }{}{}{}{}\refenvprodnewline
\refenvprodline{|}{ \llbracket  A^{\circ}  \rrbracket^{ \refenvnt{clsseq} } }{}{}{}{}}

\newcommand{\refenvclssubsts}{
\refenvrulehead{\sigma}{::=}{}\refenvprodnewline
\refenvfirstprodline{|}{\refenvnt{cls_{{\mathrm{1}}}}  \coloneqq  \refenvnt{cls_{{\mathrm{2}}}}}{}{}{}{}\refenvprodnewline
\refenvprodline{|}{ \overrightarrow{ \refenvnt{cls_{{\mathrm{1}}}} \coloneqq \refenvnt{cls_{{\mathrm{2}}}} } }{}{}{}{}\refenvprodnewline
\refenvprodline{|}{\sigma_{{\mathrm{1}}}  \refenvsym{,}  \sigma_{{\mathrm{2}}}}{}{}{}{}\refenvprodnewline
\refenvprodline{|}{ \refenvnt{cls_{{\mathrm{1}}}} \coloneqq \refenvnt{cls_{{\mathrm{2}}}} ,\dotsc, \refenvnt{cls_{{\mathrm{3}}}} \coloneqq \refenvnt{cls_{{\mathrm{4}}}} }{}{}{}{}}

\newcommand{\refenvGElm}{
\refenvrulehead{\refenvnt{GElm}}{::=}{}\refenvprodnewline
\refenvfirstprodline{|}{ \binder{ \refenvmv{x} }\mathord{:^{\binder{ \refenvnt{cls} } } } \refenvnt{A} }{}{}{}{}\refenvprodnewline
\refenvprodline{|}{ \blacktriangleright ^{ \refenvnt{cls} } }{}{}{}{}\refenvprodnewline
\refenvprodline{|}{ \blacktriangleleft _{ \refenvnt{k} }^{\binder{ \refenvnt{cls} } } }{}{}{}{}\refenvprodnewline
\refenvprodline{|}{ [  \varrho  ]^{\binder{ \refenvnt{cls} } } }{}{}{}{}}

\newcommand{\refenvG}{
\refenvrulehead{\Gamma}{::=}{}\refenvprodnewline
\refenvfirstprodline{|}{\varepsilon}{}{}{}{}\refenvprodnewline
\refenvprodline{|}{\refenvnt{GElm}}{}{}{}{}\refenvprodnewline
\refenvprodline{|}{ \overrightarrow{\binder{ \refenvnt{cls_{{\mathrm{1}}}} } \mathbin{:\succeq} \refenvnt{cls_{{\mathrm{2}}}} } }{}{}{}{}\refenvprodnewline
\refenvprodline{|}{\Gamma_{{\mathrm{1}}}  \refenvsym{,}  \Gamma_{{\mathrm{2}}}}{}{}{}{}\refenvprodnewline
\refenvprodline{|}{\Gamma  \refenvsym{[}  \sigma  \refenvsym{]}}{}{}{}{}\refenvprodnewline
\refenvprodline{|}{\refenvsym{(}  \Gamma  \refenvsym{)}}{}{}{}{}\refenvprodnewline
\refenvprodline{|}{\refenvkw{pop} \, \refenvsym{(}  \Gamma  \refenvsym{,}  \refenvnt{k}  \refenvsym{)}}{}{}{}{}\refenvprodnewline
\refenvprodline{|}{ \refenvnt{GElm_{{\mathrm{1}}}} ,\dotsc, \refenvnt{GElm_{{\mathrm{2}}}} }{}{}{}{}\refenvprodnewline
\refenvprodline{|}{ \Gamma ^{ \refenvnt{k_{{\mathrm{1}}}} , \refenvnt{k_{{\mathrm{2}}}} } }{}{}{}{}\refenvprodnewline
\refenvprodline{|}{ \Gamma ^{\pos{ \refenvnt{cls} } } }{}{}{}{}}

\newcommand{\refenvsp}{
\refenvrulehead{\refenvnt{sp}}{::=}{}\refenvprodnewline
\refenvfirstprodline{|}{\refenvmv{p}}{}{}{}{}\refenvprodnewline
\refenvprodline{|}{ \refenvmv{p} _{ \refenvnt{k} } }{}{}{}{}}

\newcommand{\refenvnhtElm}{
\refenvrulehead{\refenvnt{nhtElm}}{::=}{}\refenvprodnewline
\refenvfirstprodline{|}{ \binder{ \refenvmv{x} }: \refenvnt{A} \mathrel{@}\binder{ \refenvnt{cls_{{\mathrm{1}}}} } \mathbin{:\succeq} \refenvnt{cls_{{\mathrm{2}}}} }{}{}{}{}\refenvprodnewline
\refenvprodline{|}{ \binder{ \refenvnt{cls_{{\mathrm{1}}}} }( \mathbin{:\succeq} \; \refenvnt{cls_{{\mathrm{2}}}}  /  \mathbin{:\supseteq} \; \refenvnt{cls_{{\mathrm{3}}}} ) }{}{}{}{}\refenvprodnewline
\refenvprodline{|}{ [  \varrho  ] }{}{}{}{}\refenvprodnewline
\refenvprodline{|}{ [  \varrho_{{\mathrm{1}}}  ],\dotsc,[  \varrho_{{\mathrm{2}}}  ] }{}{}{}{}}

\newcommand{\refenvnht}{
\refenvrulehead{\Psi}{::=}{}\refenvprodnewline
\refenvfirstprodline{|}{\varepsilon}{}{}{}{}\refenvprodnewline
\refenvprodline{|}{\refenvnt{nhtElm}}{}{}{}{}\refenvprodnewline
\refenvprodline{|}{\Psi_{{\mathrm{1}}}  \refenvsym{,}  \Psi_{{\mathrm{2}}}}{}{}{}{}\refenvprodnewline
\refenvprodline{|}{ \Psi ^{ \refenvnt{sp} } }{}{}{}{}\refenvprodnewline
\refenvprodline{|}{\refenvsym{(}  \Psi  \refenvsym{)}}{}{}{}{}}

\newcommand{\refenvPSet}{
\refenvrulehead{\refenvnt{PSet}}{::=}{}\refenvprodnewline
\refenvfirstprodline{|}{\emptyset}{}{}{}{}\refenvprodnewline
\refenvprodline{|}{ \token{Dom}_{SP}( \Psi ) }{}{}{}{}}

\newcommand{\refenvLGElm}{
\refenvrulehead{\refenvnt{LGElm}}{::=}{}\refenvprodnewline
\refenvfirstprodline{|}{ \refenvnt{loc} : \refenvnt{A} }{}{}{}{}}

\newcommand{\refenvLG}{
\refenvrulehead{\Theta}{::=}{}\refenvprodnewline
\refenvfirstprodline{|}{\varepsilon}{}{}{}{}\refenvprodnewline
\refenvprodline{|}{\refenvnt{LGElm}}{}{}{}{}\refenvprodnewline
\refenvprodline{|}{\Theta_{{\mathrm{1}}}  \refenvsym{,}  \Theta_{{\mathrm{2}}}}{}{}{}{}\refenvprodnewline
\refenvprodline{|}{\refenvsym{(}  \Theta  \refenvsym{)}}{}{}{}{}\refenvprodnewline
\refenvprodline{|}{\Theta  \refenvsym{[}  \refenvnt{cls_{{\mathrm{1}}}}  \coloneqq  \refenvnt{cls_{{\mathrm{2}}}}  \refenvsym{]}}{}{}{}{}\refenvprodnewline
\refenvprodline{|}{ \refenvnt{loc_{{\mathrm{1}}}} : \refenvnt{A_{{\mathrm{1}}}} ,\dotsc, \refenvnt{loc_{{\mathrm{2}}}} : \refenvnt{A_{{\mathrm{2}}}} }{}{}{}{}}

\newcommand{\refenvCSet}{
\refenvrulehead{\refenvnt{CSet}}{::=}{\refenvcom{classifier sets}}\refenvprodnewline
\refenvfirstprodline{|}{\emptyset}{}{}{}{}\refenvprodnewline
\refenvprodline{|}{\refenvsym{\{}  \refenvnt{clsseq}  \refenvsym{\}}}{}{}{}{}\refenvprodnewline
\refenvprodline{|}{\refenvnt{CSet_{{\mathrm{1}}}}  \refenvsym{,}  \refenvnt{CSet_{{\mathrm{2}}}}}{}{}{}{}\refenvprodnewline
\refenvprodline{|}{ \token{FC}( \refenvnt{A} ) }{}{}{}{}\refenvprodnewline
\refenvprodline{|}{ \token{FC}( \varrho ) }{}{}{}{}\refenvprodnewline
\refenvprodline{|}{ \token{FC}( \Gamma ) }{}{}{}{}\refenvprodnewline
\refenvprodline{|}{\refenvnt{CSet_{{\mathrm{1}}}} \, \cap \, \refenvnt{CSet_{{\mathrm{2}}}}}{}{}{}{}\refenvprodnewline
\refenvprodline{|}{\refenvnt{CSet_{{\mathrm{1}}}} \, \cup \, \refenvnt{CSet_{{\mathrm{2}}}}}{}{}{}{}\refenvprodnewline
\refenvprodline{|}{ \refenvnt{CSet_{{\mathrm{1}}}} \cup\dotsb\cup \refenvnt{CSet_{{\mathrm{2}}}} }{}{}{}{}\refenvprodnewline
\refenvprodline{|}{\refenvnt{CSet_{{\mathrm{1}}}}  \refenvsym{-}  \refenvnt{CSet_{{\mathrm{2}}}}}{}{}{}{}\refenvprodnewline
\refenvprodline{|}{ \token{Dom}_{C}( \Gamma ) }{}{}{}{}\refenvprodnewline
\refenvprodline{|}{ \token{Dom}_{C}( \varrho ) }{}{}{}{}\refenvprodnewline
\refenvprodline{|}{ \token{Dom}_{C}( \Psi ) }{}{}{}{}\refenvprodnewline
\refenvprodline{|}{ \token{Dom}_{C}( \sigma ) }{}{}{}{}\refenvprodnewline
\refenvprodline{|}{ \token{Reg}_{C}( \sigma ) }{}{}{}{}\refenvprodnewline
\refenvprodline{|}{\refenvsym{(}  \refenvnt{CSet}  \refenvsym{)}}{}{}{}{}\refenvprodnewline
\refenvprodline{|}{ \refenvnt{CSet} }{}{}{}{}}

\newcommand{\refenvLSet}{
\refenvrulehead{\refenvnt{LSet}}{::=}{\refenvcom{location sets}}\refenvprodnewline
\refenvfirstprodline{|}{\emptyset}{}{}{}{}\refenvprodnewline
\refenvprodline{|}{ \token{Dom}_{L}( \Theta ) }{}{}{}{}\refenvprodnewline
\refenvprodline{|}{ \token{Dom}_{L}( S ) }{}{}{}{}\refenvprodnewline
\refenvprodline{|}{\refenvnt{LSet_{{\mathrm{1}}}} \, \cup \, \refenvnt{LSet_{{\mathrm{2}}}}}{}{}{}{}\refenvprodnewline
\refenvprodline{|}{\refenvsym{\{}  \refenvmv{l}  \refenvsym{\}}}{}{}{}{}}

\newcommand{\refenvAc}{
\refenvrulehead{A^{\circ}  ,\ B^{\circ}}{::=}{\refenvcom{lambda-circle types}}\refenvprodnewline
\refenvfirstprodline{|}{\iota}{}{}{}{}\refenvprodnewline
\refenvprodline{|}{A^{\circ}  \rightarrow  B^{\circ}}{}{}{}{}\refenvprodnewline
\refenvprodline{|}{ \langle  A^{\circ}  \rangle }{}{}{}{}}

\newcommand{\refenvterminals}{
\refenvrulehead{\refenvnt{terminals}}{::=}{}\refenvprodnewline
\refenvfirstprodline{|}{ \rightarrow }{}{}{}{}\refenvprodnewline
\refenvprodline{|}{ \colon }{}{}{}{}\refenvprodnewline
\refenvprodline{|}{ \coloneqq }{}{}{}{}\refenvprodnewline
\refenvprodline{|}{ \in }{}{}{}{}\refenvprodnewline
\refenvprodline{|}{ \vdash }{}{}{}{}\refenvprodnewline
\refenvprodline{|}{ \longrightarrow }{}{}{}{}\refenvprodnewline
\refenvprodline{|}{ \cap }{}{}{}{}\refenvprodnewline
\refenvprodline{|}{ \cup }{}{}{}{}\refenvprodnewline
\refenvprodline{|}{ \not\in }{}{}{}{}\refenvprodnewline
\refenvprodline{|}{ \subseteq }{}{}{}{}\refenvprodnewline
\refenvprodline{|}{ & }{}{}{}{}\refenvprodnewline
\refenvprodline{|}{ \token{ctx} }{}{}{}{}\refenvprodnewline
\refenvprodline{|}{ \token{wf} }{}{}{}{}\refenvprodnewline
\refenvprodline{|}{ \varepsilon }{}{}{}{}\refenvprodnewline
\refenvprodline{|}{ \square }{}{}{}{}\refenvprodnewline
\refenvprodline{|}{ \mathdollar }{}{}{}{}\refenvprodnewline
\refenvprodline{|}{ \langle }{}{}{}{}\refenvprodnewline
\refenvprodline{|}{ \rangle }{}{}{}{}\refenvprodnewline
\refenvprodline{|}{ \token{ref} }{}{}{}{}\refenvprodnewline
\refenvprodline{|}{ \token{set} }{}{}{}{}\refenvprodnewline
\refenvprodline{|}{ \token{get} }{}{}{}{}\refenvprodnewline
\refenvprodline{|}{ \mathcal{V}^{\mathrm{rt} } }{}{}{}{}\refenvprodnewline
\refenvprodline{|}{ \mathcal{V}^{\mathrm{fut} } }{}{}{}{}\refenvprodnewline
\refenvprodline{|}{ \mathcal{V}_{S} }{}{}{}{}\refenvprodnewline
\refenvprodline{|}{ \token{rename} }{}{}{}{}\refenvprodnewline
\refenvprodline{|}{ \token{clos} }{}{}{}{}\refenvprodnewline
\refenvprodline{|}{ \token{gensym} }{}{}{}{}\refenvprodnewline
\refenvprodline{|}{ \token{toClos} }{}{}{}{}\refenvprodnewline
\refenvprodline{|}{ \leftarrow }{}{}{}{}\refenvprodnewline
\refenvprodline{|}{ \token{return} }{}{}{}{}\refenvprodnewline
\refenvprodline{|}{ \token{lookup} }{}{}{}{}\refenvprodnewline
\refenvprodline{|}{ \token{lookupStore} }{}{}{}{}\refenvprodnewline
\refenvprodline{|}{ \token{updateStore} }{}{}{}{}\refenvprodnewline
\refenvprodline{|}{ \blacktriangleright }{}{}{}{}\refenvprodnewline
\refenvprodline{|}{ \blacktriangleleft }{}{}{}{}\refenvprodnewline
\refenvprodline{|}{ \preceq }{}{}{}{}\refenvprodnewline
\refenvprodline{|}{ \mathbin{:\succeq} }{}{}{}{}\refenvprodnewline
\refenvprodline{|}{ \mathbin{:\supseteq} }{}{}{}{}\refenvprodnewline
\refenvprodline{|}{\refenvsym{=}}{}{}{}{}\refenvprodnewline
\refenvprodline{|}{ \neq }{}{}{}{}\refenvprodnewline
\refenvprodline{|}{ \mathrel{@} }{}{}{}{}\refenvprodnewline
\refenvprodline{|}{ \emptyset }{}{}{}{}\refenvprodnewline
\refenvprodline{|}{ \token{Done} }{}{}{}{}\refenvprodnewline
\refenvprodline{|}{ \token{last} }{}{}{}{}\refenvprodnewline
\refenvprodline{|}{ \token{Result} }{}{}{}{}\refenvprodnewline
\refenvprodline{|}{ \token{alloc} }{}{}{}{}\refenvprodnewline
\refenvprodline{|}{ \token{toQuote} }{}{}{}{}\refenvprodnewline
\refenvprodline{|}{ \token{toLocation} }{}{}{}{}\refenvprodnewline
\refenvprodline{|}{ \token{len} }{}{}{}{}\refenvprodnewline
\refenvprodline{|}{ \token{deg} }{}{}{}{}}

\newcommand{\refenvsets}{
\refenvrulehead{\refenvnt{sets}  ,\ \tau}{::=}{}\refenvprodnewline
\refenvfirstprodline{|}{ \token{Var} }{}{}{}{}\refenvprodnewline
\refenvprodline{|}{ \token{Term} }{}{}{}{}\refenvprodnewline
\refenvprodline{|}{ \token{Exp}^{ \refenvnt{k} } }{}{}{}{}\refenvprodnewline
\refenvprodline{|}{ \token{Ctx}^{ \refenvnt{k_{{\mathrm{1}}}} , \refenvnt{k_{{\mathrm{2}}}} } }{}{}{}{}\refenvprodnewline
\refenvprodline{|}{ \token{Val} }{}{}{}{}\refenvprodnewline
\refenvprodline{|}{ \token{VEnv} }{}{}{}{}\refenvprodnewline
\refenvprodline{|}{ \token{REnv} }{}{}{}{}\refenvprodnewline
\refenvprodline{|}{ \token{RN} }{}{}{}{}\refenvprodnewline
\refenvprodline{|}{ \token{Int} }{}{}{}{}\refenvprodnewline
\refenvprodline{|}{ \token{Loc} }{}{}{}{}\refenvprodnewline
\refenvprodline{|}{ \token{Store} }{}{}{}{}\refenvprodnewline
\refenvprodline{|}{ \token{Result}( \refenvnt{sets} ) }{}{}{}{}\refenvprodnewline
\refenvprodline{|}{ \token{Result}( \refenvnt{sets_{{\mathrm{1}}}} \times \refenvnt{sets_{{\mathrm{2}}}} \times \refenvnt{sets_{{\mathrm{3}}}} ) }{}{}{}{}}

\newcommand{\refenvcomputation}{
\refenvrulehead{\refenvnt{computation}  ,\ r}{::=}{}\refenvprodnewline
\refenvfirstprodline{|}{\token{return} \, \refenvsym{(}  t  \refenvsym{,}  N  \refenvsym{,}  S  \refenvsym{)}}{}{}{}{}\refenvprodnewline
\refenvprodline{|}{\token{Done} \, \refenvsym{(}  t  \refenvsym{,}  N  \refenvsym{,}  S  \refenvsym{)}}{}{}{}{}\refenvprodnewline
\refenvprodline{|}{ \token{Timeout} }{}{}{}{}\refenvprodnewline
\refenvprodline{|}{ \token{Fail} }{}{}{}{}\refenvprodnewline
\refenvprodline{|}{\mathcal{V}^{\mathrm{rt} }  \refenvsym{(}  t  \refenvsym{,}  E  \refenvsym{,}  R  \refenvsym{,}  N  \refenvsym{,}  S  \refenvsym{,}  \refenvnt{k_{{\mathrm{2}}}}  \refenvsym{)}}{}{}{}{}\refenvprodnewline
\refenvprodline{|}{\refenvsym{(}  t_{{\mathrm{1}}}  \refenvsym{,}  N_{{\mathrm{1}}}  \refenvsym{,}  S_{{\mathrm{1}}}  \refenvsym{)}  \leftarrow  \mathcal{V}^{\mathrm{rt} }  \refenvsym{(}  t_{{\mathrm{2}}}  \refenvsym{,}  E  \refenvsym{,}  R  \refenvsym{,}  N_{{\mathrm{2}}}  \refenvsym{,}  S_{{\mathrm{2}}}  \refenvsym{,}  \refenvnt{k_{{\mathrm{2}}}}  \refenvsym{)}}{}{}{}{}\refenvprodnewline
\refenvprodline{|}{\mathcal{V}^{\mathrm{fut} }  \refenvsym{(}  \refenvnt{k}  \refenvsym{,}  t  \refenvsym{,}  E  \refenvsym{,}  R  \refenvsym{,}  N  \refenvsym{,}  S  \refenvsym{,}  \refenvnt{k_{{\mathrm{2}}}}  \refenvsym{)}}{}{}{}{}\refenvprodnewline
\refenvprodline{|}{\refenvsym{(}  t_{{\mathrm{1}}}  \refenvsym{,}  N_{{\mathrm{1}}}  \refenvsym{,}  S_{{\mathrm{1}}}  \refenvsym{)}  \leftarrow  \mathcal{V}^{\mathrm{fut} }  \refenvsym{(}  \refenvnt{k}  \refenvsym{,}  t_{{\mathrm{2}}}  \refenvsym{,}  E  \refenvsym{,}  R  \refenvsym{,}  N_{{\mathrm{2}}}  \refenvsym{,}  S_{{\mathrm{2}}}  \refenvsym{,}  \refenvnt{k_{{\mathrm{2}}}}  \refenvsym{)}}{}{}{}{}\refenvprodnewline
\refenvprodline{|}{t_{{\mathrm{1}}}  \leftarrow \, \token{lookup} \, \refenvsym{(}  E  \refenvsym{,}  t_{{\mathrm{2}}}  \refenvsym{)}}{}{}{}{}\refenvprodnewline
\refenvprodline{|}{\refenvmv{x}  \leftarrow \, \token{gensym} \, \refenvsym{(}  N  \refenvsym{)}}{}{}{}{}\refenvprodnewline
\refenvprodline{|}{\refenvsym{(}  E  \refenvsym{,}  R  \refenvsym{,}  \refenvmv{x}  \refenvsym{,}  t_{{\mathrm{1}}}  \refenvsym{)}  \leftarrow \, \token{toClos} \, \refenvsym{(}  t_{{\mathrm{2}}}  \refenvsym{)}}{}{}{}{}\refenvprodnewline
\refenvprodline{|}{t_{{\mathrm{1}}}  \leftarrow \, \token{toQuote} \, \refenvsym{(}  t_{{\mathrm{2}}}  \refenvsym{)}}{}{}{}{}\refenvprodnewline
\refenvprodline{|}{\refenvmv{l}  \leftarrow \, \token{alloc} \, \refenvsym{(}  S  \refenvsym{)}}{}{}{}{}\refenvprodnewline
\refenvprodline{|}{t_{{\mathrm{1}}}  \leftarrow \, \token{toLocation} \, \refenvsym{(}  t_{{\mathrm{2}}}  \refenvsym{)}}{}{}{}{}\refenvprodnewline
\refenvprodline{|}{S_{{\mathrm{1}}}  \leftarrow \, \token{updateStore} \, \refenvsym{(}  S_{{\mathrm{2}}}  \refenvsym{,}  \refenvmv{l}  \refenvsym{,}  t  \refenvsym{)}}{}{}{}{}\refenvprodnewline
\refenvprodline{|}{t  \leftarrow  \token{lookupStore}  \refenvsym{(}  S  \refenvsym{,}  \refenvmv{l}  \refenvsym{)}}{}{}{}{}\refenvprodnewline
\refenvprodline{|}{ \token{match}\  t \ \token{with} }{}{}{}{}\refenvprodnewline
\refenvprodline{|}{ \token{case}\  t \ \Rightarrow }{}{}{}{}\refenvprodnewline
\refenvprodline{|}{ \token{case}\ \_ \Rightarrow \token{Fail} }{}{}{}{}\refenvprodnewline
\refenvprodline{|}{ \token{end} }{}{}{}{}\refenvprodnewline
\refenvprodline{|}{ \hspace{1em}  \refenvnt{computation} }{}{}{}{}}

\newcommand{\refenvmembership}{
\refenvrulehead{\refenvnt{membership}}{::=}{}\refenvprodnewline
\refenvfirstprodline{|}{\refenvmv{x} \, \in \, \refenvnt{VSet}}{}{}{}{}\refenvprodnewline
\refenvprodline{|}{\refenvmv{x} \, \not\in \, \refenvnt{VSet}}{}{}{}{}\refenvprodnewline
\refenvprodline{|}{\refenvnt{cls} \, \in \, \refenvnt{CSet}}{}{}{}{}\refenvprodnewline
\refenvprodline{|}{\refenvnt{cls} \, \not\in \, \refenvnt{CSet}}{}{}{}{}\refenvprodnewline
\refenvprodline{|}{\refenvnt{loc} \, \not\in \, \refenvnt{LSet}}{}{}{}{}\refenvprodnewline
\refenvprodline{|}{\refenvnt{GElm} \, \in \, \Gamma}{}{}{}{}\refenvprodnewline
\refenvprodline{|}{\refenvnt{LGElm} \, \in \, \Theta}{}{}{}{}\refenvprodnewline
\refenvprodline{|}{\refenvnt{storeElm} \, \in \, S}{}{}{}{}\refenvprodnewline
\refenvprodline{|}{\refenvnt{nhtElm} \, \in \, \Psi}{}{}{}{}\refenvprodnewline
\refenvprodline{|}{\refenvnt{nhtElm} \, \not\in \, \Psi}{}{}{}{}\refenvprodnewline
\refenvprodline{|}{\refenvmv{p} \, \in \, \refenvnt{PSet}}{}{}{}{}\refenvprodnewline
\refenvprodline{|}{\refenvmv{p} \, \not\in \, \refenvnt{PSet}}{}{}{}{}\refenvprodnewline
\refenvprodline{|}{t \, \in \, \refenvnt{sets}}{}{}{}{}\refenvprodnewline
\refenvprodline{|}{t \, \in \, \refenvnt{sets}}{}{}{}{}\refenvprodnewline
\refenvprodline{|}{\Gamma \, \in \, \refenvnt{sets}}{}{}{}{}\refenvprodnewline
\refenvprodline{|}{\refenvnt{venvElm} \, \in \, E}{}{}{}{}}

\newcommand{\refenvmisc}{
\refenvrulehead{\refenvnt{misc}}{::=}{}\refenvprodnewline
\refenvfirstprodline{|}{t_{{\mathrm{1}}}  \refenvsym{=}  t_{{\mathrm{2}}}}{}{}{}{}\refenvprodnewline
\refenvprodline{|}{\token{lookup} \, \refenvsym{(}  \Gamma  \refenvsym{,}  \refenvmv{x}  \refenvsym{)}}{}{}{}{}\refenvprodnewline
\refenvprodline{|}{\token{Done} \, \refenvsym{(}  \refenvnt{A}  \refenvsym{,}  \refenvnt{cls}  \refenvsym{)}}{}{}{}{}\refenvprodnewline
\refenvprodline{|}{\token{lookupStore}  \refenvsym{(}  \Theta  \refenvsym{,}  \refenvmv{l}  \refenvsym{)}}{}{}{}{}\refenvprodnewline
\refenvprodline{|}{\token{Done} \, \refenvsym{(}  \refenvnt{A}  \refenvsym{)}}{}{}{}{}\refenvprodnewline
\refenvprodline{|}{\token{alloc} \, \refenvsym{(}  S  \refenvsym{)}}{}{}{}{}\refenvprodnewline
\refenvprodline{|}{\token{updateStore} \, \refenvsym{(}  S  \refenvsym{,}  \refenvnt{loc}  \refenvsym{,}  t  \refenvsym{)}}{}{}{}{}\refenvprodnewline
\refenvprodline{|}{\token{return} \, \refenvsym{(}  S  \refenvsym{)}}{}{}{}{}\refenvprodnewline
\refenvprodline{|}{\token{Done} \, \refenvsym{(}  S  \refenvsym{)}}{}{}{}{}\refenvprodnewline
\refenvprodline{|}{\refenvnt{cls_{{\mathrm{1}}}}  \refenvsym{=}  \refenvnt{cls_{{\mathrm{2}}}}}{}{}{}{}\refenvprodnewline
\refenvprodline{|}{\refenvnt{cls_{{\mathrm{1}}}}  \neq  \refenvnt{cls_{{\mathrm{2}}}}}{}{}{}{}\refenvprodnewline
\refenvprodline{|}{\token{lookup} \, \refenvsym{(}  E  \refenvsym{,}  t_{{\mathrm{2}}}  \refenvsym{)}}{}{}{}{}\refenvprodnewline
\refenvprodline{|}{\token{lookupStore}  \refenvsym{(}  S  \refenvsym{,}  \refenvmv{l}  \refenvsym{)}}{}{}{}{}\refenvprodnewline
\refenvprodline{|}{\token{Done} \, \refenvsym{(}  t  \refenvsym{)}}{}{}{}{}\refenvprodnewline
\refenvprodline{|}{ N_{{\mathrm{1}}} \subseteq N_{{\mathrm{2}}} }{}{}{}{}\refenvprodnewline
\refenvprodline{|}{ \Psi_{{\mathrm{1}}} \subseteq \Psi_{{\mathrm{2}}} }{}{}{}{}\refenvprodnewline
\refenvprodline{|}{ \Theta_{{\mathrm{1}}} \subseteq \Theta_{{\mathrm{2}}} }{}{}{}{}\refenvprodnewline
\refenvprodline{|}{\token{gensym} \, \refenvsym{(}  N  \refenvsym{)}}{}{}{}{}\refenvprodnewline
\refenvprodline{|}{\token{toClos} \, \refenvsym{(}  t  \refenvsym{)}}{}{}{}{}\refenvprodnewline
\refenvprodline{|}{\token{Done} \, \refenvsym{(}  E  \refenvsym{,}  R  \refenvsym{,}  \refenvmv{x}  \refenvsym{,}  t  \refenvsym{)}}{}{}{}{}\refenvprodnewline
\refenvprodline{|}{\token{toQuote} \, \refenvsym{(}  t  \refenvsym{)}}{}{}{}{}\refenvprodnewline
\refenvprodline{|}{\token{toLocation} \, \refenvsym{(}  t  \refenvsym{)}}{}{}{}{}\refenvprodnewline
\refenvprodline{|}{\token{return} \, \refenvsym{(}  t  \refenvsym{)}}{}{}{}{}\refenvprodnewline
\refenvprodline{|}{\token{return} \, \refenvsym{(}  \refenvmv{l}  \refenvsym{)}}{}{}{}{}\refenvprodnewline
\refenvprodline{|}{ \refenvnt{cls_{{\mathrm{1}}}} \preceq \refenvnt{cls_{{\mathrm{2}}}} }{}{}{}{}\refenvprodnewline
\refenvprodline{|}{ \refenvnt{cls_{{\mathrm{1}}}} \subseteq \refenvnt{cls_{{\mathrm{2}}}} }{}{}{}{}}

\newcommand{\refenvjudgment}{
\refenvrulehead{\refenvnt{judgment}}{::=}{}\refenvprodnewline
\refenvfirstprodline{|}{ \Gamma \vdash^{ \refenvnt{clsseq} }\token{wf} }{}{}{}{}\refenvprodnewline
\refenvprodline{|}{ \Gamma \vdash^{ \refenvnt{clsseq} } \refenvnt{A} \ \token{wf} }{}{}{}{}\refenvprodnewline
\refenvprodline{|}{ \Gamma \vdash \refenvnt{cls_{{\mathrm{1}}}} \preceq \refenvnt{cls_{{\mathrm{2}}}} }{}{}{}{}\refenvprodnewline
\refenvprodline{|}{ \Gamma \vdash \refenvnt{cls_{{\mathrm{1}}}} \subseteq \refenvnt{cls_{{\mathrm{2}}}} }{}{}{}{}\refenvprodnewline
\refenvprodline{|}{ \Gamma \vdash^{ \refenvnt{clsseq} } t : \refenvnt{A} }{}{}{}{}\refenvprodnewline
\refenvprodline{|}{ \vdash  \Psi \ \token{wf} }{}{}{}{}\refenvprodnewline
\refenvprodline{|}{\Psi  \vdash  \refenvnt{clsseq} \, \token{wf}}{}{}{}{}\refenvprodnewline
\refenvprodline{|}{ \Psi  \vdash^{ \refenvnt{clsseq} }  \refenvnt{A} \ \token{wf} }{}{}{}{}\refenvprodnewline
\refenvprodline{|}{\Psi  \vdash  \refenvnt{cls_{{\mathrm{1}}}}  \preceq  \refenvnt{cls_{{\mathrm{2}}}}}{}{}{}{}\refenvprodnewline
\refenvprodline{|}{\Psi  \vdash  \refenvnt{cls_{{\mathrm{1}}}}  \subseteq  \refenvnt{cls_{{\mathrm{2}}}}}{}{}{}{}\refenvprodnewline
\refenvprodline{|}{ \Psi \mid \Theta \vdash^{ \refenvnt{cls} } \refenvnt{v} : \refenvnt{A} \ \token{val} }{}{}{}{}\refenvprodnewline
\refenvprodline{|}{ \Psi \mid^{ \refenvnt{clsseq_{{\mathrm{1}}}} } \Gamma \vdash^{ \refenvnt{clsseq_{{\mathrm{2}}}} }\token{wf} }{}{}{}{}\refenvprodnewline
\refenvprodline{|}{ \Psi \mid^{ \refenvnt{clsseq_{{\mathrm{1}}}} }  \Gamma \vdash ^{ \refenvnt{clsseq_{{\mathrm{2}}}} } \refenvnt{A} \ \token{wf} }{}{}{}{}\refenvprodnewline
\refenvprodline{|}{ \Psi \mid^{ \refenvnt{clsseq_{{\mathrm{1}}}} }  \Gamma \vdash^{ \refenvnt{clsseq} }_{ R }  t : \refenvnt{A} }{}{}{}{}\refenvprodnewline
\refenvprodline{|}{ \Psi \mid^{ \refenvnt{clsseq_{{\mathrm{1}}}} } \Gamma \vdash \refenvnt{cls_{{\mathrm{1}}}} \preceq \refenvnt{cls_{{\mathrm{2}}}} }{}{}{}{}\refenvprodnewline
\refenvprodline{|}{ \Psi \mid^{ \refenvnt{clsseq_{{\mathrm{1}}}} } \Gamma \vdash \refenvnt{cls_{{\mathrm{1}}}} \subseteq \refenvnt{cls_{{\mathrm{2}}}} }{}{}{}{}\refenvprodnewline
\refenvprodline{|}{ \Psi \mid^{ \refenvnt{clsseq_{{\mathrm{1}}}} }{ \Gamma }^{\pos{ \refenvnt{cls_{{\mathrm{3}}}} } }\vdash \refenvnt{cls_{{\mathrm{1}}}} \preceq \refenvnt{cls_{{\mathrm{2}}}} }{}{}{}{}\refenvprodnewline
\refenvprodline{|}{ \Psi \mid^{ \refenvnt{clsseq_{{\mathrm{1}}}} }{ \Gamma }^{\pos{ \refenvnt{cls_{{\mathrm{3}}}} } }\vdash \refenvnt{cls_{{\mathrm{1}}}} \subseteq \refenvnt{cls_{{\mathrm{2}}}} }{}{}{}{}\refenvprodnewline
\refenvprodline{|}{ \Psi  \vdash  \Theta \ \token{wf} }{}{}{}{}\refenvprodnewline
\refenvprodline{|}{ \Psi \mid \Theta \vdash^{ \refenvnt{cls} } E \ \token{wf} }{}{}{}{}\refenvprodnewline
\refenvprodline{|}{\Psi  \vdash  S  \colon  \Theta}{}{}{}{}\refenvprodnewline
\refenvprodline{|}{ \Psi \vdash \refenvnt{clsseq_{{\mathrm{1}}}} \mathrel{\#} \refenvnt{clsseq_{{\mathrm{2}}}} }{}{}{}{}\refenvprodnewline
\refenvprodline{|}{ \Psi \mid \refenvnt{clsseq} \mid \Gamma \vdash \refenvnt{A_{{\mathrm{1}}}} \preceq \refenvnt{A_{{\mathrm{2}}}} }{}{}{}{}}

\newcommand{\refenvformula}{
\refenvrulehead{\refenvnt{formula}}{::=}{}\refenvprodnewline
\refenvfirstprodline{|}{\refenvnt{judgement}}{}{}{}{}}

\newcommand{\refenvjudgement}{
\refenvrulehead{\refenvnt{judgement}}{::=}{}}

\newcommand{\refenvuserXXsyntax}{
\refenvrulehead{\refenvnt{user\_syntax}}{::=}{}\refenvprodnewline
\refenvfirstprodline{|}{\refenvmv{x}}{}{}{}{}\refenvprodnewline
\refenvprodline{|}{\refenvmv{l}}{}{}{}{}\refenvprodnewline
\refenvprodline{|}{\refenvmv{p}}{}{}{}{}\refenvprodnewline
\refenvprodline{|}{\refenvmv{natliteral}}{}{}{}{}\refenvprodnewline
\refenvprodline{|}{\gamma}{}{}{}{}\refenvprodnewline
\refenvprodline{|}{\refenvnt{i}}{}{}{}{}\refenvprodnewline
\refenvprodline{|}{\refenvnt{n}}{}{}{}{}\refenvprodnewline
\refenvprodline{|}{t}{}{}{}{}\refenvprodnewline
\refenvprodline{|}{\refenvnt{VSet}}{}{}{}{}\refenvprodnewline
\refenvprodline{|}{\refenvnt{venvElm}}{}{}{}{}\refenvprodnewline
\refenvprodline{|}{E}{}{}{}{}\refenvprodnewline
\refenvprodline{|}{R}{}{}{}{}\refenvprodnewline
\refenvprodline{|}{N}{}{}{}{}\refenvprodnewline
\refenvprodline{|}{\refenvnt{loc}}{}{}{}{}\refenvprodnewline
\refenvprodline{|}{\refenvnt{storeElm}}{}{}{}{}\refenvprodnewline
\refenvprodline{|}{S}{}{}{}{}\refenvprodnewline
\refenvprodline{|}{\refenvnt{cls}}{}{}{}{}\refenvprodnewline
\refenvprodline{|}{\refenvnt{clsseq}}{}{}{}{}\refenvprodnewline
\refenvprodline{|}{\iota}{}{}{}{}\refenvprodnewline
\refenvprodline{|}{\refenvnt{polyclsdeclElm}}{}{}{}{}\refenvprodnewline
\refenvprodline{|}{\varrho}{}{}{}{}\refenvprodnewline
\refenvprodline{|}{\refenvnt{A}}{}{}{}{}\refenvprodnewline
\refenvprodline{|}{\sigma}{}{}{}{}\refenvprodnewline
\refenvprodline{|}{\refenvnt{GElm}}{}{}{}{}\refenvprodnewline
\refenvprodline{|}{\Gamma}{}{}{}{}\refenvprodnewline
\refenvprodline{|}{\refenvnt{sp}}{}{}{}{}\refenvprodnewline
\refenvprodline{|}{\refenvnt{nhtElm}}{}{}{}{}\refenvprodnewline
\refenvprodline{|}{\Psi}{}{}{}{}\refenvprodnewline
\refenvprodline{|}{\refenvnt{PSet}}{}{}{}{}\refenvprodnewline
\refenvprodline{|}{\refenvnt{LGElm}}{}{}{}{}\refenvprodnewline
\refenvprodline{|}{\Theta}{}{}{}{}\refenvprodnewline
\refenvprodline{|}{\refenvnt{CSet}}{}{}{}{}\refenvprodnewline
\refenvprodline{|}{\refenvnt{LSet}}{}{}{}{}\refenvprodnewline
\refenvprodline{|}{A^{\circ}}{}{}{}{}\refenvprodnewline
\refenvprodline{|}{\refenvnt{terminals}}{}{}{}{}\refenvprodnewline
\refenvprodline{|}{\refenvnt{sets}}{}{}{}{}\refenvprodnewline
\refenvprodline{|}{\refenvnt{computation}}{}{}{}{}\refenvprodnewline
\refenvprodline{|}{\refenvnt{membership}}{}{}{}{}\refenvprodnewline
\refenvprodline{|}{\refenvnt{misc}}{}{}{}{}\refenvprodnewline
\refenvprodline{|}{\refenvnt{judgment}}{}{}{}{}}

\newcommand{\refenvgrammar}{\refenvgrammartabular{
\refenvi\refenvinterrule
\refenvn\refenvinterrule
\refenvM\refenvinterrule
\refenvVSet\refenvinterrule
\refenvvenvElm\refenvinterrule
\refenvvenv\refenvinterrule
\refenvrenv\refenvinterrule
\refenvnh\refenvinterrule
\refenvloc\refenvinterrule
\refenvstoreElm\refenvinterrule
\refenvstore\refenvinterrule
\refenvcls\refenvinterrule
\refenvclsseq\refenvinterrule
\refenvbasetype\refenvinterrule
\refenvpolyclsdeclElm\refenvinterrule
\refenvpolyclsdecl\refenvinterrule
\refenvA\refenvinterrule
\refenvclssubsts\refenvinterrule
\refenvGElm\refenvinterrule
\refenvG\refenvinterrule
\refenvsp\refenvinterrule
\refenvnhtElm\refenvinterrule
\refenvnht\refenvinterrule
\refenvPSet\refenvinterrule
\refenvLGElm\refenvinterrule
\refenvLG\refenvinterrule
\refenvCSet\refenvinterrule
\refenvLSet\refenvinterrule
\refenvAc\refenvinterrule
\refenvterminals\refenvinterrule
\refenvsets\refenvinterrule
\refenvcomputation\refenvinterrule
\refenvmembership\refenvinterrule
\refenvmisc\refenvinterrule
\refenvjudgment\refenvinterrule
\refenvformula\refenvinterrule
\refenvjudgement\refenvinterrule
\refenvuserXXsyntax\refenvafterlastrule
}}

\newcommand{\refenvdefnss}{
}

\newcommand{\refenvall}{\refenvmetavars\\[0pt]
\refenvgrammar\\[5.0mm]
\refenvdefnss}

\title{Refined\textsuperscript{2} Environment Classifiers}

\author{Yuito Murase}
\email{murase@fos.kuis.kyoto-u.ac.jp}
\orcid{0000-0001-6038-6249}
\affiliation{%
  \institution{Kyoto University}
  \city{Kyoto}
  \country{Japan}
}

\author{Atsushi Igarashi}
\email{igrarashi@kyoto-u.ac.jp}
\orcid{0000-0002-5143-9764}
\affiliation{%
  \institution{Kyoto University}
  \city{Kyoto}
  \country{Japan}
}

\begin{abstract}
 MetaML-style multi-stage programming (MSP) supports quasi-quotation-based code generation, runtime execution of generated code, and cross-stage persistence (CSP). However, its interaction with computational effects is subtle: mutable state can cause scope extrusion, where generated code escapes the scope of variables on which it depends.

 This paper presents a type system for MetaML-style MSP with mutable state that statically rules out harmful scope extrusion while supporting multi-level code generation, runtime execution, and a variant of CSP. Our system builds on refined environment classifiers (RECs), a discipline that annotates code types with the variable scopes on which generated code depends. To scale RECs to the MetaML-style setting, we refine classifiers so that they track not only variable scopes, but also the scopes of classifiers themselves. Further, we integrated polymorphism over classifiers, enabling more general and reusable code generation patterns in a multi-level setting.

 For the resulting system, we define an operational semantics via a definitional interpreter and prove type soundness and safety of offline code generation, showing that generated code can be extracted as standalone well-typed programs. We provide working implementations and mechanized proofs in Rocq.
\end{abstract}

\maketitle

\section{Introduction}
Multi-stage programming (MSP) allows programmers to write programs that generate and execute code fragments as first-class values. In particular, MetaML-style MSP~\cite{journals/tcs/TahaS00,conf/popl/TahaN03,journals/pacmpl/XieWNY23,conf/popl/LeeXKY26} introduced quasi-quotation syntax for code generation and the \texttt{run} primitive for execution, together with cross-stage persistence (CSP), which allows variables to be used at later stages. MetaML-style MSP provides a simple and intuitive programming model and is employed in real programming languages such as MetaOCaml~\cite{conf/flops/Kiselyov14,journals/scp/Kiselyov26} and Scala 3~\cite{thesis/Stucki23}.

However, designing type systems for MetaML-style MSP has long been a central challenge for this approach. In particular, typing disciplines in the presence of runtime execution are already nontrivial~\cite{conf/esop/MoggiTBS99,conf/popl/TahaN03,journals/corr/abs-1010-3806,conf/flops/Kiselyov14}. Moreover, the interaction between code generation and computational effects is subtle because it can cause \emph{scope extrusion}, in which generated code may contain variables that are not in scope at the point of execution. At present, even semantic reasoning about effectful code generation remains an active area of research, and there is no agreed-upon typing discipline that can statically detect harmful scope extrusion~\cite{conf/icalp/CalcagnoMT00, conf/fossacs/MoggiF03, conf/esop/Rhiger12, conf/aplas/KiselyovKS16, conf/popl/LeeXKY26}.

Recently, a novel typing discipline called \emph{refined environment classifiers} (RECs) was proposed for effectful code generation while preventing harmful scope extrusion~\cite{conf/aplas/KiselyovKS16, conf/gpce/OishiK17, conf/gpce/IsodaYK24, conf/popl/LeeXKY26}.
RECs annotate code types with information about the variable scopes on which generated code depends, ensuring that generated code fragments are well-scoped.
While early proposals targeted combinator-based code generation, Lee et al.~\cite{conf/popl/LeeXKY26} designed a type system for MetaML-style MSP. However, their type system is limited to two-level staging and does not support runtime execution nor CSP.

Indeed, the original RECs discipline is inadequate for multi-level staging.
As we show in \Zcref{sec:informaldescription:r2ecs:subtlety}, it does not account for how to track variables that occur under nested quotations.
Supporting multi-level code generation therefore requires a further refinement of the typing discipline.

\subsubsection*{Our contribution}
In this paper, we present a novel typing discipline for MetaML-style MSP with
mutable state that statically detects harmful scope extrusion. We extend the RECs approach
from earlier two-level settings to a MetaML-style setting with
multi-level code generation, runtime execution, and scoped CSP.

The main technical refinement is that classifiers no longer track only variable
scope dependencies. In a multi-level setting, the type system must also track the
scopes of classifiers themselves. This additional scoping
information enables the REC approach to scale to nested code generation, and
leads to our typing discipline of \emph{refined\textsuperscript{2} environment
classifiers (R2ECs)}.

The system also integrates \emph{polymorphic classifiers}, which substantially
increase its flexibility by enabling general and reusable code generators.
Prior work on RECs discussed this idea only informally; we provide, to the best
of our knowledge, the first formal treatment of polymorphic classifiers for
REC-based typing and extend it to the multi-level setting of MetaML-style MSP.

For the full resulting calculus, we prove key metatheoretic properties,
including type soundness and the safety of offline code generation. The proofs
are mechanized in the Rocq proof assistant.

\subsubsection*{Structure of the paper}
First, \Zcref{sec:informaldescription} presents an informal overview of our approach to safe, effectful multi-stage programming through a series of examples.
\Zcref{sec:surface-typing} gives a formal definition of our surface language and type system, introducing the central ideas of our typing discipline with R2ECs.
We present polymorphic classifiers separately in \Zcref{sec:polycls}, providing the idea and formal definitions for polymorphic classifiers.
\Zcref{sec:semantics} defines the operational semantics of our language via a definitional interpreter.
\Zcref{sec:runtime-typing} introduces runtime typing, which we use in \Zcref{sec:metatheory} to prove metatheoretic properties such as type soundness and the safety of offline code generation.
Finally, \Zcref{sec:relatedwork} discusses related work, and \Zcref{sec:conclusion} concludes with future directions.

\section{Informal Description}\label{sec:informaldescription}
This section provides an informal overview of our approach to safe, effectful multi-stage programming. We begin with an operational account of untyped MetaML-style staging through a series of examples. These examples illustrate both well-behaved programs and ill-scoped ones that lead to scoping errors, such as scope extrusion, particularly in the presence of mutable state. They highlight the need for a static discipline that can distinguish harmless from harmful uses of free variables in generated code. We then outline how R2ECs provide such a discipline, enabling expressive multi-level staging while preventing the generation of ill-scoped code.

\paragraph{Implementation and Web Demo}
We provide a full implementation of the surface type system and definitional interpreter as an interactive web demo for editing, type checking, and evaluating the paper's examples and additional programs. The source code and container images are distributed as a reproducible artifact. OpenAI Codex with GPT-5.5 and Google Antigravity with Gemini 3.1 Pro were used extensively to develop the core implementation and browser interface, respectively. The authors reviewed, corrected, and tested the generated code and take full responsibility for the result.

\subsection{MetaML-style MSP by Example}\label{sec:informaldescription:metamlmsp}
We start with an informal presentation of MetaML-style multi-stage programming in an untyped setting. We first illustrate well-behaved programs that demonstrate the intended behavior of staging constructs. We then present examples that exhibit scoping errors, motivating the need for a typing discipline that enables their static detection.

\subsubsection*{Quasi-Quotation-Based Code Generation}
As a standard example of multi-stage programming, we consider \emph{specializing} the exponentiation function \lstinline|pow|. The function \lstinline|pow| takes two arguments \lstinline|n| and \lstinline|x| and computes the value \lstinline|x|\textsuperscript{\lstinline|n|}:
\begin{lstlisting}
let rec pow n x = if n == 0 then 1 else x * pow (n-1) x
\end{lstlisting}
It is often beneficial to specialize \lstinline|pow| with respect to \lstinline|n| when we know \lstinline|n| in advance. We therefore define \lstinline|spow|, a specialized variant of \lstinline|pow|: given an integer \lstinline|n|, it generates code for a function that computes \lstinline|x|\textsuperscript{\lstinline|n|} without recursive function calls or branching.
\begin{lstlisting}
let spow n = .< fun x -> .~( let rec loop m = if m == 0 then .< 1 >.
                                            else if m == 1 then .< x >.
                                            else .< x * .~(loop (m-1)) >. in
                           loop n ) >. in
(* spow 10 generates *< fun x -> x * x * ... * x >* *)
let pow10 = run (spow 10) in pow10 2 (* => 1024 *)
\end{lstlisting}
This definition illustrates quasi-quotation-based code generation. \emph{Quotation} \lstinline|.< e >.| constructs a code fragment representing the expression \lstinline|e|, while \emph{splicing} \lstinline|.~e| inserts a previously generated code fragment into a larger one. In \lstinline|spow|, the auxiliary function \lstinline|loop| builds code for the body of the power function by recursion on \lstinline|n|. Thus, \lstinline|spow 10| produces code equivalent to a function that multiplies its argument by itself ten times.
We can execute the generated code at runtime using the primitive \lstinline|run|. We apply \lstinline|run| to the code generated by \lstinline|spow 10| to obtain an executable function. We then execute \lstinline|pow10 2| to compute $2^{10}$ with optimized code.

Thus, MetaML-style MSP performs specialization by constructing code values via quotation, composing them via splicing, and executing them via \lstinline|run|. We can also emit the content of generated code for later execution: this is called \emph{offline code generation}.

\subsubsection*{Cross-Stage Persistence}
Let us optimize the code generated by \lstinline|spow| by introducing an auxiliary function \lstinline|sqr|, which squares its argument. Using \lstinline|sqr|, the generated code can reuse common subcomputations and avoid repeated multiplications.
\begin{lstlisting}
let sqr x = x * x in
let spow n = .< fun x -> .~( let rec loop m =
                             if m == 0 then .< 1 >.
                             else if m == 1 then .< x >.
                             else if m % 2 == 0 then .< sqr .~( loop (m/2) ) >.
                             else .< x * .~( loop (m-1) ) >. in
                           loop n ) >. in
(* spow 10 generates *< fun x -> sqr(x * sqr(sqr(x))) >* *)
let pow10 = run (spow 10) in pow10 2 (* => 1024 *)
\end{lstlisting}
In this program, \lstinline|sqr| is defined during code generation while being used inside a quotation. Such pattern is common in MetaML-style MSP and is called \emph{cross-stage persistence} (CSP)~\cite{journals/tcs/TahaS00, conf/popl/TahaN03, conf/flops/HanadaI14}. The gap in stages does not pose a problem because the code fragment containing \lstinline|sqr| is executed by \lstinline|run|: \lstinline|sqr| is resolved in the runtime environment at the execution point, where \lstinline|sqr| is defined.

Our language differs from the original MetaML in both the implementation and capabilities of CSP. In particular, CSP is restricted to the scope of persisting variables such as \lstinline|sqr|, so we call it \emph{scoped CSP}. We return to this point in \Zcref{sec:surface-typing, sec:semantics, sec:relatedwork}.

\paragraph{Code Generation with Mutable State}
We can also use mutable state during code generation. In the following example, \lstinline|spow| constructs the body of the function by storing an intermediate code fragment in a reference and updating it iteratively.

\begin{lstlisting}
let spow n = .< fun x -> .~( let r = ref .< 1 >. in
                           let rec loop m = if m == 0 then ()
                                            else let c = deref r in
                                                 r := .< x * .~c >.; loop (m-1) in
                           loop n; deref r ) >. in
let pow10 = run (spow 10) in pow10 2
\end{lstlisting}

Here, the reference \lstinline|r| stores a code fragment that is repeatedly extended during generation. This program is well-behaved, but in general, storing open code in mutable state can lead to \emph{scope extrusion}, as we will see shortly. While mutable state is not essential in this example, there are staged programming patterns where it is inherently valuable, such as assertion insertion~\cite{journals/scp/KameyamaKS15, conf/aplas/KiselyovKS16}.

\subsubsection*{Scoping Errors}
Allowing code fragments with free variables introduces the possibility of scoping errors, where generated code refers to variables that are not available at the point of execution. We present several examples illustrating how such errors arise.

The first example demonstrates \emph{ill-staged runtime execution}.
\begin{lstlisting}
.< fun x -> .~( run .< x >. ) >. (* error: undefined variable x *)
\end{lstlisting}
In this example, the code fragment \lstinline|< x >| refers to the variable \lstinline|x|, which is bound in a future stage inside the quotation. Hence, \lstinline|run < x >| fails to resolve \lstinline|x|, leading to a runtime error.

Another example is \emph{scope extrusion}, which illustrates the problematic interaction between code generation and mutable state.
\begin{lstlisting}
let r = ref .< 1 >. in
.< fun x -> .~( r := .< x >.; .< x >. ) >.; deref r (* returns *< x >* *)
\end{lstlisting}
Here, a code fragment with \lstinline|x| is stored in the reference \lstinline|r| during code generation. The reference itself is allocated outside the scope of \lstinline|x|, and therefore outlives the binding of \lstinline|x|. As a result, the code \lstinline|.< x >.| escapes its lexical scope and can be retrieved later via \lstinline|deref r|, even though no binding for \lstinline|x| is available.

This example highlights the subtle interaction between code generation and mutable state: code fragments may be stored and later used outside the scope of the variables they depend on. The static detection of such interactions is known to be nontrivial and has long been studied~\cite{conf/icalp/CalcagnoMT00, conf/fossacs/MoggiF03, conf/popl/KimYC06, conf/esop/Rhiger12, conf/aplas/KiselyovKS16}.

\subsection{Tracking Scopes via Refined Environment Classifiers}\label{sec:informaldescription:recs}
To prevent the scoping errors presented in the previous section, we introduce a typing discipline based on \emph{refined environment classifiers (RECs)} \cite{conf/aplas/KiselyovKS16, conf/gpce/OishiK17, conf/gpce/IsodaYK24, conf/popl/LeeXKY26}. The purpose of RECs is to statically track the variable scopes on which generated code fragments depend. Here, we follow the presentation style of Lee et al.~\cite{conf/popl/LeeXKY26} to explain the basic idea of RECs.
We illustrate the basic idea of RECs using the \lstinline|spow| example. We annotate the program with types and \emph{classifiers}.
\begin{lstlisting}
let spow (n(|:int|))(|: <int->int/!>|) = .<@@! fun x@g1 -> .~( let rec loop (m(|:int|))(|: <int/g1>|) =
                                                    if m == 0 then .<@@g1 1 >.
                                                    else if m == 1 then .<@@g1 x >.
                                                    else .<@@g1 x * .~(loop (m-1)) >. in
                                                  loop n ) >. in ...
\end{lstlisting}
The binding for \lstinline|x| is annotated with the classifier \lstinline|cg1|, which represents the scope in which \lstinline|x| is valid. The {\color{ClassifierNew} brown} color indicates that the classifier is newly introduced. We also use a distinguished classifier \lstinline|!| for the top-level scope, where no local variables are available.

A code type records both the type of a code fragment and the classifier representing the scope on which the fragment may depend. For example, the return type of \lstinline|loop| is the code type \lstinline|<int/g1>|. This indicates that the generated code has type \lstinline|int| and may depend on the scope represented by \lstinline|cg1|. Consequently, \lstinline|x| may occur free in the generated code, whereas no other variables may occur free.

A quotation switches the current scope to an existing classifier; for instance, \lstinline|.<@@g1 x >.| switches to \lstinline|cg1|, allowing \lstinline|x| to be used inside the quotation.
A splice escapes back from the quotation and moves to the scope of the past stage.

Classifiers allow us to detect potential scope extrusion. For example, consider the following annotated version of the scope extrusion example:
\begin{lstlisting}
let r@g1(|:<int@!> reft|) = ref .<@@! 1 >. in
.< fun x@h1(|:int|) -> .~( (@*\tikzmark{rassign2}*@)r := .<@@h1 x >.(@*\tikzmark{rassign2e}*@); .<@@h1 x >. ) >.;
deref r
\end{lstlisting}
\begin{tikzpicture}[overlay,remember picture]
  \draw[red,thick,dash pattern=on 4pt off 2pt]
  ([yshift=-2pt]pic cs:rassign2) -- ([yshift=-2pt]pic cs:rassign2e)
  node[pos=0,below,anchor=north west] {\texttt{error: \lstinline|<int/h1>| cannot be assigned to \lstinline|<int/!>| ref }};
\end{tikzpicture}
In this case, the type for mutable state is \lstinline|<int/!> reft|, clarifying the scope of code fragments that can be safely assigned to it. When we try to assign \lstinline|<@@h1 x >| at line 2, we get a type error because its type \lstinline|<int/h1>| is inconsistent with the type of the mutable state \lstinline|<int/!> reft|.

We impose a partial order $ \preceq $ on classifiers to reflect the nesting structure of scopes: $ \delta_{{\mathrm{1}}} \preceq \delta_{{\mathrm{2}}} $ means that the scope $\delta_{{\mathrm{2}}}$ is located within the scope $\delta_{{\mathrm{1}}}$. In the \lstinline|spow| example, we have $  \exclam  \preceq \gamma_{{\mathrm{1}}} $ because scope \lstinline|cg1| is nested within the top-level scope \lstinline|!|.
We use classifiers to track variable scopes of quasi-quotation-based code generation.

\subsection{Further Refining RECs for MetaML-style MSP}\label{sec:informaldescription:r2ecs}
We apply the RECs discipline to full-fledged MetaML-style MSP, including multi-level staging, runtime execution, and cross-stage persistence. This requires refining the original RECs discipline by elaborating the notion of variable scope.
We call the resulting discipline \emph{refined\textsuperscript{2} environment classifiers (R2ECs)}.
In this section, we explain the motivation for these refinements and illustrate them through examples.
We also show that they make it possible to type runtime execution with cross-stage persistence, which is not supported by the original RECs discipline.

\subsubsection*{Subtleties of Classifiers in Multi-Level Staging}\label{sec:informaldescription:r2ecs:subtlety}
Classifiers track variable scopes in the RECs discipline, but things become more subtle in multi-level settings. Consider the following multi-level program:
\begin{tikzpicture}[overlay,remember picture]
  \draw[blue,thick]
  ([yshift=-2pt]pic cs:rassignmultiuntyped) -- ([yshift=-2pt]pic cs:rassignmultiuntypede)
  node[pos=1,right=-3pt] {\small \ctext{1}};
\end{tikzpicture}
\begin{lstlisting}
let r = ref .< 1 >. in
.<.< fun x -> .~.~( (@*\tikzmark{rassignmultiuntyped}*@)r := .< let y = .< x >. in 10 >.;(@*\tikzmark{rassignmultiuntypede}*@)   .<.< x >.>. ) >.>.;
deref r
\end{lstlisting}
This program generates the code fragment \lstinline|.< let y = .< x >. in 10 >.| by assigning it to mutable state at \ctext{1} and retrieving it later. In this fragment, the variable \lstinline|x| appears in a nested quotation and escapes its scope. We expect such programs to be rejected, but this is not easy to detect with the discipline developed so far. Let us try annotating the program with classifiers. In multi-stage settings, all stages carry classifier annotations.
\begin{lstlisting}
let r@g1(|:<int/!> reft|) = ref .<@@! 1 >. in
.<@@!.<@@! fun x@h1(|:int|) -> .~.~( r := .<@@! let y@h2(|:<int/h1>|) = (@*\tikzmark{nesth1x}*@).<@@h1 x >.(@*\tikzmark{nesth1xe}*@)   in 10 >.; .<@@!.<@@h1 x >.>. ) >.>.;
deref r
\end{lstlisting}
\begin{tikzpicture}[overlay,remember picture]
  \draw[blue,thick]
  ([yshift=-2pt]pic cs:nesth1x) -- ([yshift=-2pt]pic cs:nesth1xe)
  node[pos=1,right=-3pt] {\small \ctext{1}};
\end{tikzpicture}
At first, this program appears well-typed.
The code fragment \lstinline|<@@h1 x >| at \ctext{1} correctly records its dependence on scope \lstinline|ch1|, but this information is not reflected in the typing of the surrounding quotation. As a result, the whole quotation is assigned the type \lstinline|<int@!>|, which no longer mentions \lstinline|ch1|. Consequently, when typing the whole \lstinline|let| binding, the type system cannot detect that \lstinline|x| may escape its scope.
We argue that the problem lies in the classifier annotation on the quotation at \ctext{1}.
This quotation is nested under the top-level scope \lstinline|c!|, which should have no knowledge of any other scope. Hence, using \lstinline|ch1| in the nested quotation appears illegal.

\subsubsection*{Scoping Structure Formed by Classifiers}
The previous example illustrates that the type system must precisely track not only where variables may be used, but also where classifiers may be used. Thus, in addition to ordinary variable scopes, we consider the scopes of classifiers themselves.

Although classifiers represent both kinds of scopes, we distinguish the relations they induce.
We call the ordinary variable-scope relation $ \preceq $ \emph{reachability}: $ \gamma_{{\mathrm{1}}} \preceq \gamma_{{\mathrm{2}}} $ (read ``$\gamma_{{\mathrm{1}}}$ is reachable from $\gamma_{{\mathrm{2}}}$'') means that variables bound under $\gamma_{{\mathrm{1}}}$ may be used under $\gamma_{{\mathrm{2}}}$. We call the classifier-scope relation $ \subseteq $ \emph{visibility}: $ \gamma_{{\mathrm{1}}} \subseteq \gamma_{{\mathrm{2}}} $ (read ``$\gamma_{{\mathrm{1}}}$ is visible to $\gamma_{{\mathrm{2}}}$'') means that the classifier $\gamma_{{\mathrm{1}}}$ may be used under $\gamma_{{\mathrm{2}}}$. Here, using a classifier $\gamma_{{\mathrm{1}}}$ under a classifier $\gamma_{{\mathrm{2}}}$ means introducing, in a context annotated with $\gamma_{{\mathrm{2}}}$, a quotation annotated with $\gamma_{{\mathrm{1}}}$, as in \lstinline|@@g2 .<@@g1 ...>.|. Both relations behaves as partial orders.

We illustrate the scoping structure induced by variables and classifiers below.
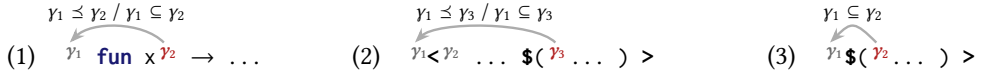
\begin{center}
  \begin{minipage}{0.30\linewidth}
\centering
\begin{tikzpicture}[
  remember picture,
  baseline=0pt,
  every node/.style={inner sep=0pt, outer sep=0pt}
]

\node[anchor=base west] (formula1) at (0,0) {%
  (1)\quad
  \lstinline|@@g1|%
  \tikzmark{gOne2}%
  \lstinline| fun x@g2|%
  \tikzmark{gTwo2}%
  \lstinline| -> ...|%
};

\path[use as bounding box]
  ([yshift=-2pt]formula1.south west)
  rectangle
  ([yshift=16pt]formula1.north east);

\coordinate (gOne2Top) at ([xshift=-0.55em,yshift=0.8em]pic cs:gOne2);
\coordinate (gTwo2Top) at ([xshift=-0.55em,yshift=0.8em]pic cs:gTwo2);

\draw[
  overlay,
  remember picture,
  gray!65,
  line width=0.9pt,
  -{Stealth[length=2mm,width=1.5mm]}
]
  (gTwo2Top)
    .. controls +(135:4mm) and +(45:4mm) ..
  node[midway, above=2pt, black]
    {\scriptsize $ \gamma_{{\mathrm{1}}} \preceq \gamma_{{\mathrm{2}}} \ /\  \gamma_{{\mathrm{1}}} \subseteq \gamma_{{\mathrm{2}}} $}
  (gOne2Top);

\end{tikzpicture}

\end{minipage}
\hfill
\begin{minipage}{0.30\linewidth}
\centering
\begin{tikzpicture}[
  remember picture,
  baseline=0pt,
  every node/.style={inner sep=0pt, outer sep=0pt}
]

\node[anchor=base west] (formula2) at (0,0) {%
  (2)\quad
  \lstinline|@@g1|%
  \tikzmark{reachSplice2G1}%
  \lstinline|.<@@g2 ... .~(@g3|%
  \tikzmark{reachSplice2G2}%
  \lstinline|... ) >.|%
};

\path[use as bounding box]
  ([yshift=-2pt]formula2.south west)
  rectangle
  ([yshift=16pt]formula2.north east);

\coordinate (reachSplice2G1Top)
  at ([xshift=-0.55em,yshift=0.8em]pic cs:reachSplice2G1);
\coordinate (reachSplice2G2Top)
  at ([xshift=-0.55em,yshift=0.8em]pic cs:reachSplice2G2);

\draw[
  overlay,
  remember picture,
  gray!65,
  line width=0.9pt,
  -{Stealth[length=2mm,width=1.5mm]}
]
  (reachSplice2G2Top)
    .. controls +(135:4mm) and +(45:4mm) ..
  node[midway, above=2pt, black]
    {\scriptsize $ \gamma_{{\mathrm{1}}} \preceq \gamma_{{\mathrm{3}}} \ /\  \gamma_{{\mathrm{1}}} \subseteq \gamma_{{\mathrm{3}}} $}
  (reachSplice2G1Top);

\end{tikzpicture}
\end{minipage}
\hfill
\begin{minipage}{0.30\linewidth}
\centering
\begin{tikzpicture}[
  remember picture,
  baseline=0pt,
  every node/.style={inner sep=0pt, outer sep=0pt}
]

\node[anchor=base west] (formula2) at (0,0) {%
  (3)\quad
  \lstinline|@@g1|%
  \tikzmark{reachSplice3G1}%
  \lstinline|.~(@g2|%
  \tikzmark{reachSplice3G2}%
  \lstinline|... ) >.|%
};

\path[use as bounding box]
  ([yshift=-2pt]formula2.south west)
  rectangle
  ([yshift=16pt]formula2.north east);

\coordinate (reachSplice3G1Top)
  at ([xshift=-0.55em,yshift=0.8em]pic cs:reachSplice3G1);
\coordinate (reachSplice3G2Top)
  at ([xshift=-0.55em,yshift=0.8em]pic cs:reachSplice3G2);

\draw[
  overlay,
  remember picture,
  gray!65,
  line width=0.9pt,
  -{Stealth[length=2mm,width=1.5mm]}
]
  (reachSplice3G2Top)
    .. controls +(135:4mm) and +(45:4mm) ..
  node[midway, above=2pt, black]
    {\scriptsize $ \gamma_{{\mathrm{1}}} \subseteq \gamma_{{\mathrm{2}}} $}
  (reachSplice3G1Top);

\end{tikzpicture}
\end{minipage}
\end{center}
\begin{enumerate}[label=(\arabic*)]
  \item When a variable $\refenvmv{x}$ is bound, a fresh classifier $\gamma_{{\mathrm{2}}}$ is
  introduced to represent its scope. This scope is nested within the surrounding
  scope $\gamma_{{\mathrm{1}}}$ as an ordinary variable scope, and hence $ \gamma_{{\mathrm{1}}} \preceq \gamma_{{\mathrm{2}}} $
  holds. The same nesting also makes the surrounding classifier available inside
  the new scope; therefore, $ \gamma_{{\mathrm{1}}} \subseteq \gamma_{{\mathrm{2}}} $ holds as well.

  \item A splice introduces a fresh classifier $\gamma_{{\mathrm{3}}}$ for the splice body.
  Since the splice escapes from the quotation back to
  the same stage as $\gamma_{{\mathrm{1}}}$, variables available under $\gamma_{{\mathrm{1}}}$ should also
  be available under $\gamma_{{\mathrm{3}}}$. Thus, $ \gamma_{{\mathrm{1}}} \preceq \gamma_{{\mathrm{3}}} $ holds. For the same
  reason, the classifier $\gamma_{{\mathrm{1}}}$ is available under $\gamma_{{\mathrm{3}}}$, so
  $ \gamma_{{\mathrm{1}}} \subseteq \gamma_{{\mathrm{3}}} $ also holds.

  \item A splice induces another visibility relation.
  In the figure, this gives $ \gamma_{{\mathrm{1}}} \subseteq \gamma_{{\mathrm{2}}} $, meaning that
  $\gamma_{{\mathrm{1}}}$ can be used under $\gamma_{{\mathrm{2}}}$. This is necessary, for example, when
  typing \lstinline|.<fun x -> .~(.<x>.)>.|: the splice body constructs the
  quotation \lstinline|.<x>.|, whose classifier is the scope of $\refenvmv{x}$.
  However, this is only a visibility relation. It does not make variables from
  $\gamma_{{\mathrm{1}}}$ available inside the splice body, so $ \gamma_{{\mathrm{1}}} \preceq \gamma_{{\mathrm{2}}} $ does not hold.
\end{enumerate}

\subsubsection*{Tracking Variables in Nested Quotations via Classifier Scopes}

We now revisit the previous example and show how R2ECs detect its scoping error.
\par\medskip
\begin{minipage}{\linewidth}
\begin{lstlisting}
let r@g1(|:<int/!> reft|) = ref .<@@! 1 >. in
.<@@!.<@@! fun x@h1(|:int|) -> .~@g2.~(@g3 r := (@*\tikzmark{nesttop}*@).<@@!(@*\tikzmark{nesttope}*@)  let y@h2(|:<int/h1>|)=(@*\tikzmark{nesth1x1}*@).<@@h1(@*\tikzmark{nesth1x1e}*@)x >. in 10 >.; .<@@!.<@@h1x >.>.)>.>.;
deref r
\end{lstlisting}
\begin{tikzpicture}[overlay,remember picture]
  \draw[blue,thick]
  ([yshift=-2pt]pic cs:nesttop) -- ([yshift=-2pt]pic cs:nesttope)
  node[pos=1,right=-3pt] {\small \ctext{1}};

  \draw[red,thick,dash pattern=on 4pt off 2pt]
  ([yshift=-2pt]pic cs:nesth1x1) -- ([yshift=-2pt]pic cs:nesth1x1e)
  node[pos=0,below,anchor=north west] {\texttt{error: \lstinline|ch1| is not visible to \lstinline|c!|}};
\end{tikzpicture}
\end{minipage}

The program contains a nested quotation annotated with \lstinline|ch1|.
However, this quotation appears inside the top-level quotation marked \ctext{1}, where \lstinline|ch1| is not visible to \lstinline|c!|.
This use of \lstinline|ch1| is therefore rejected by our scoping discipline.
One might try to repair this visibility error by changing the annotation of the surrounding quotation:
\begin{lstlisting}
let r@g1(|:<int/!> reft|) = ref .<@@! 1 >. in
.<@@!.<@@! fun x@h1(|:int|) -> .~@g2.~(@g3(@*\tikzmark{rassign}*@)r := (@*\tikzmark{rassigng2}*@).<@@g2(@*\tikzmark{rassigng2e}*@)   let y@h2(|:<int/h1>|)=(@*\tikzmark{rassignh1}*@).<@@h1(@*\tikzmark{rassignh1e}*@)  x >. in 10 >.;(@*\tikzmark{rassigne}*@) .<@@g2.<@@h1x >.>.)>.>.;
deref r
\end{lstlisting}
\begin{tikzpicture}[overlay,remember picture]
  \draw[red,thick,dash pattern=on 4pt off 2pt]
  ([yshift=-5pt]pic cs:rassign) -- ([yshift=-5pt]pic cs:rassigne)
  node[pos=0,below,anchor=north west] {\texttt{error: \lstinline|<int/g2>| cannot be assigned to \lstinline|<int/!>| ref}};

  \draw[blue,thick]
  ([yshift=-2pt]pic cs:rassigng2) -- ([yshift=-2pt]pic cs:rassigng2e)
  node[pos=1,right=-3pt,yshift=2pt] {\small \ctext{1}};

  \draw[blue,thick]
  ([yshift=-2pt]pic cs:rassignh1) -- ([yshift=-2pt]pic cs:rassignh1e)
  node[pos=1,right=-4pt,yshift=2pt] {\small \ctext{2}};
\end{tikzpicture}
Here, the classifier of the surrounding quotation is changed from \lstinline|c!| to \lstinline|cg2| in \ctext{1}.
This use of \lstinline|cg2| is well-scoped because $ \gamma_{{\mathrm{2}}} \subseteq \gamma_{{\mathrm{3}}} $ holds.
Then the use of \lstinline|ch1| in \ctext{2} is well-scoped because $ \delta_{{\mathrm{1}}} \subseteq \gamma_{{\mathrm{2}}} $ holds.
However, the assignment is now rejected instead: the mutable cell has type \lstinline|<int/!> reft|, whereas the assigned code fragment has type \lstinline|<int/g2>|. The two types record incompatible scoping information.
Thus, the type checker detects the potential scope extrusion even when the escaping variable occurs inside a nested quotation. The key point is that classifiers now track the scopes of nested quotations through the visibility relation $ \subseteq $.

\subsubsection*{Tracking Scopes across Stages}
We now show that the scoping discipline illustrated so far naturally extends to runtime execution with CSP.
In such settings, classifiers introduced at one stage may be used at later stages, because \lstinline|run| checks generated code against the current lexical scope. Interestingly, we do not need any additional machinery to support this cross-stage tracking of variable scopes: the notion of visibility already captures the structure needed to track scopes across stages.
Let us consider an annotated version of the \lstinline|spow| example with \lstinline|sqr|:
\begin{tikzpicture}[overlay,remember picture]
  \draw[blue,thick]
  ([yshift=-2pt]pic cs:tintg1) -- ([yshift=-2pt]pic cs:tintg1e)
  node[pos=1,right=-3pt] {\small \ctext{1}};

  \draw[blue,thick]
  ([yshift=-2pt]pic cs:runspow10sqr) -- ([yshift=-2pt]pic cs:runspow10sqre)
  node[pos=1,right=-3pt] {\small \ctext{2}};
\end{tikzpicture}
\begin{lstlisting}
let sqr@g1 (x@g2(|:int|)) (|: int|) = x * x in
let spow@g3 (n@g4(|:int|)) (|:<int->int/g1>|) = (@*\tikzmark{tintg1}*@).<@@g1(@*\tikzmark{tintg1e}*@)   fun x@h1:int -> ... >. in
let pow10@g6 = (@*\tikzmark{runspow10sqr}*@)run (spow 10)(@*\tikzmark{runspow10sqre}*@)   in pow10 2
\end{lstlisting}
At \ctext{1}, the body of the function \lstinline|spow| introduces a quote to switch the scope from \lstinline|cg4| to \lstinline|cg1|, which is the scope for \lstinline|sqr| at the code-generation stage.
Such a scope transition is allowed because the visibility relation $ \gamma_{{\mathrm{1}}} \subseteq \gamma_{{\mathrm{4}}} $ arises from the reachability relation $ \gamma_{{\mathrm{1}}} \preceq \gamma_{{\mathrm{4}}} $. At \ctext{2}, \lstinline|run (spow 10)| occurs in scope \lstinline|cg3|, and the classifier ordering $ \gamma_{{\mathrm{1}}} \preceq \gamma_{{\mathrm{3}}} $ holds. Therefore, the generated code is well scoped, since all free variables of the code fragment remain available in the lexical environment at the execution point. Thus, the program type-checks, and \lstinline|run| is safely applied to code with CSP.

We can also detect scoping errors in the ill-staged execution example:
\begin{tikzpicture}[overlay,remember picture]
  \draw[red,thick,dash pattern=on 4pt off 2pt]
  ([yshift=-2pt]pic cs:rqg1) -- ([yshift=-2pt]pic cs:rqg1e);
\end{tikzpicture}
\begin{lstlisting}
.<@@! fun x@g1(|:int|) -> .~(@@g2 (@*\tikzmark{rqg1}*@)run .<@@g1 x >.(@*\tikzmark{rqg1e}*@) ) >.  (e|error: cannot runt code with <int/g1> at cg2|e)
\end{lstlisting}
Here, \lstinline|run| attempts to execute a code fragment that is valid only under classifier \lstinline|cg1|. However, \lstinline|run| occurs at the scope \lstinline|cg2|, where $ \gamma_{{\mathrm{1}}} \preceq \gamma_{{\mathrm{2}}} $ does not hold. Consequently, the type system rejects this program due to inconsistency in scoping information.

\subsubsection*{Summary}
R2ECs refine the scoping discipline of RECs by managing both variable scopes and classifier scopes. This refinement allows the type system to detect scope extrusion even when it occurs through nested quotations, thereby establishing static scope safety for MetaML-style multi-stage programming. Furthermore, R2ECs naturally capture the structure needed to track scopes across stages, enabling the type system to support cross-stage persistence without additional machinery.

\section{Surface Language and Typing}\label{sec:surface-typing}
In this section, we introduce the typing discipline based on R2ECs by defining a surface type system for our multi-stage language. As its name suggests, it types only surface expressions. We later define runtime objects in \Zcref{sec:semantics}, and then introduce a runtime type system that accounts for them in \Zcref{sec:runtime-typing}.

The syntactic categories of the surface type system are defined as follows.
We have a set of integers $ \token{Int} $, ranged over by $\refenvnt{n}$, and a set of variables $ \token{Var} $, ranged over by $\refenvmv{x}, \refenvmv{y}, \dotsc$. We also have a set of classifiers, ranged over by $\gamma, \delta, \dotsc$. We reserve a special classifier $ \exclam $ to represent the top-level scope, which we call the \emph{initial classifier}.
We write variables and classifiers in \textcolor{ClassifierNew}{brown} when they are binders or newly introduced.
\begin{longtable}{cllcl}
  \multicolumn{5}{l}{\textbf{Surface Expression}}                                                                                                                                                                                   \\
  Expression       & $ \refenvnt{e} ^{ \refenvnt{k}  } $ & $\in  \token{Exp}^{ \refenvnt{k} } $    & $\Coloneqq$ & $ \refenvnt{n} \mid \refenvmv{x} \mid   \lambda \binder{ \refenvmv{x} }.\  \refenvnt{e}  ^{ \refenvnt{k}  }  \mid    \refenvnt{e_{{\mathrm{1}}}} ^{ \refenvnt{k}  }     \refenvnt{e_{{\mathrm{2}}}} ^{ \refenvnt{k}  }    \mid  \langle   \refenvnt{e} ^{  \refenvnt{k}  \refenvsym{+}  1   }   \rangle  \mid  { \mathdollar }_{ \refenvnt{k'} }\lbrace  \refenvnt{e} ^{ \refenvnt{k}  \refenvsym{-}  \refenvnt{k'}  }  \rbrace  \text{\ (where $\refenvnt{k} \ge \refenvnt{k'}$)}$ \\
                   &               &                    & $\mid$      & $  \token{rec}\ \binder{ \refenvmv{f} }(\binder{ \refenvmv{x} }).\  \refenvnt{e}  ^{ \refenvnt{k}  }  \mid \token{ref} \, \refenvsym{(}   \refenvnt{e} ^{ \refenvnt{k}  }   \refenvsym{)} \mid   \refenvnt{e_{{\mathrm{1}}}} ^{ \refenvnt{k}  }  \coloneqq  \refenvnt{e_{{\mathrm{2}}}} ^{ \refenvnt{k}  }   \mid  \token{deref}(  \refenvnt{e} ^{ \refenvnt{k}  }  ) $
  \\
  \multicolumn{5}{l}{\textbf{Type-Level Object}}                                                                                                                                                                                    \\
  Classifier Stack &               & $ \overrightarrow{ \gamma } $      & $\Coloneqq$ & $ \varepsilon  \mid  \overrightarrow{ \gamma }   \refenvsym{,}  \delta$                                                                                                                           \\
  Base Type        &               & $\iota$   & $\Coloneqq$ & $ \texttt{int}  \mid \ldots$                                                                                                                                     \\
  Type             &               & $\refenvnt{A}, \refenvnt{B}$ & $\Coloneqq$ & $\iota \mid \refenvnt{A}  \rightarrow  \refenvnt{B} \mid  \langle \refenvnt{A} \rangle^{ \gamma } \mid  \refenvnt{A} \/\ \token{ref} $                                                                                                           \\
  Typing Context   &               & $\Gamma$          & $\Coloneqq$ & $ \varepsilon  \mid \Gamma  \refenvsym{,}   \binder{ \refenvmv{x} }\mathord{:^{\binder{ \gamma } } } \refenvnt{A}  \mid \Gamma  \refenvsym{,}   \blacktriangleright ^{ \gamma }  \mid \Gamma  \refenvsym{,}   \blacktriangleleft _{ \refenvnt{k} }^{\binder{ \gamma } } $
\end{longtable}
For the surface expression, we follow Taha~\cite{conf/pepm/Taha99} in stratifying expressions into stages, where $ \refenvnt{e} ^{ \refenvnt{k}  } $ denotes an expression at stage $\refenvnt{k}$. The stage level $\refenvnt{k}$ is a non-negative integer that indicates the level of staging: $\refenvnt{k} = 0$ corresponds to the code-generating level, while higher values of $\refenvnt{k}$ correspond to deeper levels inside quotations. We omit the stage annotation and simply write $\refenvnt{e}$ when it does not matter for the discussion. It is easy to verify the following property of expressions:
\begin{lemma}\label{claim:expression-stratification}
  $ \token{Exp}^{ \refenvnt{k_{{\mathrm{1}}}} }  \subseteq  \token{Exp}^{ \refenvnt{k_{{\mathrm{2}}}} } $ if\/ $\refenvnt{k_{{\mathrm{1}}}} \le \refenvnt{k_{{\mathrm{2}}}}$.
\end{lemma}
The syntax of expressions includes integers, variables, functions, applications, quotations, and splices, as well as recursive functions and mutable-state operations. The intended behavior of these constructs is described informally in \Zcref{sec:informaldescription:metamlmsp}. Note that we unify the syntax of runtime execution and splicing: $ { \mathdollar }_{ 0 }\lbrace \refenvnt{e} \rbrace $ represents runtime execution of $\refenvnt{e}$, while $ { \mathdollar }_{ 1 }\lbrace \refenvnt{e} \rbrace $ represents a splice. For $\refenvnt{k} > 1$, $ { \mathdollar }_{ \refenvnt{k} }\lbrace \refenvnt{e} \rbrace $ represents a generalized form of splicing that embeds code fragments into deeper stages; we call this a \emph{deep splice}. In the rest of the paper, we use the term \emph{splice} for the construct $ { \mathdollar }_{ \refenvnt{k} }\lbrace \refenvnt{e} \rbrace $ at any $\refenvnt{k}$. We do not include primitive operations in the syntax, but they can be added without affecting the core ideas of our type system.

Note that expressions do not include type-level objects such as classifiers or types. As we will see in \Zcref{sec:semantics}, the dynamic semantics of our language is defined independently of the typing discipline.

We next define the type-level objects.
A classifier stack $ \overrightarrow{ \gamma } $ tracks classifiers at each stage, where the rightmost classifier corresponds to the current stage. In particular, we write $ \overrightarrow{ \gamma }^{+} $ to denote a non-empty classifier stack.
Types include base types, function types, and code types. A code type $ \langle \refenvnt{A} \rangle^{ \gamma } $ represents a code fragment of type $\refenvnt{A}$ that is valid under classifier $\gamma$.
The structure of typing contexts is more complex than that of traditional typed lambda calculi. We provide details shortly below. $ \token{Dom}_{C}( \Gamma ) $ returns the set of classifiers declared by the context $\Gamma$, and $ \token{FC}( \refenvnt{A} ) $ returns the set of classifiers that occur freely in the type $\refenvnt{A}$.

\subsection{Type-level Judgments}
We have four type-level judgments, defined in a mutually recursive manner:
\begin{description}
  \item[Well-formed Contexts $ \Gamma \vdash^{  \overrightarrow{ \gamma }^{+}  }\token{wf} $ :] the context $\Gamma$ is well-formed with $ \overrightarrow{ \gamma }^{+} $.
  \item[Well-formed Types $ \Gamma \vdash^{  \overrightarrow{ \gamma }^{+}  } \refenvnt{A} \ \token{wf} $ :] the type $\refenvnt{A}$ is well-formed under $\Gamma$ with $ \overrightarrow{ \gamma }^{+} $.
  \item[Reachability $ \Gamma \vdash \gamma_{{\mathrm{1}}} \preceq \gamma_{{\mathrm{2}}} $ :] $\gamma_{{\mathrm{1}}}$ is reachable from $\gamma_{{\mathrm{2}}}$.
  \item[Visibility $ \Gamma \vdash \gamma_{{\mathrm{1}}} \subseteq \gamma_{{\mathrm{2}}} $ :] $\gamma_{{\mathrm{1}}}$ is visible to $\gamma_{{\mathrm{2}}}$.
\end{description}
See \Zcref{sec:informaldescription:r2ecs} for the informal meaning of $ \preceq $ and $ \subseteq $.
Derivation rules are given in \Zcref{fig:surface-typelevel-judgments}.
\begin{figure}[bpt]
  \small
  \raggedright

  \fbox{$ \Gamma \vdash^{  \overrightarrow{ \gamma }^{+}  }\token{wf} $} \fbox{$ \Gamma \vdash^{  \overrightarrow{ \gamma }^{+}  } \refenvnt{A} \ \token{wf} $}\vspace{0.5em}

  \begin{center}
    \begin{prooftree}
      \caption{WF-Ctx-Emp}
      \label{rule:surface-wf-ctx-emp}
      \hypo{}
      \infer1{ \varepsilon \vdash^{  \exclam  }\token{wf} }
    \end{prooftree}
    \hspace{3em}
    \begin{prooftree}
      \caption{WF-Ctx-Var}
      \label{rule:surface-wf-ctx-var}
      \hypo{ \Gamma \vdash^{  \overrightarrow{ \gamma_{{\mathrm{1}}} }   \refenvsym{,}  \gamma_{{\mathrm{2}}} }\token{wf} }
      \hypo{ \Gamma \vdash^{  \overrightarrow{ \gamma_{{\mathrm{1}}} }   \refenvsym{,}  \gamma_{{\mathrm{2}}} } \refenvnt{A} \ \token{wf} }
      \hypo{\gamma_{{\mathrm{3}}} \, \not\in \,  \token{Dom}_{C}( \Gamma ) }
      \infer3{ \Gamma  \refenvsym{,}   \binder{ \refenvmv{x} }\mathord{:^{\binder{ \gamma_{{\mathrm{3}}} } } } \refenvnt{A}  \vdash^{  \overrightarrow{ \gamma_{{\mathrm{1}}} }   \refenvsym{,}  \gamma_{{\mathrm{3}}} }\token{wf} }
    \end{prooftree}
    \qquad
    \begin{prooftree}
      \caption{WF-Ctx-$ \blacktriangleright $}
      \label{rule:surface-wf-ctx-open}
      \hypo{ \Gamma \vdash^{  \overrightarrow{ \gamma_{{\mathrm{1}}} }   \refenvsym{,}  \gamma_{{\mathrm{2}}} }\token{wf} }
      \hypo{ \Gamma \vdash \gamma_{{\mathrm{3}}} \subseteq \gamma_{{\mathrm{2}}} }
      \infer2{ \Gamma  \refenvsym{,}   \blacktriangleright ^{ \gamma_{{\mathrm{3}}} }  \vdash^{  \overrightarrow{ \gamma_{{\mathrm{1}}} }   \refenvsym{,}  \gamma_{{\mathrm{2}}}  \refenvsym{,}  \gamma_{{\mathrm{3}}} }\token{wf} }
    \end{prooftree}
    \vspace{1em}

    \begin{prooftree}
      \caption{WF-Ctx-$ \blacktriangleleft $}
      \label{rule:surface-wf-ctx-shut}
      \hypo{ \Gamma \vdash^{  \overrightarrow{ \gamma_{{\mathrm{1}}} }   \refenvsym{,}    \delta _{ 0 }  ,\dotsc,  \delta _{ \refenvnt{k} }   }\token{wf} }
      \hypo{\gamma_{{\mathrm{2}}} \, \not\in \,  \token{Dom}_{C}( \Gamma ) }
      \infer2{ \Gamma  \refenvsym{,}   \blacktriangleleft _{ \refenvnt{k} }^{\binder{ \gamma_{{\mathrm{2}}} } }  \vdash^{  \overrightarrow{ \gamma_{{\mathrm{1}}} }   \refenvsym{,}  \gamma_{{\mathrm{2}}} }\token{wf} }
    \end{prooftree}
    \qquad
    \begin{prooftree}
      \caption{WF-T-$\langle\rangle$}
      \label{rule:surface-wf-t-code}
      \hypo{ \Gamma  \refenvsym{,}   \blacktriangleright ^{ \gamma_{{\mathrm{1}}} }  \vdash^{  \overrightarrow{ \gamma_{{\mathrm{2}}} }^{+}   \refenvsym{,}  \gamma_{{\mathrm{1}}} } \refenvnt{A} \ \token{wf} }
      \infer1{ \Gamma \vdash^{  \overrightarrow{ \gamma_{{\mathrm{2}}} }^{+}  }  \langle \refenvnt{A} \rangle^{ \gamma_{{\mathrm{1}}} }  \ \token{wf} }
    \end{prooftree}
    \vspace{1em}
  \end{center}

  \fbox{$ \Gamma \vdash \gamma_{{\mathrm{1}}} \preceq \gamma_{{\mathrm{2}}} $} \fbox{$ \Gamma \vdash \gamma_{{\mathrm{1}}} \subseteq \gamma_{{\mathrm{2}}} $}\quad Assuming that $\Gamma$ is well-formed.
  \vspace{0.5em}

  \begin{center}
    \begin{prooftree}
      \caption{$ \preceq $-Refl}
      \label{rule:surface-ltrans-refl}
      \hypo{\gamma \, \in \,  \token{Dom}_{C}( \Gamma )  \, \cup \, \refenvsym{\{}   \exclam   \refenvsym{\}}}
      \infer1{ \Gamma \vdash \gamma \preceq \gamma }
    \end{prooftree}
    \quad
    \begin{prooftree}
      \caption{$ \preceq $-Trans}
      \label{rule:surface-ltrans-trans}
      \hypo{ \Gamma \vdash \gamma_{{\mathrm{1}}} \preceq \gamma_{{\mathrm{2}}} }
      \hypo{ \Gamma \vdash \gamma_{{\mathrm{2}}} \preceq \gamma_{{\mathrm{3}}} }
      \infer2{ \Gamma \vdash \gamma_{{\mathrm{1}}} \preceq \gamma_{{\mathrm{3}}} }
    \end{prooftree}
    \quad
    \begin{prooftree}
      \caption{$ \preceq $-Var}
      \label{rule:surface-ltrans-var}
      \hypo{ \Gamma_{{\mathrm{1}}} \vdash^{  \overrightarrow{ \gamma_{{\mathrm{1}}} }   \refenvsym{,}  \gamma_{{\mathrm{2}}} }\token{wf} }
      \infer1{ \Gamma_{{\mathrm{1}}}  \refenvsym{,}   \binder{ \refenvmv{x} }\mathord{:^{\binder{ \gamma_{{\mathrm{3}}} } } } \refenvnt{A}   \refenvsym{,}  \Gamma_{{\mathrm{2}}} \vdash \gamma_{{\mathrm{2}}} \preceq \gamma_{{\mathrm{3}}} }
    \end{prooftree}
    \quad
    \begin{prooftree}
      \caption{$ \preceq $-$ \blacktriangleleft $}
      \label{rule:surface-ltrans-shut}
      \hypo{ \Gamma_{{\mathrm{1}}} \vdash^{  \overrightarrow{ \gamma_{{\mathrm{1}}} }   \refenvsym{,}   \delta _{ 0 }   \refenvsym{,}    \delta _{ 1 }  ,\dotsc,  \delta _{ \refenvnt{k} }   }\token{wf} }
      \hypo{\refenvnt{k} \ge 0}
      \infer2{ \Gamma_{{\mathrm{1}}}  \refenvsym{,}   \blacktriangleleft _{ \refenvnt{k} }^{\binder{ \gamma_{{\mathrm{2}}} } }   \refenvsym{,}  \Gamma_{{\mathrm{2}}} \vdash  \delta _{ 0 }  \preceq \gamma_{{\mathrm{2}}} }
    \end{prooftree}
    \vspace{1em}

    \begin{prooftree}
      \caption{$ \subseteq $-Lift}
      \label{rule:surface-gtrans-lift}
      \hypo{ \Gamma \vdash \gamma_{{\mathrm{1}}} \preceq \gamma_{{\mathrm{2}}} }
      \infer1{ \Gamma \vdash \gamma_{{\mathrm{1}}} \subseteq \gamma_{{\mathrm{2}}} }
    \end{prooftree}
    \qquad
    \begin{prooftree}
      \caption{$ \subseteq $-Trans}
      \label{rule:surface-gtrans-trans}
      \hypo{ \Gamma \vdash \gamma_{{\mathrm{1}}} \subseteq \gamma_{{\mathrm{2}}} }
      \hypo{ \Gamma \vdash \gamma_{{\mathrm{2}}} \subseteq \gamma_{{\mathrm{3}}} }
      \infer2{ \Gamma \vdash \gamma_{{\mathrm{1}}} \subseteq \gamma_{{\mathrm{3}}} }
    \end{prooftree}
    \qquad
    \begin{prooftree}
      \caption{$ \subseteq $-$ \blacktriangleleft $}
      \label{rule:surface-gtrans-shut}
      \hypo{ \Gamma_{{\mathrm{1}}} \vdash^{  \overrightarrow{ \gamma_{{\mathrm{1}}} }   \refenvsym{,}  \gamma_{{\mathrm{2}}} }\token{wf} }
      \infer1{ \Gamma_{{\mathrm{1}}}  \refenvsym{,}   \blacktriangleleft _{ \refenvnt{k} }^{\binder{ \gamma_{{\mathrm{3}}} } }   \refenvsym{,}  \Gamma_{{\mathrm{2}}} \vdash \gamma_{{\mathrm{2}}} \subseteq \gamma_{{\mathrm{3}}} }
    \end{prooftree}
  \end{center}

  \captionsetup{skip=4pt}
  \caption{Curated rules for type-level judgments in surface typing.}\label{fig:surface-typelevel-judgments}
  \vspace{-0.75\baselineskip}
\end{figure}
We begin with a description of the well-formedness of typing contexts, along with the structure of contexts themselves.
A typing context reflects the structure of the surrounding program: $ \binder{ \refenvmv{x} }\mathord{:^{\binder{ \gamma } } } \refenvnt{A} $ corresponds to a variable binding, $ \blacktriangleright ^{ \gamma } $ corresponds to a quotation, and $ \blacktriangleleft _{ \refenvnt{k} }^{\binder{ \gamma } } $ corresponds to splices (when $\refenvnt{k} > 1 $) or runtime execution (when $\refenvnt{k} = 0$). A typing context provides three kinds of information:
\begin{enumerate}
  \item\label{item:contextinfo:var} The mapping of variables, types and classifiers, which are provided by the item $ \binder{ \refenvmv{x} }\mathord{:^{\binder{ \gamma } } } \refenvnt{A} $.
  \item\label{item:contextinfo:clsstack} The current classifiers at each stage, provided by a classifier stack.
  \item\label{item:contextinfo:birelationalstructure} Birelational structure over classifiers with $ \preceq $ and $ \subseteq $.
\end{enumerate}
The information in \Zcref{item:contextinfo:var} is already provided in standard typed lambda-calculi. The remaining two points are less immediate. We discuss \Zcref{item:contextinfo:clsstack} next, and return to \Zcref{item:contextinfo:birelationalstructure} when introducing the reachability/visibility judgments.
\Zcref{table:context-elements} summarizes the structure induced by each context element; the
corresponding constraints are formalized by the context well-formedness and
reachability/visibility judgments.
\begin{table}[bt]
  \caption{Context Elements and Their Induced Structure}\label{table:context-elements}
  \vspace{-0.7em}
  \centering
  \small
  \begin{tabular}{ccccccc}
    \hline
    Context&
    Corresponding &
    \multicolumn{3}{c}{Effect on Classifier Stack} &
    Constraints &
    Constraints \\
    \cline{3-5}
    Element &
    Form &
    (before) &
    $\to$ &
    (after) &
    Introduced &
    Required \\
    \hline
    $ \binder{ \refenvmv{x} }\mathord{:^{\binder{ \gamma } } } \refenvnt{A} $ &
    $ \lambda \binder{ \refenvmv{x} }.\   \ldots  $&
    $ \overrightarrow{ \delta_{{\mathrm{1}}} }   \refenvsym{,}  \delta_{{\mathrm{2}}}$ &
    $\to$ &
    $ \overrightarrow{ \delta_{{\mathrm{1}}} }   \refenvsym{,}  \gamma$ &
    $ \delta_{{\mathrm{2}}} \preceq \gamma $ &
    \\
    $ \blacktriangleright ^{ \gamma } $ &
    $ \langle   \ldots   \rangle $ &
    $ \overrightarrow{ \delta_{{\mathrm{1}}} }   \refenvsym{,}  \delta_{{\mathrm{2}}}$ &
    $\to$ &
    $ \overrightarrow{ \delta_{{\mathrm{1}}} }   \refenvsym{,}  \delta_{{\mathrm{2}}}  \refenvsym{,}  \gamma$ &
    &
    $ \gamma \subseteq \delta_{{\mathrm{2}}} $\\
    $ \blacktriangleleft _{ \refenvnt{k} }^{\binder{ \gamma } } $ &
    $ { \mathdollar }_{ \refenvnt{k} }\lbrace  \ldots  \rbrace $ &
    $ \overrightarrow{ \delta }   \refenvsym{,}    \delta _{ 0 }  ,\dotsc,  \delta _{ \refenvnt{k} }  $ &
    $\to$ &
    $ \overrightarrow{ \delta }   \refenvsym{,}  \gamma$ &
    $  \delta _{ 0 }  \preceq \gamma $, $  \delta _{ \refenvnt{k} }  \subseteq \gamma $ &
    \\
    \hline
  \end{tabular}
  \vspace{-0.5\baselineskip}
\end{table}

A well-formed context judgment $ \Gamma \vdash^{  \overrightarrow{ \gamma }^{+}  }\token{wf} $ states that the context $\Gamma$ is well-formed and classifiers at each stage are given by the classifier stack $ \overrightarrow{ \gamma }^{+} $. We can confirm that the classifier stack in a well-formed context is unique:
\begin{lemma}
  If\/ $ \Gamma \vdash^{  \overrightarrow{ \gamma_{{\mathrm{1}}} }^{+}  }\token{wf} $ and\/ $ \Gamma \vdash^{  \overrightarrow{ \gamma_{{\mathrm{2}}} }^{+}  }\token{wf} $, then $ \overrightarrow{ \gamma_{{\mathrm{1}}} }^{+}  =  \overrightarrow{ \gamma_{{\mathrm{2}}} }^{+} $.
\end{lemma}
\noindent In this sense, the information in the classifier stack is already included in the context itself.

We review each rule for well-formed contexts in \Zcref{fig:surface-typelevel-judgments}.
\ref{rule:surface-wf-ctx-emp} states that the empty context is well-formed. It corresponds to a single stage with the initial classifier.
\ref{rule:surface-wf-ctx-var} states that adding an item $ \binder{ \refenvmv{x} }\mathord{:^{\binder{ \gamma_{{\mathrm{3}}} } } } \refenvnt{A} $ requires that $\gamma_{{\mathrm{3}}}$ is fresh and that $\refenvnt{A}$ is well-formed. This item also replaces the classifier of the current stage with $\gamma_{{\mathrm{3}}}$, thereby introducing a reachability relation $ \gamma_{{\mathrm{2}}} \preceq \gamma_{{\mathrm{3}}} $ (corresponding to\ref{rule:surface-ltrans-var}). Here, $\refenvmv{x}$ does not have to be fresh and can shadow another variable in $\Gamma$ with the same name.
\ref{rule:surface-wf-ctx-open} states that entering a quotation with the classifier $\gamma_{{\mathrm{3}}}$ requires that $\gamma_{{\mathrm{3}}}$ is visible from the current classifier $\gamma_{{\mathrm{2}}}$.
\ref{rule:surface-wf-ctx-shut} pops $\refenvnt{k}$ elements from the classifier stack: this represents the effect of splicing that escapes $\refenvnt{k}$ levels of quotations. It also introduces a new classifier $\gamma_{{\mathrm{2}}}$ replacing $\delta_{{\mathrm{0}}}$, thereby introducing a reachability relation $ \delta_{{\mathrm{0}}} \preceq \gamma_{{\mathrm{2}}} $ (corresponding to \ref{rule:surface-ltrans-shut}) and a visibility relation $  \delta _{ \refenvnt{k} }  \subseteq \gamma_{{\mathrm{2}}} $ (corresponding to \ref{rule:surface-gtrans-shut}).

The rules for well-formed types $ \Gamma \vdash^{  \overrightarrow{ \gamma }^{+}  } \refenvnt{A} \ \token{wf} $ are designed so that $\Gamma$ is well formed.
These rules are mostly standard and are therefore omitted. An exception is the rule for code types, \ref{rule:surface-wf-t-code}: it requires $\refenvnt{A}$ to be well formed under $\Gamma  \refenvsym{,}   \blacktriangleright ^{ \gamma_{{\mathrm{1}}} } $, meaning that code types introduce the same scoping structure as quotations.
Here, the well-formedness of $\Gamma  \refenvsym{,}   \blacktriangleright ^{ \gamma_{{\mathrm{1}}} } $ entails that $\gamma_{{\mathrm{1}}}$ is visible to the current scope of $\Gamma$.

Judgments for the reachability and visibility relations derive these relations from a well-formed context, as described in \zcref{table:context-elements}.

\subsection{Expression Typing}
We now turn to the typing rules for expressions, given by the judgment $ \Gamma \vdash^{  \overrightarrow{ \gamma }^{+}  } \refenvnt{e} : \refenvnt{A} $.
This judgment states that the expression $\refenvnt{e}$ has the type $\refenvnt{A}$ under the context $\Gamma$ and the classifier stack $ \overrightarrow{ \gamma }^{+} $. We present important rules in \Zcref{fig:surface-typing-judgments}.
\begin{figure}[bpt]
  \small
  \begin{center}
    \begin{prooftree}
      \caption{T-Int}
      \label{rule:surface-type-int}
      \hypo{ \Gamma \vdash^{  \overrightarrow{ \gamma }  }\token{wf} }
      \infer1{ \Gamma \vdash^{  \overrightarrow{ \gamma }  } \refenvnt{n} :  \texttt{int}  }
    \end{prooftree}
    \qquad
    \begin{prooftree}
      \caption{T-Var}
      \label{rule:surface-type-var}
      \hypo{ \Gamma \vdash^{  \overrightarrow{ \gamma_{{\mathrm{1}}} }   \refenvsym{,}  \gamma_{{\mathrm{2}}} }\token{wf} }
      \hypo{\token{lookup} \, \refenvsym{(}  \Gamma  \refenvsym{,}  \refenvmv{x}  \refenvsym{)} = \token{Done} \, \refenvsym{(}  \refenvnt{A}  \refenvsym{,}  \gamma_{{\mathrm{3}}}  \refenvsym{)}}
      \hypo{ \Gamma \vdash \gamma_{{\mathrm{3}}} \preceq \gamma_{{\mathrm{2}}} }
      \infer3{ \Gamma \vdash^{  \overrightarrow{ \gamma_{{\mathrm{1}}} }   \refenvsym{,}  \gamma_{{\mathrm{2}}} } \refenvmv{x} : \refenvnt{A} }
    \end{prooftree}
    \vspace{1em}

    \begin{prooftree}
      \caption{T-Func}
      \label{rule:surface-type-func}
      \hypo{ \Gamma  \refenvsym{,}   \binder{ \refenvmv{x} }\mathord{:^{\binder{ \gamma_{{\mathrm{1}}} } } } \refenvnt{A_{{\mathrm{1}}}}  \vdash^{  \overrightarrow{ \gamma_{{\mathrm{2}}} }   \refenvsym{,}  \gamma_{{\mathrm{1}}} } \refenvnt{e} : \refenvnt{A_{{\mathrm{2}}}} }
      \hypo{\gamma_{{\mathrm{1}}} \, \not\in \,  \token{FC}( \refenvnt{A_{{\mathrm{2}}}} ) }
      \hypo{ \Gamma \vdash^{  \overrightarrow{ \gamma_{{\mathrm{2}}} }   \refenvsym{,}  \gamma_{{\mathrm{3}}} }\token{wf} }
      \infer3{ \Gamma \vdash^{  \overrightarrow{ \gamma_{{\mathrm{2}}} }   \refenvsym{,}  \gamma_{{\mathrm{3}}} }  \lambda \binder{ \refenvmv{x} }.\  \refenvnt{e}  : \refenvnt{A_{{\mathrm{1}}}}  \rightarrow  \refenvnt{A_{{\mathrm{2}}}} }
    \end{prooftree}
    \qquad
    \begin{prooftree}
      \caption{T-Ref}
      \label{rule:surface-type-ref}
      \hypo{ \Gamma \vdash^{  \overrightarrow{ \gamma }^{+}  } \refenvnt{e} : \refenvnt{A} }
      \infer1{ \Gamma \vdash^{  \overrightarrow{ \gamma }^{+}  } \token{ref} \, \refenvsym{(}  \refenvnt{e}  \refenvsym{)} :  \refenvnt{A} \/\ \token{ref}  }
    \end{prooftree}
    \vspace{1em}

    \begin{prooftree}
      \caption{T-Quo}
      \label{rule:surface-type-quo}
      \hypo{ \Gamma  \refenvsym{,}   \blacktriangleright ^{ \gamma_{{\mathrm{1}}} }  \vdash^{  \overrightarrow{ \gamma_{{\mathrm{2}}} }^{+}   \refenvsym{,}  \gamma_{{\mathrm{1}}} } \refenvnt{e} : \refenvnt{A} }
      \infer1{ \Gamma \vdash^{  \overrightarrow{ \gamma_{{\mathrm{2}}} }^{+}  }  \langle  \refenvnt{e}  \rangle  :  \langle \refenvnt{A} \rangle^{ \gamma_{{\mathrm{1}}} }  }
    \end{prooftree}
    \qquad
    \begin{prooftree}
      \caption{T-Splice}
      \label{rule:surface-type-splice}
      \hypo{ \Gamma  \refenvsym{,}   \blacktriangleleft _{ \refenvnt{k} }^{\binder{ \gamma_{{\mathrm{1}}} } }  \vdash^{  \overrightarrow{ \gamma_{{\mathrm{3}}} }   \refenvsym{,}  \gamma_{{\mathrm{1}}} } \refenvnt{e} :  \langle \refenvnt{A} \rangle^{ \gamma_{{\mathrm{2}}} }  }
      \hypo{ \Gamma \vdash^{  \overrightarrow{ \gamma_{{\mathrm{3}}} }   \refenvsym{,}    \delta _{ 0 }  ,\dotsc,  \delta _{ \refenvnt{k} }   }\token{wf} }
      \hypo{ \Gamma \vdash \gamma_{{\mathrm{2}}} \preceq  \delta _{ \refenvnt{k} }  }
      \hypo{\gamma_{{\mathrm{1}}} \neq \gamma_{{\mathrm{2}}}}
      \infer4{ \Gamma \vdash^{  \overrightarrow{ \gamma_{{\mathrm{3}}} }   \refenvsym{,}    \delta _{ 0 }  ,\dotsc,  \delta _{ \refenvnt{k} }   }  { \mathdollar }_{ \refenvnt{k} }\lbrace \refenvnt{e} \rbrace  : \refenvnt{A} }
    \end{prooftree}
  \end{center}
  \captionsetup{skip=4pt}
  \caption{Curated rules for the expression typing judgment $ \Gamma \vdash^{  \overrightarrow{ \gamma }^{+}  } \refenvnt{e} : \refenvnt{A} $, from surface typing.}\label{fig:surface-typing-judgments}
  \vspace{-0.75\baselineskip}
\end{figure}
\ref{rule:surface-type-var} states that a variable $\refenvmv{x}$ can be used if its scope $\gamma_{{\mathrm{3}}}$ is reachable from the current scope $\gamma_{{\mathrm{2}}}$. Note that we overload the function $ \token{lookup} $ so that it looks up the type and classifier of $\refenvmv{x}$ in the context $\Gamma$; this function tries to find the closest binding for $\refenvmv{x}$ in $\Gamma$, and otherwise returns $ \token{Fail} $.
Due to this behavior of $ \token{lookup} $, shadowed variables cannot be used.
\ref{rule:surface-type-func} gives the standard typing rule for functions, except for classifier annotations. The side condition $\gamma_{{\mathrm{1}}} \, \not\in \,  \token{FC}( \refenvnt{A_{{\mathrm{2}}}} ) $ in \ref{rule:surface-type-func} states that the classifier $\gamma_{{\mathrm{1}}}$ of the parameter cannot appear in the return type, which ensures that parameter variables in quoted code fragments do not escape their scope.
\ref{rule:surface-type-quo} establishes a correspondence between the context element $ \blacktriangleright ^{ \gamma_{{\mathrm{1}}} } $ and quotation syntax, introducing the code type $ \langle \refenvnt{A} \rangle^{ \gamma_{{\mathrm{1}}} } $. Similarly, \ref{rule:surface-type-splice} establishes a correspondence between the context element $ \blacktriangleleft _{ \refenvnt{k} }^{\binder{ \gamma_{{\mathrm{1}}} } } $ and splice syntax, eliminating the code type $ \langle \refenvnt{A} \rangle^{ \gamma_{{\mathrm{2}}} } $. The side condition $ \Gamma \vdash \gamma_{{\mathrm{2}}} \preceq  \delta _{ \refenvnt{k} }  $ ensures that the scope of the spliced code is consistent with the current scope.
The other side condition $\gamma_{{\mathrm{1}}} \neq \gamma_{{\mathrm{2}}}$ ensures that use of $\gamma_{{\mathrm{1}}}$ in typing $\refenvnt{e}$ does not escape its scope. Note that $\gamma_{{\mathrm{1}}} \, \not\in \,  \token{FC}( \refenvnt{A} ) $ is not required because it is already guaranteed by the well-formedness of $\refenvnt{A}$. The rules for mutable state are standard.

\subsection{Typing Scoped Cross-Stage Persistence}\label{sec:surface-typing:csp}
We now discuss the subtlety of typing CSP in our type system. As we discussed in \Zcref{sec:informaldescription:metamlmsp}, our type system allows scoped CSP, a form of CSP in which a code fragment containing runtime variables can be safely evaluated at runtime. For example, the expression $ \lambda \binder{ \refenvmv{x} }.\   { \mathdollar }_{ 0 }\lbrace  \langle  \refenvmv{x}  \rangle  \rbrace  $ is well-typed in our type system:
\begin{center}
  \begin{prooftree}\small
    \hypo{\vdots}
    \infer[no rule]1{  \binder{ \refenvmv{x} }\mathord{:^{\binder{ \gamma_{{\mathrm{1}}} } } }  \texttt{int}    \refenvsym{,}   \blacktriangleleft _{ 0 }^{\binder{ \gamma_{{\mathrm{2}}} } }  \vdash^{ \gamma_{{\mathrm{2}}} }  \langle  \refenvmv{x}  \rangle  :  \langle  \texttt{int}  \rangle^{ \gamma_{{\mathrm{1}}} }  }
    \hypo{  \binder{ \refenvmv{x} }\mathord{:^{\binder{ \gamma_{{\mathrm{1}}} } } }  \texttt{int}   \vdash \gamma_{{\mathrm{1}}} \preceq \gamma_{{\mathrm{1}}} }
    \hypo{\gamma_{{\mathrm{2}}} \neq \gamma_{{\mathrm{1}}}}
    \infer[left label=\ref{rule:surface-type-splice}]3{  \binder{ \refenvmv{x} }\mathord{:^{\binder{ \gamma_{{\mathrm{1}}} } } }  \texttt{int}   \vdash^{  \gamma _{ 1 }  }  { \mathdollar }_{ 0 }\lbrace  \langle  \refenvmv{x}  \rangle  \rbrace  :  \texttt{int}  }
    \hypo{\gamma_{{\mathrm{1}}} \, \not\in \,  \token{FC}(  \texttt{int}  ) }
    \infer[left label=\ref{rule:surface-type-func}]2{ \varepsilon \vdash^{  \exclam  }  \lambda \binder{ \refenvmv{x} }.\   { \mathdollar }_{ 0 }\lbrace  \langle  \refenvmv{x}  \rangle  \rbrace   :  \texttt{int}   \rightarrow   \texttt{int}  }
  \end{prooftree}
\end{center}
We next consider a slightly different expression, $ \lambda \binder{ \refenvmv{x} }.\   \langle  \refenvmv{x}  \rangle  $, which returns a code fragment containing an occurrence of $\refenvmv{x}$ without executing it.
Traditional CSP accepts such an expression by embedding the value of $\refenvmv{x}$ into a code fragment. However, our type system rejects it because of the side condition of \ref{rule:surface-type-func}:
\begin{center}
  \begin{prooftree}\small
    \hypo{\vdots}
    \infer[no rule]1{  \binder{ \refenvmv{x} }\mathord{:^{\binder{ \gamma_{{\mathrm{1}}} } } }  \texttt{int}   \vdash^{ \gamma_{{\mathrm{1}}} }  \langle  \refenvmv{x}  \rangle  :  \langle  \texttt{int}  \rangle^{ \gamma_{{\mathrm{1}}} }  }
    \hypo{\gamma_{{\mathrm{1}}} \, \not\in \,  \token{FC}(  \langle  \texttt{int}  \rangle^{ \gamma_{{\mathrm{1}}} }  ) \text{\ does NOT hold}}
    \infer[left label=\ref{rule:surface-type-func} cannot be applied]2{( \lambda \binder{ \refenvmv{x} }.\   \langle  \refenvmv{x}  \rangle   \text{ fails to type }) }
  \end{prooftree}
\end{center}
In this type derivation, \ref{rule:surface-type-func} requires the side condition $\gamma_{{\mathrm{1}}} \, \not\in \,  \token{FC}(  \langle  \texttt{int}  \rangle^{ \gamma_{{\mathrm{1}}} }  ) $, which does not hold. The question, then, is how this side condition is justified.

While traditional CSP is achieved by embedding values into code fragments, scoped CSP in our language is achieved by deferring variable resolution until execution. In the expression $ \lambda \binder{ \refenvmv{x} }.\   { \mathdollar }_{ 0 }\lbrace  \langle  \refenvmv{x}  \rangle  \rbrace  $, the variable $\refenvmv{x}$ is not resolved during code generation; instead, it is resolved when the code fragment is executed. If we do not perform runtime execution, as in $ \lambda \binder{ \refenvmv{x} }.\   \langle  \refenvmv{x}  \rangle  $, then the code fragment $ \langle  \refenvmv{x}  \rangle $ is returned and escapes the scope of $\refenvmv{x}$. In \Zcref{sec:semantics}, we formalize this intuition in the operational semantics.

Assuming this behavior of scoped CSP, the side condition in \ref{rule:surface-type-func} effectively rejects programs that may cause code fragments to escape their scope. In this sense, the typing discipline with R2ECs has strong synergy with the semantics of scoped CSP, which is why we include scoped CSP in our language design. We further discuss this design choice in \Zcref{sec:relatedwork}.

\section{Polymorphic Classifiers}\label{sec:polycls}
In this section, we extend our type system with polymorphic classifiers, which allow us to type a wider range of staged programs in our language. In fact, this extension is essential for typing realistic staged programs. We first present motivating examples and then give the formal definition of the surface type system.

The idea of polymorphic classifiers already appears in the first paper on RECs~\cite{conf/aplas/KiselyovKS16}, but that work did not provide a formal treatment.
Polymorphic classifiers were not easy to accommodate in earlier RECs-based type systems and were not discussed in the subsequent work on RECs~\cite{conf/gpce/OishiK17,conf/gpce/IsodaYK24,conf/popl/LeeXKY26}.
Our contributions on polymorphic classifiers are twofold. First, we redesigned the RECs-based type system so that polymorphic classifiers can be integrated much more easily. Second, we show how to accommodate polymorphic classifiers in a multi-level setting, thereby enabling more flexible typing of multi-stage programs.
In this section, we first explain the informal idea of polymorphic classifiers and then give the formal definition of the corresponding surface type system.

\subsection{Informal Description}
Let us refactor the \lstinline|spow| example with \lstinline|sqr| by lifting the \lstinline|loop| function to the top level for readability.
\begin{lstlisting}
let sqr z = z * z in
let rec loop n yc = if n == 0 then .< 1 >. else if n == 1 then yc
                    else if n%2 == 0 then .< sqr .~(loop (n/2) yc) >.
                    else .< .~yc * .~(loop (n-1) yc) >. in
let spow n = .< fun x -> .~(loop n .< x >.) >. in ...
\end{lstlisting}
Since \lstinline|loop| is defined outside the scope of \lstinline|x|, we add a new parameter \lstinline|yc|. Then \lstinline|spow| passes the code fragment \lstinline|.< x >.| as the argument \lstinline|yc|, thereby injecting \lstinline|x| into the code generated by \lstinline|loop|.
However, this program is not typable in our current type system: we expect \lstinline|yc| to have the code type \lstinline|<int/h1>|, where $\delta_{{\mathrm{1}}}$ is the classifier for \lstinline|x|, but $\delta_{{\mathrm{1}}}$ is not in scope at the point where \lstinline|loop| is defined.
To type this program, we introduce \emph{polymorphism over classifiers}. (We omit unnecessary classifier annotations for readability in the following examples)
\begin{tikzpicture}[overlay,remember picture]
  \draw[blue,thick]
  ([yshift=-2pt]pic cs:polyclsi) -- ([yshift=-2pt]pic cs:polyclsie)
  node[pos=1,right=-3pt] {\small \ctext{1}};

  \draw[blue,thick]
  ([yshift=-2pt]pic cs:polycls_intch1) -- ([yshift=-2pt]pic cs:polycls_intch1e)
  node[pos=1,right=-3pt] {\small \ctext{2}};

  \draw[blue,thick]
  ([yshift=-2pt]pic cs:polycls_qh1) -- ([yshift=-2pt]pic cs:polycls_qh1e)
  node[pos=1,right=-3pt] {\small \ctext{3}};

  \draw[blue,thick]
  ([yshift=-2pt]pic cs:polycls_sqr) -- ([yshift=-2pt]pic cs:polycls_sqre)
  node[pos=1,right=-3pt] {\small \ctext{4}};

  \draw[blue,thick]
  ([yshift=-2pt]pic cs:polycls_apph2) -- ([yshift=-2pt]pic cs:polycls_apph2e)
  node[pos=1,right=-3pt] {\small \ctext{5}};
\end{tikzpicture}
\begin{lstlisting}
let sqr@g1 (z(|:int|))(|:int|) = z * z in
let rec loop (@*\tikzmark{polyclsi}*@)(|[dh1 :> cg1]@g2|)(@*\tikzmark{polyclsie}*@)   (n(|:int|)) (yc(|:(@*\tikzmark{polycls_intch1}*@)<int@ch1>(@*\tikzmark{polycls_intch1e}*@)|)  )(| : <int@ch1>|) =
  if n == 0 then .<@@h1 1 >. else if n == 1 then yc
  else if n%2 == 0 then (@*\tikzmark{polycls_qh1}*@).<@@h1(@*\tikzmark{polycls_qh1e}*@)   (@*\tikzmark{polycls_sqr}*@)sqr(@*\tikzmark{polycls_sqre}*@)   .~(loop (|[ch1]|) (n/2) yc) >. else ... in
let spow (n(|:int|))(|: <int->int@cg1>|) = .<@@g1fun x@h2(|:int|) -> .~(@g3loop (|(@*\tikzmark{polycls_apph2}*@)[ch2](@*\tikzmark{polycls_apph2e}*@)|)   n .<@@h2x>.) >. in ...
\end{lstlisting}
Here, \lstinline|[dh1 :> cg1]@g2| at \ctext{1} introduces a \emph{polymorphic classifier} \lstinline|ch1|, which is declared to satisfy $ \gamma_{{\mathrm{1}}} \preceq \delta_{{\mathrm{1}}} $.
It also introduces a new scope \lstinline|cg2|, where $\delta_{{\mathrm{1}}}$ is available. Hence, we can use the classifier \lstinline|ch1| in a type such as \lstinline|<int@ch1>| at \ctext{2}, and introduce a quotation with \lstinline|ch1| at \ctext{3}.
At \ctext{4}, we can use \lstinline|sqr| because \lstinline|cg1| is reachable from \lstinline|ch1|.
As a result, the type of the \lstinline|loop| function is \lstinline|[dh1 :> cg1](int -> <int@ch1> -> <int@ch1>)|, and we can instantiate \lstinline|ch1| later.

At \ctext{5}, \lstinline|loop [ch2]| instantiates the polymorphic classifier with \lstinline|ch2|. The type checker confirms that \lstinline|ch2| satisfies two constraints. First, $ \gamma_{{\mathrm{1}}} \preceq \delta_{{\mathrm{2}}} $ is required by the declaration of \lstinline|ch1|. Second, $\delta_{{\mathrm{2}}}$ should be available at the current scope: namely, $ \delta_{{\mathrm{2}}} \subseteq \gamma_{{\mathrm{3}}} $ is required, where \lstinline|cg3| is the classifier of the current scope. Since both constraints are satisfied, this instantiation is valid, and \lstinline|loop [ch2]| has type \lstinline|int -> <int@ch2> -> <int@ch2>|, which can be used to generate code fragments that include \lstinline|x|.
Thus, polymorphic classifiers allow us to name classifiers that are not available at the point of definition and instantiate them at the point of use.
This feature is particularly useful for defining reusable code generators, which are often defined outside the scope of the variables they manipulate.

Since our language is a multi-level calculus, we can also introduce polymorphic classifiers for multi-level code generation.
For example, consider a function \lstinline|lift_code| that takes a code fragment and returns its code representation. Note that \lstinline|.~2| is a deep splice, as explained in \Zcref{sec:surface-typing}.
\begin{lstlisting}
let lift_code@g1 (|[dh1 :> !, dh2 :> !]|)@h0 (x(|: <int@ch2>|))(|: <<int@ch2>@ch1>|) = .<@h1 .<@h2 .~2(x) >. >.
\end{lstlisting}
Its type is \lstinline|[dh1:>!, dh2:>!] <int@ch2> -> <<int@ch2>@ch1>|, where the classifiers \lstinline|ch1| and \lstinline|ch2| are introduced simultaneously.
This declaration introduces reachability $  \exclam  \preceq \delta_{{\mathrm{1}}} $ and $  \exclam  \preceq \delta_{{\mathrm{2}}} $, and also introduces a new scope with classifier \lstinline|ch0|, where $ \delta_{{\mathrm{1}}} \subseteq \delta_{{\mathrm{0}}} $ and $ \delta_{{\mathrm{2}}} \subseteq \delta_{{\mathrm{1}}} $ hold. These visibility constraints allow us to use a quotation with \lstinline|ch2| under \lstinline|ch1|. In this sense, the polymorphic classifier declaration \lstinline|[dh1:>!, dh2:>!]| differs from two separate declarations \lstinline|[dh1:>!][dh2:>!]|, since the latter does not introduce the relation $ \delta_{{\mathrm{2}}} \subseteq \delta_{{\mathrm{1}}} $.

\subsection{Introduction-Form Restriction}
Polymorphic classifiers must be used carefully with mutable state: if they are used without restriction, they can cause scope extrusion and lead to unsoundness. The core issue is that a polymorphic classifier may be instantiated differently for the same mutable location, even though the location's contents are not themselves polymorphic. This mismatch between the instantiated classifier and the actual classifier of the stored state leads to unsoundness. For example, the following program exhibits this problem:
\begin{lstlisting}
let r: (|[dh1:>c!](<int@ch1> reft)|) = (|[dh1:>c!]@h2|) ref .<@@h1 1 >. in
.<@@! fun x@g1:int -> .~( r (|[cg1]|) := .<@@g1 x >.; .<@@g1 x >. ) >.
deref (r (|[c!]|)) (* results in .< x >. *)
\end{lstlisting}
To avoid this unsoundness, one approach is to apply the usual value restriction for polymorphic types~\cite{journals/iandc/WrightF94} to polymorphic classifiers, restricting their use to values that do not perform computation. However, this restriction can be relaxed in certain cases.
\begin{lstlisting}
let gen_add_1 (|[dg1:>!,dg2:>!]|)@g3 (x(|:<<int@cg2>@cg1>|))(|:<<int@cg2>@cg1>|) = .<@g1 .<@g2 .~.~x + 1 >. >. in
.<@@! let f = (@*\tikzmark{polyif1}*@)(|[dg4:>!]|) fun (y(|:<int@cg4>|)@g5)(|:<int@cg4>|) -> .~(gen_add_1 (|[cg5,cg4]|) .<@@g5y>.)(@*\tikzmark{polyif1e}*@)   in ... >.
\end{lstlisting}
\begin{tikzpicture}[overlay,remember picture]
  \draw[blue,thick]
  ([yshift=-2pt]pic cs:polyif1) -- ([yshift=-2pt]pic cs:polyif1e)
  node[pos=1,right=-3pt] {\small \ctext{1}};
\end{tikzpicture}
Here, the function at \ctext{1} is typed with polymorphic classifiers, but it is not a value: there is a splice in the function body, yielding computation. Nevertheless, such computation in splices does not cause unsoundness, as we confirm later.
In order to allow such cases, we restrict the use of polymorphic classifiers to \emph{introduction forms}:
polymorphic classifiers are only introduced to functions, recursive functions and quotations, while computation is permitted in their bodies. We give a formal definition of introduction forms in the next section.

\subsection{Surface Typing}
Based on the intuition from the examples above, we extend the syntax of the surface type system.
We extend type syntax with polymorphic classifier types, and extend typing contexts with polymorphic classifier declarations.
We also introduce a new syntactic category \emph{introduction forms}.
They are defined as functions, recursive functions, and quotations. These introduction forms may still contain computation via splices: for example, $ \langle   { \mathdollar }_{ 1 }\lbrace  \refenvnt{e_{{\mathrm{1}}}} \refenvnt{e_{{\mathrm{2}}}}  \rbrace   \rangle $ is a valid introduction form at level 0, in which the splice evaluates the computation $ \refenvnt{e_{{\mathrm{1}}}} \refenvnt{e_{{\mathrm{2}}}} $.
\begin{center}
  \begin{tabular}{cllcl}
    Type            &  & $\refenvnt{A}, \refenvnt{B}$ & $\Coloneqq$ & $\cdots \mid  \forall    \binder{  \gamma _{ 1 }  }  \mathbin{:\succeq}    \delta _{ 1 }  ,\dotsc,\binder{  \gamma _{ \refenvnt{k} }  }  \mathbin{:\succeq}   \delta  _{ \refenvnt{k} }  . \refenvnt{A} $       \\
    Typing Context  &  & $\Gamma$          & $\Coloneqq$ & $\cdots \mid \Gamma  \refenvsym{,}   [    \binder{  \gamma _{ 1 }  }  \mathbin{:\succeq}    \delta _{ 1 }  ,\dotsc,\binder{  \gamma _{ \refenvnt{k} }  }  \mathbin{:\succeq}   \delta  _{ \refenvnt{k} }   ]^{\binder{  \gamma _{ 0 }  } } $        \\
    Introduction Forms &  & $ {e_{I} } ^{ \refenvnt{k}  } $    & $\Coloneqq$ & $  \lambda \binder{ \refenvmv{x} }.\  \refenvnt{e}  ^{ \refenvnt{k}  }  \mid   \token{rec}\ \binder{ \refenvmv{f} }(\binder{ \refenvmv{x} }).\  \refenvnt{e}  ^{ \refenvnt{k}  }   \mid  \langle   \refenvnt{e} ^{  \refenvnt{k}  \refenvsym{+}  1   }   \rangle $
  \end{tabular}
\end{center}
Derivation rules for the new type form and context form are presented in \Zcref{fig:surface-type-polycls}.
\Zcref{table:polycls} summarizes the induced structure of polymorphic classifiers, which is formally described in the derivation rules.
\begin{figure}[bpt]\small
  \begin{center}
    \begin{prooftree}
      \caption{WF-Ctx-Polycls}\label{rule:surface-wf-ctx-polycls}
      \hypo{ \Gamma \vdash^{  \overrightarrow{ \delta }   \refenvsym{,}   \gamma' _{ 0 }  }\token{wf} }
      \hypo{ \gamma _{ \refenvnt{i} }  \, \not\in \,  \token{Dom}_{C}( \Gamma )  \text{\ for\ } 0 \le \refenvnt{i} \le \refenvnt{k}}
      \hypo{ \Gamma \vdash  \gamma' _{ \refenvnt{i}  \refenvsym{+}  1 }  \subseteq  \gamma' _{ \refenvnt{i} }   \text{\ for \ } 0 \le \refenvnt{i} < \refenvnt{k}}
      \infer3{ \Gamma  \refenvsym{,}   [    \binder{  \gamma _{ 1 }  }  \mathbin{:\succeq}    \gamma' _{ 1 }  ,\dotsc,\binder{  \gamma _{ \refenvnt{k} }  }  \mathbin{:\succeq}   \gamma'  _{ \refenvnt{k} }   ]^{\binder{  \gamma _{ 0 }  } }  \vdash^{  \overrightarrow{ \delta_{{\mathrm{1}}} }   \refenvsym{,}   \gamma _{ 0 }  }\token{wf} }
    \end{prooftree}

    \begin{prooftree}
      \caption{WF-T-Polycls}\label{rule:surface-wf-t-polycls}
      \hypo{ \Gamma  \refenvsym{,}   [    \binder{  \gamma _{ 1 }  }  \mathbin{:\succeq}    \gamma' _{ 1 }  ,\dotsc,\binder{  \gamma _{ \refenvnt{k} }  }  \mathbin{:\succeq}   \gamma'  _{ \refenvnt{k} }   ]^{\binder{  \gamma _{ 0 }  } }  \vdash^{  \overrightarrow{ \delta_{{\mathrm{1}}} }   \refenvsym{,}   \gamma _{ 0 }  } \refenvnt{A} \ \token{wf} }
      \hypo{ \gamma _{ 0 }  \, \not\in \,  \token{FC}( \refenvnt{A} ) }
      \hypo{ \Gamma \vdash^{  \overrightarrow{ \delta_{{\mathrm{1}}} }   \refenvsym{,}   \gamma' _{ 0 }  }\token{wf} }
      \infer3{ \Gamma \vdash^{  \overrightarrow{ \delta_{{\mathrm{1}}} }   \refenvsym{,}   \gamma' _{ 0 }  }  \forall    \binder{  \gamma _{ 1 }  }  \mathbin{:\succeq}    \gamma' _{ 1 }  ,\dotsc,\binder{  \gamma _{ \refenvnt{k} }  }  \mathbin{:\succeq}   \gamma'  _{ \refenvnt{k} }  . \refenvnt{A}  \ \token{wf} }
    \end{prooftree}
    \vspace{0.5em}

    \begin{prooftree}
      \caption{$ \preceq $-Polycls}\label{rule:surface-ltrans-polycls}
      \hypo{ \Gamma_{{\mathrm{1}}} \vdash^{  \overrightarrow{ \delta }   \refenvsym{,}   \gamma' _{ 0 }  }\token{wf} }
      \hypo{0 \le \refenvnt{i} \le \refenvnt{k}}
      \infer2{ \Gamma_{{\mathrm{1}}}  \refenvsym{,}   [    \binder{  \gamma _{ 1 }  }  \mathbin{:\succeq}    \gamma' _{ 1 }  ,\dotsc,\binder{  \gamma _{ \refenvnt{k} }  }  \mathbin{:\succeq}   \gamma'  _{ \refenvnt{k} }   ]^{\binder{  \gamma _{ 0 }  } }   \refenvsym{,}  \Gamma_{{\mathrm{2}}} \vdash  \gamma' _{ \refenvnt{i} }  \preceq  \gamma _{ \refenvnt{i} }  }
    \end{prooftree}
    \qquad
    \begin{prooftree}
      \caption{$ \subseteq $-Polycls}\label{rule:surface-gtrans-polycls}
      \hypo{0 \le \refenvnt{i} < \refenvnt{k}}
      \infer1{ \Gamma_{{\mathrm{1}}}  \refenvsym{,}   [    \binder{  \gamma _{ 1 }  }  \mathbin{:\succeq}    \gamma' _{ 1 }  ,\dotsc,\binder{  \gamma _{ \refenvnt{k} }  }  \mathbin{:\succeq}   \gamma'  _{ \refenvnt{k} }   ]^{\binder{  \gamma _{ 0 }  } }   \refenvsym{,}  \Gamma_{{\mathrm{2}}} \vdash  \gamma _{ \refenvnt{i}  \refenvsym{+}  1 }  \subseteq  \gamma _{ \refenvnt{i} }  }
    \end{prooftree}
    \vspace{0.5em}

    \begin{prooftree}
      \caption{T-Polycls}\label{rule:surface-type-polycls}
      \hypo{ \Gamma  \refenvsym{,}   [    \binder{  \gamma _{ 1 }  }  \mathbin{:\succeq}    \gamma' _{ 1 }  ,\dotsc,\binder{  \gamma _{ \refenvnt{k} }  }  \mathbin{:\succeq}   \gamma'  _{ \refenvnt{k} }   ]^{\binder{  \gamma _{ 0 }  } }  \vdash^{  \overrightarrow{ \delta_{{\mathrm{1}}} }   \refenvsym{,}   \gamma _{ 0 }  } {e_{I} } : \refenvnt{A} }
      \hypo{ \gamma _{ 0 }  \, \not\in \,  \token{FC}( \refenvnt{A} ) }
      \hypo{ \Gamma \vdash^{  \overrightarrow{ \delta_{{\mathrm{1}}} }   \refenvsym{,}   \gamma' _{ 0 }  }\token{wf} }
      \infer3{ \Gamma \vdash^{  \overrightarrow{ \delta_{{\mathrm{1}}} }   \refenvsym{,}   \gamma' _{ 0 }  } {e_{I} } :  \forall    \binder{  \gamma _{ 1 }  }  \mathbin{:\succeq}    \gamma' _{ 1 }  ,\dotsc,\binder{  \gamma _{ \refenvnt{k} }  }  \mathbin{:\succeq}   \gamma'  _{ \refenvnt{k} }  . \refenvnt{A}  }
    \end{prooftree}

    \begin{prooftree}
      \caption{T-Appcls}\label{rule:surface-type-appcls}
      \hypo{ \Gamma \vdash^{  \overrightarrow{ \delta' }   \refenvsym{,}   \delta _{ 0 }  } \refenvnt{e} :  \forall    \binder{  \gamma _{ 1 }  }  \mathbin{:\succeq}    \gamma' _{ 1 }  ,\dotsc,\binder{  \gamma _{ \refenvnt{k} }  }  \mathbin{:\succeq}   \gamma'  _{ \refenvnt{k} }  . \refenvnt{A}  }
      \hypo{ \Gamma \vdash  \gamma' _{ \refenvnt{i} }  \preceq  \delta _{ \refenvnt{i} }   \text{\ for\ }  1 \le \refenvnt{i} \le \refenvnt{k}}
      \hypo{ \Gamma \vdash  \delta _{ \refenvnt{i}  \refenvsym{+}  1 }  \subseteq  \delta _{ \refenvnt{i} }   \text{\ for\ }  0 \le \refenvnt{i} < \refenvnt{k}}
      \infer3{ \Gamma \vdash^{  \overrightarrow{ \delta' }   \refenvsym{,}   \delta _{ 0 }  } \refenvnt{e} : \refenvnt{A}  \refenvsym{[}    \gamma _{ 1 }  \coloneqq  \delta _{ 1 }  ,\dotsc,  \gamma _{ \refenvnt{k} }  \coloneqq  \delta _{ \refenvnt{k} }    \refenvsym{]} }
    \end{prooftree}
  \end{center}
  \captionsetup{skip=4pt}
  \caption{Surface typing for polymorphic classifiers.}\label{fig:surface-type-polycls}
  \vspace{-0.75\baselineskip}
\end{figure}
\begin{table}[b]
  \caption{The Induced Structure of Polymorphic Classifiers}\label{table:polycls}
  \vspace{-0.75\baselineskip}
  \centering
  \small
  \begin{tabular}{cccccc}
    \hline
    \multirow{2}{*}{\shortstack[c]{Context Element}}&
    \multicolumn{3}{c}{Effect on Classifier Stack} &
    Constraints &
    Constraints \\
    \cline{2-4}
    &
    (before) &
    $\to$ &
    (after) &
    Introduced &
    Required \\
    \hline
    \multirow{2}{*}{$ [    \binder{  \gamma _{ 1 }  }  \mathbin{:\succeq}    \gamma' _{ 1 }  ,\dotsc,\binder{  \gamma _{ \refenvnt{k} }  }  \mathbin{:\succeq}   \gamma'  _{ \refenvnt{k} }   ]^{\binder{  \gamma _{ 0 }  } } $} &
    \multirow{2}{*}{$ \overrightarrow{ \delta_{{\mathrm{1}}} }   \refenvsym{,}   \gamma' _{ 0 } $} &
    \multirow{2}{*}{$\to$} &
    \multirow{2}{*}{$ \overrightarrow{ \delta_{{\mathrm{1}}} }   \refenvsym{,}   \gamma _{ 0 } $} &
    $  \gamma' _{ \refenvnt{i} }  \preceq  \gamma _{ \refenvnt{i} }  \ (0\le i\le k)$&
    \multirow{2}{*}{$  \gamma' _{ \refenvnt{i}  \refenvsym{+}  1 }  \subseteq  \gamma' _{ \refenvnt{i} }  \ (0 \le i < k)$}\\
    &&&&$  \gamma _{ \refenvnt{i}  \refenvsym{+}  1 }  \subseteq  \gamma _{ \refenvnt{i} }  \ (0 \le i < k)$&\\
    \hline
  \end{tabular}
\end{table}
\ref{rule:surface-wf-ctx-polycls}, \ref{rule:surface-ltrans-polycls}, and \ref{rule:surface-gtrans-polycls} describe relations between classifiers that are declared or used in a polymorphic classifier declaration. Given a declaration $ [    \binder{  \gamma _{ 1 }  }  \mathbin{:\succeq}    \gamma' _{ 1 }  ,\dotsc,\binder{  \gamma _{ \refenvnt{k} }  }  \mathbin{:\succeq}   \gamma'  _{ \refenvnt{k} }   ]^{\binder{  \gamma _{ 0 }  } } $ at $ \gamma' _{ 0 } $, we have the following relations. \Scope and \Stage represent $ \preceq $ and $ \subseteq $, respectively.
\begin{center}
  \begin{tikzpicture}[
      >=Latex,
      snakearrow/.style={
          <-,
          thick,
          decorate,
          decoration={
              snake,
              amplitude=0.5mm,
              segment length=2.5mm,
              pre length=1em
            }
        },
      varrow/.style={<-,thick}
    ]

    \matrix (m) [matrix of nodes,
    nodes={
        draw,
        circle,
        minimum size=6mm,
        inner sep=1pt,
        font=\small
      },
    row sep=4mm,
    column sep=12mm,
    nodes={anchor=center},
    ampersand replacement=\&
    ] {
    $\gamma_0'$ \& $\gamma_1'$ \& |[draw=none]| $\cdots$ \& $\gamma_k'$ \\
    $\gamma_0$  \& $\gamma_1$  \& |[draw=none]| $\cdots$ \& $\widetilde{\gamma_k}$ \\
    };

    \draw[snakearrow] (m-1-1) -- (m-1-2)
    node[midway,above,draw=none] {$ \subseteq $};;
    \draw[snakearrow] (m-1-2) -- (m-1-3);
    \draw[snakearrow] (m-1-3) -- (m-1-4);

    \draw[snakearrow] (m-2-1) -- (m-2-2);
    \draw[snakearrow] (m-2-2) -- (m-2-3);
    \draw[snakearrow] (m-2-3) -- (m-2-4);

    \draw[varrow] (m-1-1) -- (m-2-1)
    node[midway,left,draw=none] {$ \preceq $};;
    \draw[varrow] (m-1-2) -- (m-2-2);
    \draw[varrow] (m-1-4) -- (m-2-4);

  \end{tikzpicture}
\end{center}
When introducing a polymorphic classifier type, \ref{rule:surface-wf-t-polycls} requires the side condition $ \gamma _{ 0 }  \, \not\in \,  \token{FC}( \refenvnt{A} ) $, which safely hides $ \gamma _{ 0 } $ from the type. The introduction rule, \ref{rule:surface-type-polycls}, applies only to introduction forms: functions, recursive functions, and quotations, which may contain splices of level-0 expressions. Eliminating polymorphic classifier types requires that the instantiated classifiers satisfy the constraints declared by the polymorphic classifier type.

\section{Operational Semantics}\label{sec:semantics}
This section presents the operational semantics of our MetaML-style multi-level calculus, given as a definitional interpreter. The runtime syntax and operational semantics are type-erased. Hence, the typing information presented in \Zcref{sec:surface-typing} is used purely for static checking and does not affect the runtime behavior of programs.
Our operational semantics is characterized by three key design choices.

\textbf{Global name generation}: Our semantics generates fresh names and reserves them to manage variable names that appear in generated code.
This approach is common in multi-stage languages to avoid name conflicts in generated code fragments~\cite{conf/fossacs/MoggiF03,conf/gpce/CalcagnoTHL03,conf/aplas/KiselyovKS16}.
It is useful to model the behavior of effectful code generation because it allows us to capture situations in which variables temporarily escape their lexical scope while open code fragments are stored in a global store.

\textbf{Semantics via a definitional interpreter}: We formalize the semantics using a definitional interpreter rather than a substitution-based reduction semantics.
There are two main reasons for this choice.
First, globally reserved names make $\alpha$-equivalence awkward in a substitution-based reduction semantics. Our definitional interpreter instead deterministically renames binders, yielding a simpler formalization without the need for $\alpha$-equivalence.
Second, the definitional interpreter fits well with Siek's \textit{three-easy-lemmas} methodology~\cite{blog/Siek13,conf/popl/AminR17}, which we use to structure the type soundness proof in later sections.

\textbf{Scoped CSP}: Our semantics realizes scoped CSP, in which runtime execution of code fragments respects lexical scoping. This design choice is motivated by the synergy between the semantics of scoped CSP and our typing discipline with R2ECs, as discussed in \Zcref{sec:surface-typing:csp}. \Zcref{sec:semantics:examples} provides evaluation examples to illustrate how our semantics ensures lexical scoping by variable renaming.

\subsection{Runtime Objects}
We start by defining syntactic categories of runtime objects. We have a set of locations $ \token{Loc} $, ranged over by $\refenvmv{l}$. Locations represent pointers to mutable reference cells.
\begin{longtable}{cllcl}
  Value          & $\refenvnt{v}$     & $\in  \token{Val} $ & $\Coloneqq$ & $\refenvnt{n} \mid \token{clos} \, \refenvsym{(}  E  \refenvsym{,}  R  \refenvsym{,}    \lambda \binder{ \refenvmv{x} }.\  \refenvnt{e}  ^{ 0  }   \refenvsym{)} \mid \token{clos} \, \refenvsym{(}  E  \refenvsym{,}  R  \refenvsym{,}    \token{rec}\ \binder{ \refenvmv{f} }(\binder{ \refenvmv{x} }).\  \refenvnt{e}  ^{ 0  }   \refenvsym{)} \mid  \langle   \refenvnt{e} ^{ 0  }   \rangle  \mid \refenvmv{l}$ \\
  Reserved Names & $N$    & $\in  \token{RN} $    & $\Coloneqq$ & $ \varepsilon  \mid N  \refenvsym{,}  \refenvmv{x}$                                                                                   \\
  Value Env.     & $E$  & $\in  \token{VEnv} $  & $\Coloneqq$ & $ \varepsilon  \mid E  \refenvsym{,}  \refenvmv{x}  \coloneqq  \refenvnt{v}$                                                                              \\
  Renaming Env.  & $R$  & $\in  \token{REnv} $  & $\Coloneqq$ & $ \token{id}  \mid R  \refenvsym{,}  \refenvmv{x}  \coloneqq  \refenvmv{y}$                                                                            \\
  Store          & $S$ & $\in  \token{Store} $ & $\Coloneqq$ & $ \varepsilon  \mid S  \refenvsym{,}  \refenvmv{l}  \coloneqq  \refenvnt{v}$
\end{longtable}
Values include integers, closures for both non-recursive and recursive functions, quoted code fragments, and locations.
Reserved names track the names generated during execution, ensuring that fresh names can be generated without conflicts.
Value environments map variables to values, and stores model mutable state by mapping locations to values.
Renaming environments are specific to our multi-stage setting: they track variable renaming to avoid unintended name capture.
$ \token{id} $ is the identity renaming, which maps each variable to itself.
We explain how these components work through examples in \Zcref{sec:semantics:examples}.

\subsection{Definitional Interpreter}
We define the operational semantics of our language using a definitional interpreter, following the style of Siek~\cite{blog/Siek13}.
We use a pseudo-functional language similar to ML to write the interpreter.
First, we define an auxiliary type $ \token{Result} $ to represent the possible outcomes of evaluation.
\[
  \token{type}\  \token{Result}( \tau ) \ \Coloneqq  \token{Timeout}  \mid  \token{Fail}  \mid  \token{Done} (\tau)
\]
Here, $ \token{Timeout} $ represents a timeout due to exceeding the step limit, $ \token{Fail} $ represents a runtime error (e.g., an undefined variable or application of a non-function), and $ \token{Done} (\tau)$ represents a successful outcome, yielding a value of type $\tau$.
We then introduce monadic notation to describe operations on $ \token{Result} $:
\begin{align*}
   \token{return}  (M) \equiv\        &  \token{Done}  (M)                                                \\
  X \leftarrow M_1;\ M_2 \equiv\  & \token{match}\ M_1\ \token{with}                              \\
                                  & \token{case}\  \token{Timeout}  \Rightarrow  \token{Timeout}          \\
                                  & \token{case}\   \token{Fail}  \Rightarrow  \token{Fail}               \\
                                  & \token{case}\   \token{Done}  (R) \Rightarrow [ X \coloneq R ]M_2
\end{align*}
We next introduce auxiliary functions. For brevity, we omit the definitions of these functions here, and provide their type signatures and brief descriptions:
{\small
\begin{align*}
   \token{lookup}        & :  \token{VEnv}  \to  \token{Var}  \to  \token{Result}(  \token{Val}  )                   &  & \text{(look up variable in value environment)} \\
   \token{gensym}        & :  \token{RN}  \to  \token{Result}(  \token{Var}  )                                   &  & \text{(generate a fresh variable name)}        \\
   \token{rename}        & :  \token{REnv}  \to  \token{Var}  \to  \token{Var}                           &  & \text{(apply renaming)}                        \\
   \token{alloc}         & :  \token{Store}  \to  \token{Result}(  \token{Loc}  )                                &  & \text{(allocate a new location)}               \\
   \token{lookupStore}  & :  \token{Store}  \to  \token{Loc}  \to  \token{Result}(  \token{Val}  )                  &  & \text{(look up value at location)}             \\
   \token{updateStore}        & :  \token{Store}  \to  \token{Loc}  \to  \token{Val}  \to  \token{Result}(  \token{Store}  )  &  & \text{(update value at location)}              \\
   \token{toQuote}       & :  \token{Val}  \to  \token{Result}(  \token{Exp}^{ 0 }  )                                &  & \text{(decompose value as quoted code)}      \\
   \token{toLocation}    & :  \token{Val}  \to  \token{Result}(  \token{Loc}  )                                  &  & \text{(decompose value as a location)}
\end{align*}
}

\begin{figure}[bpt]
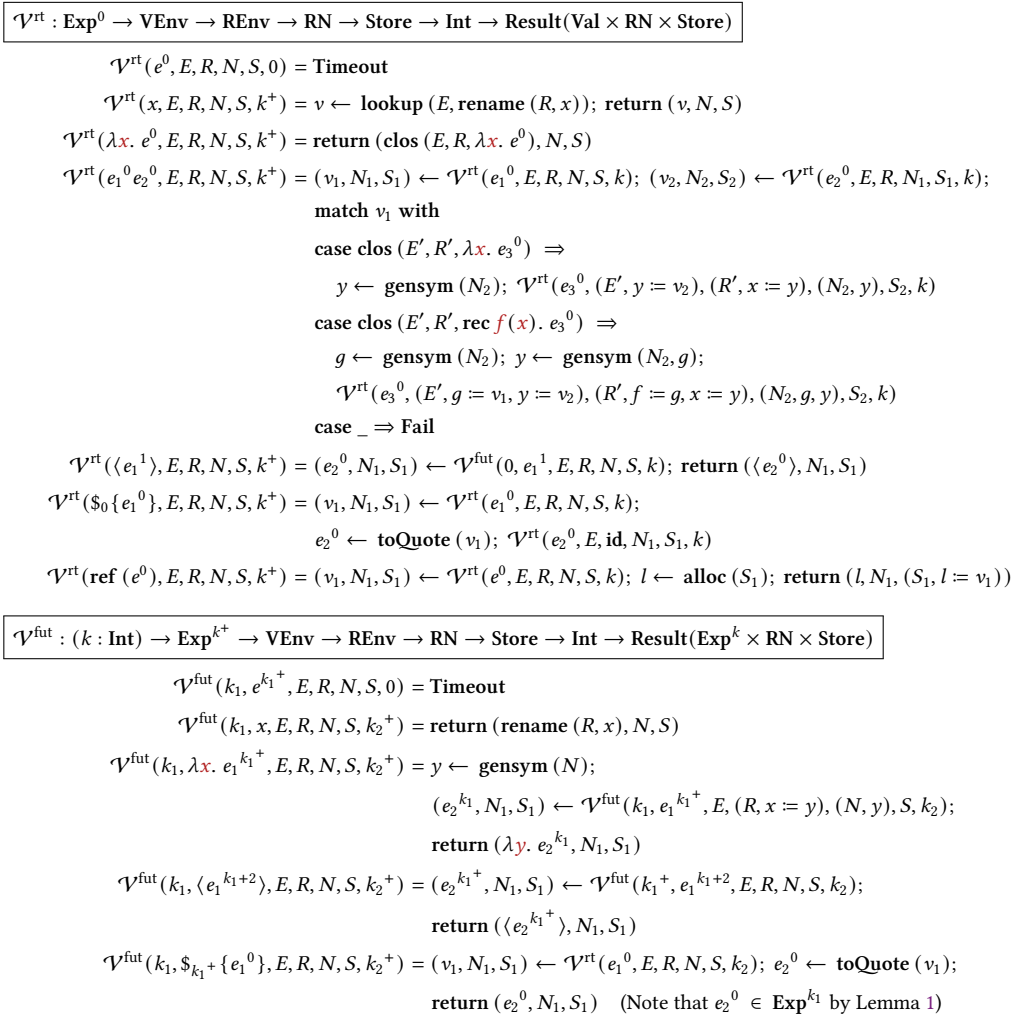

  \raggedright
  \footnotesize
  \noindent\fbox{$ \mathcal{V}^{\mathrm{rt} }   :  \token{Exp}^{ 0 }  \to  \token{VEnv}  \to  \token{REnv}  \to  \token{RN}  \to  \token{Store}  \to  \token{Int}  \to  \token{Result}(  \token{Val}  \times  \token{RN}  \times  \token{Store}  ) $}
  \begin{align*}
    \mathcal{V}^{\mathrm{rt} }  \refenvsym{(}   \refenvnt{e} ^{ 0  }   \refenvsym{,}  E  \refenvsym{,}  R  \refenvsym{,}  N  \refenvsym{,}  S  \refenvsym{,}  0  \refenvsym{)}           & =  \token{Timeout}                                                                                                                                 \\
    \mathcal{V}^{\mathrm{rt} }  \refenvsym{(}  \refenvmv{x}  \refenvsym{,}  E  \refenvsym{,}  R  \refenvsym{,}  N  \refenvsym{,}  S  \refenvsym{,}   \refenvnt{k} ^{+}   \refenvsym{)}            & = \refenvnt{v}  \leftarrow \, \token{lookup} \, \refenvsym{(}  E  \refenvsym{,}  \token{rename} \, \refenvsym{(}  R  \refenvsym{,}  \refenvmv{x}  \refenvsym{)}  \refenvsym{)};\ \token{return} \, \refenvsym{(}  \refenvnt{v}  \refenvsym{,}  N  \refenvsym{,}  S  \refenvsym{)}                                                                        \\
    \mathcal{V}^{\mathrm{rt} }  \refenvsym{(}    \lambda \binder{ \refenvmv{x} }.\  \refenvnt{e}  ^{ 0  }   \refenvsym{,}  E  \refenvsym{,}  R  \refenvsym{,}  N  \refenvsym{,}  S  \refenvsym{,}   \refenvnt{k} ^{+}   \refenvsym{)}  & = \token{return} \, \refenvsym{(}  \token{clos} \, \refenvsym{(}  E  \refenvsym{,}  R  \refenvsym{,}    \lambda \binder{ \refenvmv{x} }.\  \refenvnt{e}  ^{ 0  }   \refenvsym{)}  \refenvsym{,}  N  \refenvsym{,}  S  \refenvsym{)}                                                                                          \\
    \mathcal{V}^{\mathrm{rt} }  \refenvsym{(}     \refenvnt{e_{{\mathrm{1}}}} ^{ 0  }  \refenvnt{e_{{\mathrm{2}}}}  ^{ 0  }   \refenvsym{,}  E  \refenvsym{,}  R  \refenvsym{,}  N  \refenvsym{,}  S  \refenvsym{,}   \refenvnt{k} ^{+}   \refenvsym{)}  & = \refenvsym{(}  \refenvnt{v_{{\mathrm{1}}}}  \refenvsym{,}  N_{{\mathrm{1}}}  \refenvsym{,}  S_{{\mathrm{1}}}  \refenvsym{)}  \leftarrow  \mathcal{V}^{\mathrm{rt} }  \refenvsym{(}   \refenvnt{e_{{\mathrm{1}}}} ^{ 0  }   \refenvsym{,}  E  \refenvsym{,}  R  \refenvsym{,}  N  \refenvsym{,}  S  \refenvsym{,}  \refenvnt{k}  \refenvsym{)};\ \refenvsym{(}  \refenvnt{v_{{\mathrm{2}}}}  \refenvsym{,}  N_{{\mathrm{2}}}  \refenvsym{,}  S_{{\mathrm{2}}}  \refenvsym{)}  \leftarrow  \mathcal{V}^{\mathrm{rt} }  \refenvsym{(}   \refenvnt{e_{{\mathrm{2}}}} ^{ 0  }   \refenvsym{,}  E  \refenvsym{,}  R  \refenvsym{,}  N_{{\mathrm{1}}}  \refenvsym{,}  S_{{\mathrm{1}}}  \refenvsym{,}  \refenvnt{k}  \refenvsym{)}; \\
                                                          & \:\quad  \token{match}\  \refenvnt{v_{{\mathrm{1}}}} \ \token{with}                                                                                                                        \\
                                                          & \:\quad  \token{case}\  \token{clos} \, \refenvsym{(}  E'  \refenvsym{,}  R'  \refenvsym{,}    \lambda \binder{ \refenvmv{x} }.\  \refenvnt{e_{{\mathrm{3}}}}  ^{ 0  }   \refenvsym{)} \ \Rightarrow                                                                                             \\
                                                          & \:\qquad \refenvmv{y}  \leftarrow \, \token{gensym} \, \refenvsym{(}  N_{{\mathrm{2}}}  \refenvsym{)};\ \mathcal{V}^{\mathrm{rt} }  \refenvsym{(}   \refenvnt{e_{{\mathrm{3}}}} ^{ 0  }   \refenvsym{,}  \refenvsym{(}  E'  \refenvsym{,}  \refenvmv{y}  \coloneqq  \refenvnt{v_{{\mathrm{2}}}}  \refenvsym{)}  \refenvsym{,}  \refenvsym{(}  R'  \refenvsym{,}  \refenvmv{x}  \coloneqq  \refenvmv{y}  \refenvsym{)}  \refenvsym{,}  \refenvsym{(}  N_{{\mathrm{2}}}  \refenvsym{,}  \refenvmv{y}  \refenvsym{)}  \refenvsym{,}  S_{{\mathrm{2}}}  \refenvsym{,}  \refenvnt{k}  \refenvsym{)}                                         \\
                                                          & \:\quad  \token{case}\  \token{clos} \, \refenvsym{(}  E'  \refenvsym{,}  R'  \refenvsym{,}    \token{rec}\ \binder{ \refenvmv{f} }(\binder{ \refenvmv{x} }).\  \refenvnt{e_{{\mathrm{3}}}}  ^{ 0  }   \refenvsym{)} \ \Rightarrow                                                                                          \\
                                                          & \:\qquad g  \leftarrow \, \token{gensym} \, \refenvsym{(}  N_{{\mathrm{2}}}  \refenvsym{)};\ \refenvmv{y}  \leftarrow \, \token{gensym} \, \refenvsym{(}  N_{{\mathrm{2}}}  \refenvsym{,}  g  \refenvsym{)};                                                                                  \\
                                                          & \:\qquad \mathcal{V}^{\mathrm{rt} }  \refenvsym{(}   \refenvnt{e_{{\mathrm{3}}}} ^{ 0  }   \refenvsym{,}  \refenvsym{(}  E'  \refenvsym{,}  g  \coloneqq  \refenvnt{v_{{\mathrm{1}}}}  \refenvsym{,}  \refenvmv{y}  \coloneqq  \refenvnt{v_{{\mathrm{2}}}}  \refenvsym{)}  \refenvsym{,}  \refenvsym{(}  R'  \refenvsym{,}  \refenvmv{f}  \coloneqq  g  \refenvsym{,}  \refenvmv{x}  \coloneqq  \refenvmv{y}  \refenvsym{)}  \refenvsym{,}  \refenvsym{(}  N_{{\mathrm{2}}}  \refenvsym{,}  g  \refenvsym{,}  \refenvmv{y}  \refenvsym{)}  \refenvsym{,}  S_{{\mathrm{2}}}  \refenvsym{,}  \refenvnt{k}  \refenvsym{)}                                        \\
                                                          & \:\quad  \token{case}\ \_ \Rightarrow \token{Fail}                                                                                                                        \\
    \mathcal{V}^{\mathrm{rt} }  \refenvsym{(}   \langle   \refenvnt{e_{{\mathrm{1}}}} ^{ 1  }   \rangle   \refenvsym{,}  E  \refenvsym{,}  R  \refenvsym{,}  N  \refenvsym{,}  S  \refenvsym{,}   \refenvnt{k} ^{+}   \refenvsym{)} & = \refenvsym{(}   \refenvnt{e_{{\mathrm{2}}}} ^{ 0  }   \refenvsym{,}  N_{{\mathrm{1}}}  \refenvsym{,}  S_{{\mathrm{1}}}  \refenvsym{)}  \leftarrow  \mathcal{V}^{\mathrm{fut} }  \refenvsym{(}  0  \refenvsym{,}   \refenvnt{e_{{\mathrm{1}}}} ^{ 1  }   \refenvsym{,}  E  \refenvsym{,}  R  \refenvsym{,}  N  \refenvsym{,}  S  \refenvsym{,}  \refenvnt{k}  \refenvsym{)};\ \token{return} \, \refenvsym{(}   \langle   \refenvnt{e_{{\mathrm{2}}}} ^{ 0  }   \rangle   \refenvsym{,}  N_{{\mathrm{1}}}  \refenvsym{,}  S_{{\mathrm{1}}}  \refenvsym{)}                                  \\
    \mathcal{V}^{\mathrm{rt} }  \refenvsym{(}   { \mathdollar }_{ 0 }\lbrace  \refenvnt{e_{{\mathrm{1}}}} ^{ 0  }  \rbrace   \refenvsym{,}  E  \refenvsym{,}  R  \refenvsym{,}  N  \refenvsym{,}  S  \refenvsym{,}   \refenvnt{k} ^{+}   \refenvsym{)}
                                                          & = \refenvsym{(}  \refenvnt{v_{{\mathrm{1}}}}  \refenvsym{,}  N_{{\mathrm{1}}}  \refenvsym{,}  S_{{\mathrm{1}}}  \refenvsym{)}  \leftarrow  \mathcal{V}^{\mathrm{rt} }  \refenvsym{(}   \refenvnt{e_{{\mathrm{1}}}} ^{ 0  }   \refenvsym{,}  E  \refenvsym{,}  R  \refenvsym{,}  N  \refenvsym{,}  S  \refenvsym{,}  \refenvnt{k}  \refenvsym{)};                                                                        \\
                                                          & \:\quad  \refenvnt{e_{{\mathrm{2}}}} ^{ 0  }   \leftarrow \, \token{toQuote} \, \refenvsym{(}  \refenvnt{v_{{\mathrm{1}}}}  \refenvsym{)};\ \mathcal{V}^{\mathrm{rt} }  \refenvsym{(}   \refenvnt{e_{{\mathrm{2}}}} ^{ 0  }   \refenvsym{,}  E  \refenvsym{,}   \token{id}   \refenvsym{,}  N_{{\mathrm{1}}}  \refenvsym{,}  S_{{\mathrm{1}}}  \refenvsym{,}  \refenvnt{k}  \refenvsym{)}                                                         \\
    \mathcal{V}^{\mathrm{rt} }  \refenvsym{(}  \token{ref} \, \refenvsym{(}   \refenvnt{e} ^{ 0  }   \refenvsym{)}  \refenvsym{,}  E  \refenvsym{,}  R  \refenvsym{,}  N  \refenvsym{,}  S  \refenvsym{,}   \refenvnt{k} ^{+}   \refenvsym{)}
                                                          & = \refenvsym{(}  \refenvnt{v_{{\mathrm{1}}}}  \refenvsym{,}  N_{{\mathrm{1}}}  \refenvsym{,}  S_{{\mathrm{1}}}  \refenvsym{)}  \leftarrow  \mathcal{V}^{\mathrm{rt} }  \refenvsym{(}   \refenvnt{e} ^{ 0  }   \refenvsym{,}  E  \refenvsym{,}  R  \refenvsym{,}  N  \refenvsym{,}  S  \refenvsym{,}  \refenvnt{k}  \refenvsym{)};\ \refenvmv{l}  \leftarrow \, \token{alloc} \, \refenvsym{(}  S_{{\mathrm{1}}}  \refenvsym{)};\ \token{return} \, \refenvsym{(}  \refenvmv{l}  \refenvsym{,}  N_{{\mathrm{1}}}  \refenvsym{,}  \refenvsym{(}  S_{{\mathrm{1}}}  \refenvsym{,}  \refenvmv{l}  \coloneqq  \refenvnt{v_{{\mathrm{1}}}}  \refenvsym{)}  \refenvsym{)}
  \end{align*}

  \vspace{0.5em}
  \noindent\fbox{$ \mathcal{V}^{\mathrm{fut} }  : (\refenvnt{k}: \token{Int} ) \to  \token{Exp}^{  \refenvnt{k} ^{+}  }  \to  \token{VEnv}  \to  \token{REnv}  \to  \token{RN}  \to  \token{Store}  \to  \token{Int}  \to  \token{Result}(  \token{Exp}^{ \refenvnt{k} }  \times  \token{RN}  \times  \token{Store}  ) $}
  \begin{align*}
    \mathcal{V}^{\mathrm{fut} }  \refenvsym{(}  \refenvnt{k_{{\mathrm{1}}}}  \refenvsym{,}   \refenvnt{e} ^{  \refenvnt{k_{{\mathrm{1}}}} ^{+}   }   \refenvsym{,}  E  \refenvsym{,}  R  \refenvsym{,}  N  \refenvsym{,}  S  \refenvsym{,}  0  \refenvsym{)}                & =  \token{Timeout}                                                                                                                     \\
    \mathcal{V}^{\mathrm{fut} }  \refenvsym{(}  \refenvnt{k_{{\mathrm{1}}}}  \refenvsym{,}  \refenvmv{x}  \refenvsym{,}  E  \refenvsym{,}  R  \refenvsym{,}  N  \refenvsym{,}  S  \refenvsym{,}   \refenvnt{k_{{\mathrm{2}}}} ^{+}   \refenvsym{)}                  & = \token{return} \, \refenvsym{(}  \token{rename} \, \refenvsym{(}  R  \refenvsym{,}  \refenvmv{x}  \refenvsym{)}  \refenvsym{,}  N  \refenvsym{,}  S  \refenvsym{)}                                                                                          \\
    \mathcal{V}^{\mathrm{fut} }  \refenvsym{(}  \refenvnt{k_{{\mathrm{1}}}}  \refenvsym{,}    \lambda \binder{ \refenvmv{x} }.\  \refenvnt{e_{{\mathrm{1}}}}  ^{  \refenvnt{k_{{\mathrm{1}}}} ^{+}   }   \refenvsym{,}  E  \refenvsym{,}  R  \refenvsym{,}  N  \refenvsym{,}  S  \refenvsym{,}   \refenvnt{k_{{\mathrm{2}}}} ^{+}   \refenvsym{)}     & = \refenvmv{y}  \leftarrow \, \token{gensym} \, \refenvsym{(}  N  \refenvsym{)};                                                                                                           \\
                                                                     & \:\quad \refenvsym{(}   \refenvnt{e_{{\mathrm{2}}}} ^{ \refenvnt{k_{{\mathrm{1}}}}  }   \refenvsym{,}  N_{{\mathrm{1}}}  \refenvsym{,}  S_{{\mathrm{1}}}  \refenvsym{)}  \leftarrow  \mathcal{V}^{\mathrm{fut} }  \refenvsym{(}  \refenvnt{k_{{\mathrm{1}}}}  \refenvsym{,}   \refenvnt{e_{{\mathrm{1}}}} ^{  \refenvnt{k_{{\mathrm{1}}}} ^{+}   }   \refenvsym{,}  E  \refenvsym{,}  \refenvsym{(}  R  \refenvsym{,}  \refenvmv{x}  \coloneqq  \refenvmv{y}  \refenvsym{)}  \refenvsym{,}  \refenvsym{(}  N  \refenvsym{,}  \refenvmv{y}  \refenvsym{)}  \refenvsym{,}  S  \refenvsym{,}  \refenvnt{k_{{\mathrm{2}}}}  \refenvsym{)};                                  \\
                                                                     & \:\quad \token{return} \, \refenvsym{(}    \lambda \binder{ \refenvmv{y} }.\  \refenvnt{e_{{\mathrm{2}}}}  ^{ \refenvnt{k_{{\mathrm{1}}}}  }   \refenvsym{,}  N_{{\mathrm{1}}}  \refenvsym{,}  S_{{\mathrm{1}}}  \refenvsym{)}                                                                                        \\
    \mathcal{V}^{\mathrm{fut} }  \refenvsym{(}  \refenvnt{k_{{\mathrm{1}}}}  \refenvsym{,}   \langle   \refenvnt{e_{{\mathrm{1}}}} ^{  \refenvnt{k_{{\mathrm{1}}}}  \refenvsym{+}  2   }   \rangle   \refenvsym{,}  E  \refenvsym{,}  R  \refenvsym{,}  N  \refenvsym{,}  S  \refenvsym{,}   \refenvnt{k_{{\mathrm{2}}}} ^{+}   \refenvsym{)} & = \refenvsym{(}   \refenvnt{e_{{\mathrm{2}}}} ^{  \refenvnt{k_{{\mathrm{1}}}} ^{+}   }   \refenvsym{,}  N_{{\mathrm{1}}}  \refenvsym{,}  S_{{\mathrm{1}}}  \refenvsym{)}  \leftarrow  \mathcal{V}^{\mathrm{fut} }  \refenvsym{(}   \refenvnt{k_{{\mathrm{1}}}} ^{+}   \refenvsym{,}   \refenvnt{e_{{\mathrm{1}}}} ^{  \refenvnt{k_{{\mathrm{1}}}}  \refenvsym{+}  2   }   \refenvsym{,}  E  \refenvsym{,}  R  \refenvsym{,}  N  \refenvsym{,}  S  \refenvsym{,}  \refenvnt{k_{{\mathrm{2}}}}  \refenvsym{)};                                           \\
                                                                     & \:\quad \token{return} \, \refenvsym{(}   \langle   \refenvnt{e_{{\mathrm{2}}}} ^{  \refenvnt{k_{{\mathrm{1}}}} ^{+}   }   \rangle   \refenvsym{,}  N_{{\mathrm{1}}}  \refenvsym{,}  S_{{\mathrm{1}}}  \refenvsym{)}                                                                                        \\
    \mathcal{V}^{\mathrm{fut} }  \refenvsym{(}  \refenvnt{k_{{\mathrm{1}}}}  \refenvsym{,}   { \mathdollar }_{  \refenvnt{k_{{\mathrm{1}}}} ^{+}  }\lbrace  \refenvnt{e_{{\mathrm{1}}}} ^{ 0  }  \rbrace   \refenvsym{,}  E  \refenvsym{,}  R  \refenvsym{,}  N  \refenvsym{,}  S  \refenvsym{,}   \refenvnt{k_{{\mathrm{2}}}} ^{+}   \refenvsym{)}
                                                                     & = \refenvsym{(}  \refenvnt{v_{{\mathrm{1}}}}  \refenvsym{,}  N_{{\mathrm{1}}}  \refenvsym{,}  S_{{\mathrm{1}}}  \refenvsym{)}  \leftarrow  \mathcal{V}^{\mathrm{rt} }  \refenvsym{(}   \refenvnt{e_{{\mathrm{1}}}} ^{ 0  }   \refenvsym{,}  E  \refenvsym{,}  R  \refenvsym{,}  N  \refenvsym{,}  S  \refenvsym{,}  \refenvnt{k_{{\mathrm{2}}}}  \refenvsym{)};\  \refenvnt{e_{{\mathrm{2}}}} ^{ 0  }   \leftarrow \, \token{toQuote} \, \refenvsym{(}  \refenvnt{v_{{\mathrm{1}}}}  \refenvsym{)};                             \\
                                                                     & \:\quad \token{return} \, \refenvsym{(}   \refenvnt{e_{{\mathrm{2}}}} ^{ 0  }   \refenvsym{,}  N_{{\mathrm{1}}}  \refenvsym{,}  S_{{\mathrm{1}}}  \refenvsym{)} \quad\text{(Note that $ \refenvnt{e_{{\mathrm{2}}}} ^{ 0  }  \, \in \,  \token{Exp}^{ \refenvnt{k_{{\mathrm{1}}}} } $ by \Zcref{claim:expression-stratification})}
  \end{align*}
  \captionsetup{skip=4pt}
  \caption{Curated definitions of evaluator functions. $ \refenvnt{k} ^{+} $ is a shorthand for $\refenvnt{k} + 1$.}\label{fig:definitional-interpreter}
  \vspace{-0.75\baselineskip}
\end{figure}

We are now ready to define the main evaluation functions, $ \mathcal{V}^{\mathrm{rt} } $ and $ \mathcal{V}^{\mathrm{fut} } $, which evaluate expressions at runtime and within quotations, respectively. Their type signatures and curated definitions are given in \Zcref{fig:definitional-interpreter}.
Both functions take a value environment, a renaming environment, reserved names, a store, and a step limit to ensure termination. The results of these functions include reserved names and a store, modeling the global effects of name generation and mutable state.
$ \mathcal{V}^{\mathrm{rt} } $ evaluates expressions at stage 0 and returns a value.
$ \mathcal{V}^{\mathrm{fut} } $ is a dependent function that takes an integer $\refenvnt{k}$ representing a staging level. It evaluates expressions at the future-stage level $\refenvnt{k}  \refenvsym{+}  1$ and returns an expression at the lower level $\refenvnt{k}$, indicating that evaluation for the top-level stage is complete.
We highlight key aspects of these definitions:

\noindent \textbf{Runtime evaluator} ($ \mathcal{V}^{\mathrm{rt} } $)\nopagebreak\begin{itemize}
  \item A variable ($\refenvmv{x}$) is first renamed and then looked up in the value environment.
  \item Function application ($   \refenvnt{e_{{\mathrm{1}}}} ^{ 0  }  \refenvnt{e_{{\mathrm{2}}}}  ^{ 0  } $) evaluates the function and argument, and then evaluates the function body under appropriately extended environments.
        Instead of adding a mapping directly from $\refenvmv{x}$ to $\refenvnt{v_{{\mathrm{2}}}}$, we generate a fresh name $\refenvmv{y}$, add the mapping from $\refenvmv{y}$ to $\refenvnt{v_{{\mathrm{2}}}}$ to the value environment, and add the mapping from $\refenvmv{x}$ to $\refenvmv{y}$ to the renaming environment. This is necessary to ensure lexical scoping during runtime execution of code fragments with scoped CSP, as illustrated by the examples in \Zcref{sec:semantics:examples}. For recursive functions, we also add a binding for the function name to the environments.
  \item A quotation ($ \langle   \refenvnt{e_{{\mathrm{1}}}} ^{ 1  }   \rangle $) is evaluated by evaluating the inner expression $ \refenvnt{e_{{\mathrm{1}}}} ^{ 1  } $ at the future stage via $ \mathcal{V}^{\mathrm{fut} } $, and then returning the resulting expression $ \refenvnt{e_{{\mathrm{2}}}} ^{ 0  } $ as a quoted code fragment.
  \item Runtime execution of generated code ($ { \mathdollar }_{ 0 }\lbrace  \refenvnt{e_{{\mathrm{1}}}} ^{ 0  }  \rbrace $) involves evaluating $ \refenvnt{e_{{\mathrm{1}}}} ^{ 0  } $ at runtime to obtain a quoted code fragment, and then evaluating the expression inside it under the same runtime environment. This allows evaluation of open code fragments that may contain free variables: this is how we realize scoped CSP. Free variables in the code fragment have already been renamed; hence, evaluation is performed under the identity renaming environment $ \token{id} $ to avoid further renaming.
\end{itemize}

\noindent \textbf{Future-stage evaluator} ($ \mathcal{V}^{\mathrm{fut} } $)
\begin{itemize}
  \item A variable ($\refenvmv{x}$) is renamed and returned as an expression. This allows generated code fragments to be well-scoped under future-stage bindings or runtime environments.
  \item For generation of a function ($  \lambda \binder{ \refenvmv{x} }.\  \refenvnt{e}  ^{  \refenvnt{k} ^{+}   } $), a fresh name $\refenvmv{y}$ is generated for the parameter, and $\refenvmv{x}$ in the body is renamed to $\refenvmv{y}$ by extending the renaming environment. This ensures that the generated code fragment does not contain duplicate binder names. After evaluating $ \refenvnt{e} ^{  \refenvnt{k} ^{+}   } $, the reserved name $\refenvmv{y}$ binds the occurrences of $\refenvmv{y}$ in the generated code fragment $ \refenvnt{e_{{\mathrm{2}}}} ^{ \refenvnt{k_{{\mathrm{1}}}}  } $. We then construct the function $  \lambda \binder{ \refenvmv{y} }.\  \refenvnt{e_{{\mathrm{2}}}}  ^{ \refenvnt{k_{{\mathrm{1}}}}  } $, where the parameter $\refenvmv{y}$ binds those same occurrences in the body. At this point, the binding structure of $ \refenvnt{e_{{\mathrm{2}}}} ^{ \refenvnt{k_{{\mathrm{1}}}}  } $ is transferred from the reserved name to the new binder. This transfer is in fact the most nontrivial part of the type soundness proof and is captured later by \Zcref{claim:demotion}.
  \item When splicing generated code from runtime level ($ { \mathdollar }_{  \refenvnt{k_{{\mathrm{1}}}} ^{+}  }\lbrace  \refenvnt{e_{{\mathrm{1}}}} ^{ 0  }  \rbrace $), the code is evaluated at runtime via $ \mathcal{V}^{\mathrm{rt} } $, and the expression in the resulting quoted code fragment is returned. Here, the returned expression $ \refenvnt{e_{{\mathrm{2}}}} ^{ 0  } $ is a level-0 expression, but it can be safely treated as a level-$\refenvnt{k_{{\mathrm{1}}}}$ expression thanks to \Zcref{claim:expression-stratification}.
\end{itemize}

\subsection{Evaluation Examples}\label{sec:semantics:examples}
We show several examples to illustrate how our semantics ensures lexical scoping for generated code fragments by systematically renaming variables.

\subsubsection*{Example 1: Hygienic Code Generation} Let us consider the following program:
\begin{tikzpicture}[overlay,remember picture]

  \draw[blue,thick]
  ([yshift=-2pt]pic cs:f1) -- ([yshift=-2pt]pic cs:f1e)
  node[pos=1,right=-3pt] {\small \ctext{1}};

  \draw[blue,thick]
  ([yshift=-2pt]pic cs:xq) -- ([yshift=-2pt]pic cs:xqe)
  node[pos=1,right=-3pt] {\small \ctext{2}};

  \draw[blue,thick]
  ([yshift=-2pt]pic cs:f2) -- ([yshift=-2pt]pic cs:f2e)
  node[pos=1,right=-3pt] {\small \ctext{3}};

  \draw[blue,thick]
  ([yshift=-2pt]pic cs:y) -- ([yshift=-2pt]pic cs:ye)
  node[pos=1,right=-3pt] {\small \ctext{4}};

\end{tikzpicture}
\begin{lstlisting}
.< (@*\tikzmark{f1}*@)fun x(@*\tikzmark{f1e}*@)   -> .~( let y = (@*\tikzmark{xq}*@).< x >.(@*\tikzmark{xqe}*@)   in .< (@*\tikzmark{f2}*@)fun x(@*\tikzmark{f2e}*@)   -> (@*\tikzmark{y}*@).~y(@*\tikzmark{ye}*@)   >. ) >.
\end{lstlisting}
This example generates a code fragment \lstinline|.< x >.| and splices it under another binding for $\refenvmv{x}$ (\lstinline|let| is syntactic sugar for a function and an application). Without renaming, this program would result in \lstinline|.< fun x -> fun x -> x >.|, where the use of \lstinline|x| is captured by the inner binding, which violates lexical scoping. Our semantics avoids this by systematic renaming with $ \token{gensym} $ and renaming environments. We highlight the evaluation steps for each underlined part of the code:
\begin{enumerate}[label=\ctext{\arabic*}]
  \item Evaluating the outer \lstinline|fun x| generates a fresh name, say $\refenvmv{x_{{\mathrm{1}}}}$, for $\refenvmv{x}$, and adds $\refenvmv{x}  \coloneqq  \refenvmv{x_{{\mathrm{1}}}}$ to the renaming environment.
  \item Evaluating \lstinline|.< x >.| then renames $\refenvmv{x}$ to $\refenvmv{x_{{\mathrm{1}}}}$ and returns \lstinline|.< x1 >.|.
  \item Evaluating the inner \lstinline|fun x| generates another fresh name, say $\refenvmv{x_{{\mathrm{2}}}}$, and adds $\refenvmv{x}  \coloneqq  \refenvmv{x_{{\mathrm{2}}}}$ to the renaming environment.
  \item Splicing \lstinline|y| embeds \lstinline|.< x1 >.| directly. Since it is already a value, no further renaming is performed, and \lstinline|x1| correctly refers to the outer binding.
\end{enumerate}
As a result, the final outcome of evaluating this program is \lstinline|.< fun x1 -> fun x2 -> x1 >.|, where the use of \lstinline|x1| is correctly captured by the outer binding.

\subsubsection*{Example 2: Scoped CSP with Lexical Scoping}
We consider another staged program that executes a code fragment with scoped CSP.
\begin{tikzpicture}[overlay,remember picture]
  \draw[blue,thick]
  ([yshift=-2pt]pic cs:a1) -- ([yshift=-2pt]pic cs:a1e)
  node[pos=1,right=-3pt] {\small \ctext{1}};

  \draw[blue,thick]
  ([yshift=-2pt]pic cs:a2) -- ([yshift=-2pt]pic cs:a2e)
  node[pos=1,right=-3pt] {\small \ctext{2}};

  \draw[blue,thick]
  ([yshift=-2pt]pic cs:a3) -- ([yshift=-2pt]pic cs:a3e)
  node[pos=1,right=-3pt] {\small \ctext{3}};

  \draw[blue,thick]
  ([yshift=-2pt]pic cs:a4) -- ([yshift=-2pt]pic cs:a4e)
  node[pos=1,right=-3pt] {\small \ctext{4}};
\end{tikzpicture}
\begin{lstlisting}
(@*\tikzmark{a1}*@)let x = 1(@*\tikzmark{a1e}*@)   in
let run_under_x_10 code = (@*\tikzmark{a3}*@)let x = 10(@*\tikzmark{a3e}*@)   in (@*\tikzmark{a4}*@)run code(@*\tikzmark{a4e}*@)   in
run_under_x_10 (@*\tikzmark{a2}*@).< x >.(@*\tikzmark{a2e}*@)
\end{lstlisting}
In this example, we generate a code fragment \lstinline|.< x >.| that contains a free variable $\refenvmv{x}$ and execute it under another binding of $\refenvmv{x}$.
Without renaming, \lstinline|run_under_x_10 .< x >.| would execute the code fragment under the inner binding $\refenvmv{x} = 10$, yielding $10$ instead of $1$.
Our semantics preserves lexical scoping by systematic renaming and thereby prevents dynamic capture of free variables.
Consequently, evaluating this program yields $1$.
We highlight the evaluation steps corresponding to each underlined part of the program:
\begin{enumerate}[label=\ctext{\arabic*}]
  \item Evaluating \lstinline|let x = 1 in ...| generates a fresh name, say $\refenvmv{x_{{\mathrm{1}}}}$, for $\refenvmv{x}$, adds $\refenvmv{x_{{\mathrm{1}}}}  \coloneqq  1$ to the value environment, and adds $\refenvmv{x}  \coloneqq  \refenvmv{x_{{\mathrm{1}}}}$ to the renaming environment.
  \item Since the renaming environment contains $\refenvmv{x}  \coloneqq  \refenvmv{x_{{\mathrm{1}}}}$, evaluating \lstinline|.< x >.| yields \lstinline|.< x1 >.|.
  \item Evaluating \lstinline|run_under_x_10| generates another fresh name, say $\refenvmv{x_{{\mathrm{2}}}}$, for $\refenvmv{x}$, adds $\refenvmv{x_{{\mathrm{2}}}}  \coloneqq  10$ to the value environment, and adds $\refenvmv{x}  \coloneqq  \refenvmv{x_{{\mathrm{2}}}}$ to the renaming environment.
  \item Finally, \lstinline|run code| executes the code fragment \lstinline|.< x1 >.|. Since this is already a value, no further renaming is needed, and \lstinline|x1| is resolved to $1$, correctly referring to the outer binding.
\end{enumerate}
Thus, fresh name generation together with the renaming environment performs systematic renaming during evaluation, ensuring lexical scoping of generated code and preventing unintended variable capture. In particular, generated code always refers to globally fresh names, ensuring that binders introduced later cannot capture them.

\section{Runtime Typing}\label{sec:runtime-typing}
The surface type system presented in \Zcref{sec:surface-typing,,sec:polycls} is not sufficient to prove type soundness with respect to the operational semantics defined in \Zcref{sec:semantics}. This is because the surface type system does not account for runtime objects such as values, reserved names, and stores. To bridge this gap, we introduce a runtime type system that reasons about these runtime objects.

First, we define syntactic categories for typing runtime objects.

\vspace{0.5em}
\begin{tabular}{clcl}
  \textbf{Reserved Names Typing} & $\Psi$ & $\Coloneqq$ & $ \varepsilon  \mid \Psi  \refenvsym{,}   \binder{ \refenvmv{x} }: \refenvnt{A} \mathrel{@}\binder{ \gamma_{{\mathrm{1}}} } \mathbin{:\succeq} \gamma_{{\mathrm{2}}}  \mid \Psi  \refenvsym{,}   \binder{ \gamma_{{\mathrm{1}}} }( \mathbin{:\succeq} \; \gamma_{{\mathrm{2}}}  /  \mathbin{:\supseteq} \; \gamma_{{\mathrm{3}}} ) $ \\
  && $\mid$ & $\Psi  \refenvsym{,}   [    \binder{  \gamma _{ 0 }  }  \mathbin{:\succeq}    \delta _{ 0 }  ,\dotsc,\binder{  \gamma _{ \refenvnt{k} }  }  \mathbin{:\succeq}   \delta  _{ \refenvnt{k} }   ] $ \\
  \textbf{Location Typing}       & $\Theta$  & $\Coloneqq$ & $ \varepsilon  \mid \Theta  \refenvsym{,}   \refenvmv{l} : \refenvnt{A} $
\end{tabular}
\vspace{0.5em}

A reserved names typing $\Psi$ assigns types to globally reserved names and maintains reachability and visibility constraints among their classifiers.
An element $ \binder{ \refenvmv{x} }: \refenvnt{A} \mathrel{@}\binder{ \gamma_{{\mathrm{1}}} } \mathbin{:\succeq} \gamma_{{\mathrm{2}}} $ states that the reserved name $\refenvmv{x}$ has type $\refenvnt{A}$, is declared at classifier $\gamma_{{\mathrm{1}}}$, and that $\gamma_{{\mathrm{2}}}$ is reachable from $\gamma_{{\mathrm{1}}}$. This element is generated during evaluation of variable binding forms. The element $ \binder{ \gamma_{{\mathrm{1}}} }( \mathbin{:\succeq} \; \gamma_{{\mathrm{2}}}  /  \mathbin{:\supseteq} \; \gamma_{{\mathrm{3}}} ) $ declares a new classifier $\gamma_{{\mathrm{1}}}$ such that $ \gamma_{{\mathrm{2}}} \preceq \gamma_{{\mathrm{1}}} $ and $ \gamma_{{\mathrm{3}}} \subseteq \gamma_{{\mathrm{1}}} $. This element is generated during evaluation of splices. The other element $ [    \binder{  \gamma _{ 0 }  }  \mathbin{:\succeq}    \delta _{ 0 }  ,\dotsc,\binder{  \gamma _{ \refenvnt{k} }  }  \mathbin{:\succeq}   \delta  _{ \refenvnt{k} }   ] $ declares a sequence of classifiers $  \gamma _{ 0 }  ,\dotsc,  \gamma _{ \refenvnt{k} }  $ with their bases $  \delta _{ 0 }  ,\dotsc,  \delta _{ \refenvnt{k} }  $, introducing reachability constraints $  \delta _{ \refenvnt{i} }  \preceq  \gamma _{ \refenvnt{i} }  $ and visibility constraints $  \gamma _{ \refenvnt{i}  \refenvsym{+}  1 }  \subseteq  \gamma _{ \refenvnt{i} }  $. This element is generated during evaluation of polymorphic classifiers. The structure of location typing $\Theta$ is standard, mapping locations to their types.

The runtime type system has eleven judgments:
\begin{longtable}{c l l}
  (1)  & Well-formed Reserved Names Typing & $ \vdash  \Psi \ \token{wf} $                                                                                          \\
  (2)  & Well-formed Classifier Stack      & $\Psi  \vdash   \overrightarrow{ \gamma }^{+}  \, \token{wf}$                                                                                   \\
  (3)  & Well-formed Contexts              & $ \Psi \mid^{  \overrightarrow{ \gamma_{{\mathrm{1}}} }^{+}  } \Gamma \vdash^{  \overrightarrow{ \gamma_{{\mathrm{2}}} }^{+}  }\token{wf} $                                                                    \\
  (4)  & Well-formed Types                 & $ \Psi \mid^{  \overrightarrow{ \gamma_{{\mathrm{1}}} }^{+}  }  \Gamma \vdash ^{  \overrightarrow{ \gamma_{{\mathrm{2}}} }^{+}  } \refenvnt{A} \ \token{wf} $ (or, $ \Psi  \vdash^{  \overrightarrow{ \gamma_{{\mathrm{1}}} }^{+}  }  \refenvnt{A} \ \token{wf} $ if $\Gamma=  \varepsilon $ ) \\
  (5)  & Reachability                      & $ \Psi \mid^{  \overrightarrow{ \gamma_{{\mathrm{1}}} }^{+}  } \Gamma \vdash \gamma_{{\mathrm{2}}} \preceq \gamma_{{\mathrm{3}}} $ (or, $\Psi  \vdash  \gamma_{{\mathrm{2}}}  \preceq  \gamma_{{\mathrm{3}}}$ if $\Gamma=  \varepsilon $ )             \\
  (6)  & Visibility                        & $ \Psi \mid^{  \overrightarrow{ \gamma_{{\mathrm{1}}} }^{+}  } \Gamma \vdash \gamma_{{\mathrm{2}}} \subseteq \gamma_{{\mathrm{3}}} $ (or, $\Psi  \vdash  \gamma_{{\mathrm{2}}}  \subseteq  \gamma_{{\mathrm{3}}}$ if $\Gamma=  \varepsilon $ )             \\
  (7)  & Expression Typing                 & $ \Psi \mid^{  \overrightarrow{ \gamma_{{\mathrm{1}}} }^{+}  }  \Gamma \vdash^{  \overrightarrow{ \gamma_{{\mathrm{2}}} }^{+}  }_{ R }  \refenvnt{e} : \refenvnt{A} $                                                    \\
  (8)  & Well-formed Location Typing       & $ \Psi  \vdash  \Theta \ \token{wf} $                                                                                       \\
  (9)  & Store Typing                      & $\Psi  \vdash  S  \colon  \Theta$                                                                                  \\
  (10) & Value Typing                      & $ \Psi \mid \Theta \vdash^{ \gamma } \refenvnt{v} : \refenvnt{A} \ \token{val} $                                                                          \\
  (11) & Well-formed Value Environments    & $ \Psi \mid \Theta \vdash^{ \gamma } E \ \token{wf} $
\end{longtable}
\noindent $ \vdash  \Psi \ \token{wf} $ defines well-formed reserved names typing, and $\Psi  \vdash   \overrightarrow{ \gamma }^{+}  \, \token{wf}$ defines a well-formed classifier stack under $\Psi$. Their derivation rules are straightforward and omitted here.
The other judgments largely divide into two groups: judgments extended from surface typing with reserved names typing (3--7), and new judgments for runtime objects (8--11).

For the first group, we present important rules in \Zcref{fig:rules-extended-surface-typing} for brevity, and the other rules are mostly the same as those in surface typing. These judgments extend those from surface typing with a reserved names typing $\Psi$ and a classifier stack $ \overrightarrow{ \gamma_{{\mathrm{1}}} }^{+} $.
Here, $ \overrightarrow{ \gamma_{{\mathrm{1}}} }^{+} $ represents the classifier stack at the point where evaluation takes place; hence, the classifier stack for the empty context will be $ \overrightarrow{ \gamma_{{\mathrm{1}}} }^{+} $ instead of $ \exclam $, as stated in \ref{rule:runtime-wf-ctx-nil}.
We also introduce five additional rules that deduce reachability and visibility from reserved names typing: \ref{rule:runtime-ltrans-nht1}, \ref{rule:runtime-ltrans-nht2}, \ref{rule:runtime-gtrans-shut-nht}, \ref{rule:runtime-ltrans-nht3} and \ref{rule:runtime-gtrans-nht-polycls}.
\begin{figure}[bpt]
  \small
  \raggedright

  \fbox{$ \Psi \mid^{  \overrightarrow{ \gamma_{{\mathrm{1}}} }^{+}  } \Gamma \vdash^{  \overrightarrow{ \gamma_{{\mathrm{2}}} }^{+}  }\token{wf} $} \fbox{$ \Psi \mid^{  \overrightarrow{ \gamma_{{\mathrm{1}}} }^{+}  } \Gamma \vdash \gamma_{{\mathrm{2}}} \preceq \gamma_{{\mathrm{3}}} $} \fbox{$ \Psi \mid^{  \overrightarrow{ \gamma_{{\mathrm{1}}} }^{+}  } \Gamma \vdash \gamma_{{\mathrm{2}}} \subseteq \gamma_{{\mathrm{3}}} $}
  \vspace{0.5em}

  \begin{mathpar}
    \myinferrule[R-WF-Ctx-Nil]
    {\Psi  \vdash   \overrightarrow{ \gamma }^{+}  \, \token{wf}}
    { \Psi \mid^{  \overrightarrow{ \gamma }^{+}  } \varepsilon \vdash^{  \overrightarrow{ \gamma }^{+}  }\token{wf} }
    \label{rule:runtime-wf-ctx-nil}
    \and
    \myinferrule[R-$ \preceq $-NHT-Name]
    { \binder{ \refenvmv{x} }: \refenvnt{A} \mathrel{@}\binder{ \gamma_{{\mathrm{1}}} } \mathbin{:\succeq} \gamma_{{\mathrm{2}}}  \, \in \, \Psi}
    { \Psi \mid^{  \overrightarrow{ \gamma_{{\mathrm{3}}} }^{+}  } \Gamma \vdash \gamma_{{\mathrm{2}}} \preceq \gamma_{{\mathrm{1}}} }
    \label{rule:runtime-ltrans-nht1}
    \and
    \myinferrule[R-$ \preceq $-NHT-$ \blacktriangleleft $]
    { \binder{ \gamma_{{\mathrm{1}}} }( \mathbin{:\succeq} \; \gamma_{{\mathrm{2}}}  /  \mathbin{:\supseteq} \; \gamma_{{\mathrm{3}}} )  \, \in \, \Psi}
    { \Psi \mid^{  \overrightarrow{ \gamma_{{\mathrm{4}}} }^{+}  } \Gamma \vdash \gamma_{{\mathrm{2}}} \preceq \gamma_{{\mathrm{1}}} }
    \label{rule:runtime-ltrans-nht2}
    \and
    \myinferrule[R-$ \subseteq $-NHT-$ \blacktriangleleft $]
    { \binder{ \gamma_{{\mathrm{1}}} }( \mathbin{:\succeq} \; \gamma_{{\mathrm{2}}}  /  \mathbin{:\supseteq} \; \gamma_{{\mathrm{3}}} )  \, \in \, \Psi}
    { \Psi \mid^{  \overrightarrow{ \gamma_{{\mathrm{4}}} }^{+}  } \Gamma \vdash \gamma_{{\mathrm{3}}} \subseteq \gamma_{{\mathrm{1}}} }
    \label{rule:runtime-gtrans-shut-nht}
    \and
    \myinferrule[R-$ \preceq $-NHT-Polycls]
    { [    \binder{  \gamma _{ 0 }  }  \mathbin{:\succeq}    \gamma' _{ 0 }  ,\dotsc,\binder{  \gamma _{ \refenvnt{k} }  }  \mathbin{:\succeq}   \gamma'  _{ \refenvnt{k} }   ]  \, \in \, \Psi\\
      0 \le \refenvnt{i} \le \refenvnt{k}
    }
    { \Psi \mid^{  \overrightarrow{ \delta }^{+}  } \Gamma \vdash  \gamma' _{ \refenvnt{i} }  \preceq  \gamma _{ \refenvnt{i} }  }
    \label{rule:runtime-ltrans-nht3}
    \and
    \myinferrule[R-$ \subseteq $-NHT-Polycls]
    { [    \binder{  \gamma _{ 0 }  }  \mathbin{:\succeq}    \gamma' _{ 0 }  ,\dotsc,\binder{  \gamma _{ \refenvnt{k} }  }  \mathbin{:\succeq}   \gamma'  _{ \refenvnt{k} }   ]  \, \in \, \Psi\\
      0 \le \refenvnt{i} < \refenvnt{k}
    }
    { \Psi \mid^{  \overrightarrow{ \delta }^{+}  } \Gamma \vdash  \gamma _{ \refenvnt{i}  \refenvsym{+}  1 }  \subseteq  \gamma _{ \refenvnt{i} }  }
    \label{rule:runtime-gtrans-nht-polycls}
  \end{mathpar}

  \fbox{$ \Psi \mid^{  \overrightarrow{ \gamma_{{\mathrm{1}}} }^{+}  }  \Gamma \vdash^{  \overrightarrow{ \gamma_{{\mathrm{2}}} }^{+}  }_{ R }  \refenvnt{e} : \refenvnt{A} $}
  \begin{mathpar}
    \myinferrule[R-T-Name]
    {\token{lookup} \, \refenvsym{(}  \Gamma  \refenvsym{,}  \refenvmv{x}  \refenvsym{)} =  \token{Fail}  \\
     \binder{ \refenvmv{y} }: \refenvnt{A} \mathrel{@}\binder{ \gamma_{{\mathrm{1}}} } \mathbin{:\succeq} \gamma_{{\mathrm{2}}}  \, \in \, \Psi \\
     \Psi \mid^{  \overrightarrow{ \gamma_{{\mathrm{3}}} }^{+}  } \Gamma \vdash^{  \overrightarrow{ \gamma_{{\mathrm{4}}} }   \refenvsym{,}  \gamma_{{\mathrm{5}}} }\token{wf}  \\
     \Psi \mid^{  \overrightarrow{ \gamma_{{\mathrm{3}}} }^{+}  } \Gamma \vdash \gamma_{{\mathrm{1}}} \preceq \gamma_{{\mathrm{5}}}  \\
    \token{rename} \, \refenvsym{(}  R  \refenvsym{,}  \refenvmv{x}  \refenvsym{)}  \refenvsym{=}  \refenvmv{y}}
    { \Psi \mid^{  \overrightarrow{ \gamma_{{\mathrm{3}}} }^{+}  }  \Gamma \vdash^{  \overrightarrow{ \gamma_{{\mathrm{4}}} }   \refenvsym{,}  \gamma_{{\mathrm{5}}} }_{ R }  \refenvmv{x} : \refenvnt{A} }
    \label{rule:runtime-type-reserved-var}
  \end{mathpar}
  \captionsetup{skip=4pt}
  \caption{Curated rules for judgments that are extended from surface typing.}\label{fig:rules-extended-surface-typing}
  \vspace{-0.75\baselineskip}
\end{figure}
The judgment for expression typing $ \Psi \mid^{  \overrightarrow{ \gamma_{{\mathrm{1}}} }^{+}  }  \Gamma \vdash^{  \overrightarrow{ \gamma_{{\mathrm{2}}} }^{+}  }_{ R }  \refenvnt{e} : \refenvnt{A} $ is also extended with a renaming environment $R$. We use it to type reserved names in \ref{rule:runtime-type-reserved-var}. This rule states that a reserved name $\refenvmv{y}$ can be used as a variable $\refenvmv{x}$ if
\begin{enumerate*}
  \item the scope of $\refenvmv{y}$ is reachable from the current classifier,
  \item $\refenvmv{x}$ renames to $\refenvmv{y}$ under the renaming environment, and
  \item $\refenvmv{x}$ is not shadowed in the context $\Gamma$.
\end{enumerate*}

\begin{figure}[bpt]
  \small
  \raggedright
  \fbox{$ \Psi \mid \Theta \vdash^{ \gamma } \refenvnt{v} : \refenvnt{A} \ \token{val} $}
  \begin{mathpar}
    \myinferrule[VT-Clos-Func]
    { \Psi \mid^{ \gamma_{{\mathrm{1}}} }  \varepsilon \vdash^{ \gamma_{{\mathrm{1}}} }_{ R }    \lambda \binder{ \refenvmv{x} }.\  \refenvnt{e}  ^{ 0  }  : \refenvnt{A}  \\
     \Psi \mid \Theta \vdash^{ \gamma_{{\mathrm{1}}} } E \ \token{wf}  \\
     \Psi  \vdash^{ \delta }  \refenvnt{A} \ \token{wf} 
    }
    { \Psi \mid \Theta \vdash^{ \delta } \token{clos} \, \refenvsym{(}  E  \refenvsym{,}  R  \refenvsym{,}    \lambda \binder{ \refenvmv{x} }.\  \refenvnt{e}  ^{ 0  }   \refenvsym{)} : \refenvnt{A} \ \token{val} }
    \label{rule:value-type-clos-func}
    \and
    \myinferrule[VT-Code]
    { \Psi \mid^{ \gamma_{{\mathrm{1}}} }  \varepsilon \vdash^{ \gamma_{{\mathrm{1}}} }_{  \token{id}  }   \langle   \refenvnt{e} ^{ 0  }   \rangle  : \refenvnt{A}  \\
     \Psi  \vdash  \Theta \ \token{wf} 
    }
    { \Psi \mid \Theta \vdash^{ \gamma_{{\mathrm{1}}} }  \langle   \refenvnt{e} ^{ 0  }   \rangle  : \refenvnt{A} \ \token{val} }
    \label{rule:value-type-code}
    \and
    \myinferrule[VT-Loc]
    { \Psi  \vdash  \Theta \ \token{wf}  \\
    \token{lookupStore}  \refenvsym{(}  \Theta  \refenvsym{,}  \refenvmv{l}  \refenvsym{)} = \token{Done} \, \refenvsym{(}  \refenvnt{A}  \refenvsym{)} \\
     \Psi  \vdash^{ \gamma }  \refenvnt{A} \ \token{wf} }
    { \Psi \mid \Theta \vdash^{ \gamma } \refenvmv{l} :  \refenvnt{A} \/\ \token{ref}  \ \token{val} }
    \label{rule:value-type-loc}
  \end{mathpar}

  \fbox{$ \Psi \mid \Theta \vdash^{ \gamma } E \ \token{wf} $}   \fbox{$\Psi  \vdash  S  \colon  \Theta$}
  \begin{mathpar}
    \myinferrule[WF-VEnv-Empty]
    { \Psi  \vdash  \Theta \ \token{wf} }
    { \Psi \mid \Theta \vdash^{  \exclam  } \varepsilon \ \token{wf} }
    \label{rule:runtime-wf-venv-empty}
    \and
    \myinferrule[WF-VEnv-Var]
    { \Psi \mid \Theta \vdash^{ \gamma_{{\mathrm{1}}} } E \ \token{wf}  \\
     \Psi \mid \Theta \vdash^{ \gamma_{{\mathrm{1}}} } \refenvnt{v} : \refenvnt{A} \ \token{val}  \\
     \binder{ \refenvmv{x} }: \refenvnt{A} \mathrel{@}\binder{ \gamma_{{\mathrm{2}}} } \mathbin{:\succeq} \gamma_{{\mathrm{1}}}  \, \in \, \Psi}
    { \Psi \mid \Theta \vdash^{ \gamma_{{\mathrm{2}}} } E  \refenvsym{,}  \refenvmv{x}  \coloneqq  \refenvnt{v} \ \token{wf} }
    \label{rule:runtime-wf-venv-var}
    \and
    \myinferrule[WF-VEnv-Shut]
    { \Psi \mid \Theta \vdash^{ \gamma_{{\mathrm{1}}} } E \ \token{wf}  \\
     \binder{ \gamma_{{\mathrm{2}}} }( \mathbin{:\succeq} \; \gamma_{{\mathrm{1}}}  /  \mathbin{:\supseteq} \; \gamma_{{\mathrm{3}}} )  \, \in \, \Psi}
    { \Psi \mid \Theta \vdash^{ \gamma_{{\mathrm{2}}} } E \ \token{wf} }
    \label{rule:runtime-wf-venv-cls}
    \and
    \myinferrule[WF-VEnv-Polycls]
    { \Psi \mid \Theta \vdash^{  \delta _{ 0 }  } E \ \token{wf}  \\
     [    \binder{  \gamma _{ 0 }  }  \mathbin{:\succeq}    \delta _{ 0 }  ,\dotsc,\binder{  \gamma _{ \refenvnt{k} }  }  \mathbin{:\succeq}   \delta  _{ \refenvnt{k} }   ]  \, \in \, \Psi}
    { \Psi \mid \Theta \vdash^{  \gamma _{ 0 }  } E \ \token{wf} }
    \label{rule:runtime-wf-venv-polycls}
    \and
    \myinferrule[T-Store]
    {S = \refenvsym{(}    { \refenvmv{l} _{ 1 } }  \coloneqq  { \refenvnt{v} _{ 1 } }  ,\dotsc,  { \refenvmv{l} _{ \refenvnt{k} } }  \coloneqq  { \refenvnt{v} _{ \refenvnt{k} } }    \refenvsym{)} \\
    \Theta = \refenvsym{(}    { \refenvmv{l} _{ 1 } }  :  { \refenvnt{A} _{ 1 } }  ,\dotsc,  { \refenvmv{l} _{ \refenvnt{k} } }  :  { \refenvnt{A} _{ \refenvnt{k} } }    \refenvsym{)} \\
     \Psi  \vdash  \Theta \ \token{wf}  \\\\
     \Psi \mid \Theta \vdash^{  \gamma _{ \refenvnt{i} }  }  { \refenvnt{v} _{ \refenvnt{i} } }  :  { \refenvnt{A} _{ \refenvnt{i} } }  \ \token{val} \ \text{with some $ \gamma _{ \refenvnt{i} } $ for}\ 1 \le \refenvnt{i} \le \refenvnt{k}}
    {\Psi  \vdash  S  \colon  \Theta}
    \label{rule:type-store}
  \end{mathpar}
  \captionsetup{skip=4pt}
  \caption{Curated rules for judgments with regard to runtime objects.}\label{fig:rules-runtime-obj}
  \vspace{-0.75\baselineskip}
\end{figure}
For judgments on runtime objects, we present the relevant rules in \Zcref{fig:rules-runtime-obj}. These judgments type values, stores, and value environments.
A value typing judgment $ \Psi \mid \Theta \vdash^{ \gamma } \refenvnt{v} : \refenvnt{A} \ \token{val} $ states that a value $\refenvnt{v}$ has type $\refenvnt{A}$ at classifier $\gamma$ under reserved names typing $\Psi$ and location typing $\Theta$. Unlike standard typed lambda calculi, where values are typed only at the top-level scope, this judgment types values at a local scope $\gamma$, whose visibility information is needed to type open code fragments.
\ref{rule:value-type-clos-func} states that a closure value $\token{clos} \, \refenvsym{(}  E  \refenvsym{,}  R  \refenvsym{,}    \lambda \binder{ \refenvmv{x} }.\  \refenvnt{e}  ^{ 0  }   \refenvsym{)}$ is well-typed if the underlying function is well-typed at some classifier under the given renaming environment, the captured value environment is well-formed at that classifier, and the closure type is well-formed at the classifier where the closure itself is typed. We have a similar rule for typing closures with recursive functions.
\ref{rule:value-type-code} states that a code value $ \langle   \refenvnt{e} ^{ 0  }   \rangle $ is well-typed if the quotation itself is well-typed under the identity renaming environment and the location typing $\Theta$ is well-formed.
\ref{rule:value-type-loc} states that a location value $\refenvmv{l}$ has type $ \refenvnt{A} \/\ \token{ref} $ if $ \refenvmv{l} : \refenvnt{A} $ appears in $\Theta$ and $\refenvnt{A}$ is well-formed at the current classifier $\gamma$. Here, we overload $ \token{lookupStore} $ to look up the type assigned to $\refenvmv{l}$ in $\Theta$.
One key aspect of value typing is that a value can be typed at any classifier as long as its type is well-formed at that classifier, as stated in the following lemma:
\begin{lemma}[Value Portability]\label{claim:value-portability}
  If\/ $ \Psi \mid \Theta \vdash^{ \gamma_{{\mathrm{1}}} } \refenvnt{v} : \refenvnt{A} \ \token{val} $ and\/ $ \Psi  \vdash^{ \gamma_{{\mathrm{2}}} }  \refenvnt{A} \ \token{wf} $, then $ \Psi \mid \Theta \vdash^{ \gamma_{{\mathrm{2}}} } \refenvnt{v} : \refenvnt{A} \ \token{val} $.
\end{lemma}
\noindent We leverage this fact for typing stores later.

The value-environment well-formedness judgment $ \Psi \mid \Theta \vdash^{ \gamma } E \ \token{wf} $ states that a value environment $E$ is well-formed at the classifier $\gamma$. The empty environment is well-formed at the initial classifier, as specified by \ref{rule:runtime-wf-venv-empty}.
\ref{rule:runtime-wf-venv-var} extends the environment with a new binding $\refenvmv{x}  \coloneqq  \refenvnt{v}$ while advancing the classifier along the reachability witnessed by an entry $ \binder{ \refenvmv{x} }: \refenvnt{A} \mathrel{@}\binder{ \gamma_{{\mathrm{2}}} } \mathbin{:\succeq} \gamma_{{\mathrm{1}}}  \, \in \, \Psi$.
\ref{rule:runtime-wf-venv-cls} and \ref{rule:runtime-wf-venv-polycls} similarly advance the classifier along the reachability witnessed by the entries, leaving the value environment unchanged.

For stores, we have two judgments: well-formedness of location typing $ \Psi  \vdash  \Theta \ \token{wf} $ and store typing $\Psi  \vdash  S  \colon  \Theta$.
In $\Theta$, the type assigned to each location is assumed to be well-formed at some classifier, but that classifier is not recorded in $\Theta$. We can observe a similar discipline in \ref{rule:type-store}. This is justified by \Zcref{claim:value-portability}: since any classifier may be chosen as long as the relevant types are well-formed under it, we do not need to track a specific classifier for each location.

\section{Metatheoretic Properties}\label{sec:metatheory}
We now have all the machinery needed to state and prove metatheoretic properties of our language.
The goal of this section is to prove two properties: soundness of our surface and runtime type systems, and safety of offline code generation.

\subsection{Type Soundness}
To state type soundness with respect to the definitional interpreter in \Zcref{sec:semantics}, we employ Siek's three-easy-lemmas style of type soundness proof~\cite{blog/Siek13, conf/popl/AminR17}.
First, we prove the safety of auxiliary functions used in the definitional interpreter.
\begin{lemma}[Safety of $ \token{lookup} $]\label{claim:safety-lookup}
  If\/ $ \Psi \mid^{ \gamma }  \varepsilon \vdash^{ \gamma }_{ R }  \refenvmv{x} : \refenvnt{A} $ and\/ $ \Psi \mid \Theta \vdash^{ \gamma } E \ \token{wf} $, then there exists\/ $\refenvnt{v}$ such that\/ $\token{lookup} \, \refenvsym{(}  E  \refenvsym{,}  \token{rename} \, \refenvsym{(}  R  \refenvsym{,}  \refenvmv{x}  \refenvsym{)}  \refenvsym{)} = \token{Done} \, \refenvsym{(}  \refenvnt{v}  \refenvsym{)}$ and\/ $ \Psi \mid \Theta \vdash^{ \gamma } \refenvnt{v} : \refenvnt{A} \ \token{val} $.
\end{lemma}
\begin{lemma}[Safety of $ \token{lookupStore} $]\label{claim:safety-lookup-store}
  If\/ $ \Psi \mid \Theta \vdash^{ \gamma } \refenvmv{l} :  \refenvnt{A} \/\ \token{ref}  \ \token{val} $ and\/ $\Psi  \vdash  S  \colon  \Theta$, then there exists\/ $\refenvnt{v}$ such that\/ $\token{lookupStore}  \refenvsym{(}  S  \refenvsym{,}  \refenvmv{l}  \refenvsym{)} = \token{Done} \, \refenvsym{(}  \refenvnt{v}  \refenvsym{)}$ and\/ $ \Psi \mid \Theta \vdash^{ \gamma } \refenvnt{v} : \refenvnt{A} \ \token{val} $.
\end{lemma}
\begin{lemma}[Safety of $ \token{updateStore} $]\label{claim:safety-assignment}
  If\/ $ \Psi \mid \Theta \vdash^{ \gamma } \refenvnt{v} : \refenvnt{A} \ \token{val} $, $\Psi  \vdash  S  \colon  \Theta$ and\/ $ \refenvmv{l} : \refenvnt{A}  \, \in \, \Theta$, then\/ $\token{updateStore} \, \refenvsym{(}  S  \refenvsym{,}  \refenvmv{l}  \refenvsym{,}  \refenvnt{v}  \refenvsym{)} = \token{Done} \, \refenvsym{(}  S_{{\mathrm{1}}}  \refenvsym{)}$ for some\/ $S_{{\mathrm{1}}}$ where\/ $\Psi  \vdash  S_{{\mathrm{1}}}  \colon  \Theta$.
\end{lemma}
We then prove \emph{expression persistency}, which ensures that an expression well-typed at the top-level stage remains well-typed at any future stage as long as classifiers are consistent. This property is used when a generated code fragment is spliced into larger code fragments.
\begin{lemma}[Expression Persistency]\label{claim:monotonicity}
  If\/ $ \Psi \mid^{ \gamma_{{\mathrm{1}}} }  \varepsilon \vdash^{ \gamma_{{\mathrm{1}}} }_{  \token{id}  }   \refenvnt{e} ^{ 0  }  : \refenvnt{A} $,\/ $\Psi  \vdash   \overrightarrow{ \gamma_{{\mathrm{2}}} }   \refenvsym{,}  \gamma_{{\mathrm{3}}} \, \token{wf}$ and\/ $\Psi  \vdash  \gamma_{{\mathrm{1}}}  \preceq  \gamma_{{\mathrm{3}}}$, then $ \Psi \mid^{  \overrightarrow{ \gamma_{{\mathrm{2}}} }   \refenvsym{,}  \gamma_{{\mathrm{3}}} }  \varepsilon \vdash^{  \overrightarrow{ \gamma_{{\mathrm{2}}} }   \refenvsym{,}  \gamma_{{\mathrm{3}}} }_{  \token{id}  }   \refenvnt{e} ^{ 0  }  : \refenvnt{A} $.
\end{lemma}
The following promotion and demotion results move variables and classifiers between reserved names typing and contexts.
The demotion result is a difficult and important part of the metatheory, because it requires careful handling of binding relations.
\begin{lemma}[Promotion]\label{claim:promotion}\leavevmode\samepage
  \begin{enumerate}
    \item If\/ $ \Psi \mid^{  \overrightarrow{ \gamma_{{\mathrm{1}}} }^{+}  }   \blacktriangleright ^{ \gamma_{{\mathrm{2}}} }  \vdash^{  \overrightarrow{ \gamma_{{\mathrm{1}}} }^{+}   \refenvsym{,}  \gamma_{{\mathrm{2}}} }_{ R }  \refenvnt{e} : \refenvnt{A} $,
          then
          $ \Psi \mid^{  \overrightarrow{ \gamma_{{\mathrm{1}}} }^{+}   \refenvsym{,}  \gamma_{{\mathrm{2}}} }  \varepsilon \vdash^{  \overrightarrow{ \gamma_{{\mathrm{1}}} }^{+}   \refenvsym{,}  \gamma_{{\mathrm{2}}} }_{ R }  \refenvnt{e} : \refenvnt{A} $.
    \item If\/
          $ \Psi \mid^{  \overrightarrow{ \gamma_{{\mathrm{1}}} }   \refenvsym{,}    \delta _{ 0 }  ,\dotsc,  \delta _{ \refenvnt{k} }   }   \blacktriangleleft _{ \refenvnt{k} }^{\binder{ \gamma_{{\mathrm{2}}} } }  \vdash^{  \overrightarrow{ \gamma_{{\mathrm{1}}} }   \refenvsym{,}  \gamma_{{\mathrm{2}}} }_{ R }  \refenvnt{e} : \refenvnt{A} $,
          then
          $ \Psi  \refenvsym{,}   \binder{ \gamma_{{\mathrm{2}}} }( \mathbin{:\succeq} \;  \delta _{ 0 }   /  \mathbin{:\supseteq} \;  \delta _{ \refenvnt{k} }  )  \mid^{  \overrightarrow{ \gamma_{{\mathrm{1}}} }   \refenvsym{,}  \gamma_{{\mathrm{2}}} }  \varepsilon \vdash^{  \overrightarrow{ \gamma_{{\mathrm{1}}} }   \refenvsym{,}  \gamma_{{\mathrm{2}}} }_{ R }  \refenvnt{e} : \refenvnt{A} $.
    \item If\/
          $ \Psi \mid^{  \overrightarrow{ \gamma_{{\mathrm{1}}} }   \refenvsym{,}  \gamma_{{\mathrm{2}}} }   \binder{ \refenvmv{x} }\mathord{:^{\binder{ \gamma_{{\mathrm{3}}} } } } \refenvnt{A_{{\mathrm{1}}}}  \vdash^{  \overrightarrow{ \gamma_{{\mathrm{1}}} }   \refenvsym{,}  \gamma_{{\mathrm{3}}} }_{ R }  \refenvnt{e} : \refenvnt{A_{{\mathrm{2}}}} $
          and $\refenvmv{y} \, \not\in \,  \token{Dom}_{V}( \Psi ) $,
          then
          $ \Psi  \refenvsym{,}   \binder{ \refenvmv{y} }: \refenvnt{A_{{\mathrm{1}}}} \mathrel{@}\binder{ \gamma_{{\mathrm{3}}} } \mathbin{:\succeq} \gamma_{{\mathrm{2}}}  \mid^{  \overrightarrow{ \gamma_{{\mathrm{1}}} }   \refenvsym{,}  \gamma_{{\mathrm{3}}} }  \varepsilon \vdash^{  \overrightarrow{ \gamma_{{\mathrm{1}}} }   \refenvsym{,}  \gamma_{{\mathrm{3}}} }_{ R  \refenvsym{,}  \refenvmv{x}  \coloneqq  \refenvmv{y} }  \refenvnt{e} : \refenvnt{A_{{\mathrm{2}}}} $.
    \item If\/
          $ \Psi \mid^{  \overrightarrow{ \delta_{{\mathrm{1}}} }   \refenvsym{,}   \gamma' _{ 0 }  }   [    \binder{  \gamma _{ 1 }  }  \mathbin{:\succeq}    \gamma' _{ 1 }  ,\dotsc,\binder{  \gamma _{ \refenvnt{k} }  }  \mathbin{:\succeq}   \gamma'  _{ \refenvnt{k} }   ]^{\binder{  \gamma _{ 0 }  } }  \vdash^{  \overrightarrow{ \delta_{{\mathrm{1}}} }   \refenvsym{,}   \gamma _{ 0 }  }_{ R }  \refenvnt{e} : \refenvnt{A} $,
          then
          $ \Psi  \refenvsym{,}   [    \binder{  \gamma _{ 0 }  }  \mathbin{:\succeq}    \gamma' _{ 0 }  ,\dotsc,\binder{  \gamma _{ \refenvnt{k} }  }  \mathbin{:\succeq}   \gamma'  _{ \refenvnt{k} }   ]  \mid^{  \overrightarrow{ \delta_{{\mathrm{1}}} }   \refenvsym{,}   \gamma _{ 0 }  }  \varepsilon \vdash^{  \overrightarrow{ \delta_{{\mathrm{1}}} }   \refenvsym{,}   \gamma _{ 0 }  }_{ R }  \refenvnt{e} : \refenvnt{A} $.
  \end{enumerate}
\end{lemma}

\begin{definition}[Disjointness of Classifiers]
  We write $ \Psi \vdash  \overrightarrow{ \gamma_{{\mathrm{1}}} }  \mathrel{\#}  \overrightarrow{ \gamma_{{\mathrm{2}}} }  $ if\/ $\Psi  \vdash  \gamma'_{{\mathrm{2}}}  \subseteq  \gamma'_{{\mathrm{1}}}$ does not hold
  for any $\gamma'_{{\mathrm{1}}} \, \in \, \refenvsym{\{}   \overrightarrow{ \gamma_{{\mathrm{1}}} }   \refenvsym{\}}$ and $\gamma'_{{\mathrm{2}}} \, \in \, \refenvsym{\{}   \overrightarrow{ \gamma_{{\mathrm{2}}} }   \refenvsym{\}}$.
\end{definition}

\begin{lemma}[Demotion]\label{claim:demotion}\leavevmode
  \begin{enumerate}
    \item Suppose
          $ \Psi \mid^{  \overrightarrow{ \gamma_{{\mathrm{1}}} }^{+}   \refenvsym{,}  \gamma_{{\mathrm{2}}}  \refenvsym{,}  \gamma_{{\mathrm{3}}} }  \varepsilon \vdash^{  \overrightarrow{ \gamma_{{\mathrm{1}}} }^{+}   \refenvsym{,}  \gamma_{{\mathrm{2}}}  \refenvsym{,}  \gamma_{{\mathrm{3}}} }_{  \token{id}  }  \refenvnt{e} : \refenvnt{A} $.
          If\/ $\Psi  \vdash  \gamma_{{\mathrm{3}}}  \subseteq  \gamma_{{\mathrm{2}}}$, then
          $ \Psi \mid^{  \overrightarrow{ \gamma_{{\mathrm{1}}} }^{+}   \refenvsym{,}  \gamma_{{\mathrm{2}}} }   \blacktriangleright ^{ \gamma_{{\mathrm{3}}} }  \vdash^{  \overrightarrow{ \gamma_{{\mathrm{1}}} }^{+}   \refenvsym{,}  \gamma_{{\mathrm{2}}}  \refenvsym{,}  \gamma_{{\mathrm{3}}} }_{  \token{id}  }  \refenvnt{e} : \refenvnt{A} $.
    \item Suppose
          $ \Psi \mid^{  \overrightarrow{ \gamma_{{\mathrm{1}}} }   \refenvsym{,}  \gamma_{{\mathrm{2}}} }  \varepsilon \vdash^{  \overrightarrow{ \gamma_{{\mathrm{1}}} }   \refenvsym{,}  \gamma_{{\mathrm{2}}} }_{  \token{id}  }  \refenvnt{e} : \refenvnt{A} $
          and $ \binder{ \gamma_{{\mathrm{2}}} }( \mathbin{:\succeq} \;  \delta _{ 0 }   /  \mathbin{:\supseteq} \;  \delta _{ \refenvnt{k} }  )  \, \in \, \Psi$
          and $ \Psi \vdash  \overrightarrow{ \gamma_{{\mathrm{1}}} }  \mathrel{\#} \gamma_{{\mathrm{2}}} $.
          If\/ $\gamma_{{\mathrm{4}}} \, \not\in \,  \token{Dom}_{C}( \Psi ) $, then
          $ \Psi \mid^{  \overrightarrow{ \gamma_{{\mathrm{1}}} }   \refenvsym{,}    \delta _{ 0 }  ,\dotsc,  \delta _{ \refenvnt{k} }   }   \blacktriangleleft _{ \refenvnt{k} }^{\binder{ \gamma_{{\mathrm{4}}} } }  \vdash^{  \overrightarrow{ \gamma_{{\mathrm{1}}} }   \refenvsym{,}  \gamma_{{\mathrm{4}}} }_{  \token{id}  }  \refenvnt{e} : \refenvnt{A}  \refenvsym{[}  \gamma_{{\mathrm{2}}}  \coloneqq  \gamma_{{\mathrm{4}}}  \refenvsym{]} $.
    \item Suppose
          $ \Psi \mid^{  \overrightarrow{ \gamma_{{\mathrm{1}}} }   \refenvsym{,}  \gamma_{{\mathrm{2}}} }  \varepsilon \vdash^{  \overrightarrow{ \gamma_{{\mathrm{1}}} }   \refenvsym{,}  \gamma_{{\mathrm{2}}} }_{  \token{id}  }  \refenvnt{e} : \refenvnt{A_{{\mathrm{2}}}} $
          and $ \binder{ \refenvmv{x} }: \refenvnt{A_{{\mathrm{1}}}} \mathrel{@}\binder{ \gamma_{{\mathrm{2}}} } \mathbin{:\succeq} \gamma_{{\mathrm{4}}}  \, \in \, \Psi$
          and $ \Psi \vdash  \overrightarrow{ \gamma_{{\mathrm{1}}} }  \mathrel{\#} \gamma_{{\mathrm{2}}} $.
          If\/ $\gamma_{{\mathrm{5}}} \, \not\in \,  \token{Dom}_{C}( \Psi ) $, then
          $ \Psi \mid^{  \overrightarrow{ \gamma_{{\mathrm{1}}} }   \refenvsym{,}  \gamma_{{\mathrm{4}}} }   \binder{ \refenvmv{x} }\mathord{:^{\binder{ \gamma_{{\mathrm{5}}} } } } \refenvnt{A_{{\mathrm{1}}}}  \refenvsym{[}  \gamma_{{\mathrm{2}}}  \coloneqq  \gamma_{{\mathrm{5}}}  \refenvsym{]}  \vdash^{  \overrightarrow{ \gamma_{{\mathrm{1}}} }   \refenvsym{,}  \gamma_{{\mathrm{5}}} }_{  \token{id}  }  \refenvnt{e} : \refenvnt{A_{{\mathrm{2}}}}  \refenvsym{[}  \gamma_{{\mathrm{2}}}  \coloneqq  \gamma_{{\mathrm{5}}}  \refenvsym{]} $.
    \item Let $\sigma =   \gamma _{ 0 }  \coloneqq  \delta' _{ 0 }  ,\dotsc,  \gamma _{ \refenvnt{k} }  \coloneqq  \delta' _{ \refenvnt{k} }  $.
          Suppose
          $ \Psi \mid^{  \overrightarrow{ \delta_{{\mathrm{1}}} }^{+}   \refenvsym{,}   \gamma _{ 0 }  }  \varepsilon \vdash^{  \overrightarrow{ \delta_{{\mathrm{1}}} }^{+}   \refenvsym{,}   \gamma _{ 0 }  }_{  \token{id}  }  \refenvnt{e} : \refenvnt{A} $,
          and $ [    \binder{  \gamma _{ 0 }  }  \mathbin{:\succeq}    \gamma' _{ 0 }  ,\dotsc,\binder{  \gamma _{ \refenvnt{k} }  }  \mathbin{:\succeq}   \gamma'  _{ \refenvnt{k} }   ]  \, \in \, \Psi$
          and $ \Psi \vdash  \overrightarrow{ \delta_{{\mathrm{1}}} }^{+}  \mathrel{\#}   \gamma _{ 0 }  ,\dotsc,  \gamma _{ \refenvnt{k} }   $.
          If\/ $ \delta' _{ \refenvnt{i} }  \, \not\in \,  \token{Dom}_{C}( \Psi ) $ for $0 \le \refenvnt{i} \le \refenvnt{k}$,
          then
          $ \Psi \mid^{  \overrightarrow{ \delta_{{\mathrm{1}}} }^{+}   \refenvsym{,}   \gamma' _{ 0 }  }   [    \binder{  \delta' _{ 1 }  }  \mathbin{:\succeq}    \gamma' _{ 1 }  ,\dotsc,\binder{  \delta' _{ \refenvnt{k} }  }  \mathbin{:\succeq}   \gamma'  _{ \refenvnt{k} }   ]^{\binder{  \delta' _{ 0 }  } }  \vdash^{  \overrightarrow{ \delta_{{\mathrm{1}}} }^{+}   \refenvsym{,}   \delta' _{ 0 }  }_{  \token{id}  }  \refenvnt{e} : \refenvnt{A}  \refenvsym{[}  \sigma  \refenvsym{]} $.
  \end{enumerate}
\end{lemma}
With these lemmas, we can prove type soundness of the runtime type system with respect to the definitional interpreter. The proof proceeds by induction on the derivation of the given expression typing and the step limit $\refenvnt{k_{{\mathrm{2}}}}$.
\begin{theorem}[Soundness of Runtime Typing]\label{claim:type-soundness}\leavevmode\samepage
  \begin{enumerate}
    \item If\/ $ \Psi_{{\mathrm{1}}} \mid^{ \gamma_{{\mathrm{1}}} }  \varepsilon \vdash^{ \gamma_{{\mathrm{1}}} }_{ R }   \refenvnt{e} ^{ 0  }  : \refenvnt{A} $, $ \Psi_{{\mathrm{1}}} \mid \Theta_{{\mathrm{1}}} \vdash^{ \gamma_{{\mathrm{1}}} } E \ \token{wf} $ and\/ $\Psi_{{\mathrm{1}}}  \vdash  S_{{\mathrm{1}}}  \colon  \Theta_{{\mathrm{1}}}$, then either of the following holds.
          \begin{itemize}
            \item $\mathcal{V}^{\mathrm{rt} }  \refenvsym{(}   \refenvnt{e} ^{ 0  }   \refenvsym{,}  E  \refenvsym{,}  R  \refenvsym{,}   \token{Dom}_{V}( \Psi_{{\mathrm{1}}} )   \refenvsym{,}  S_{{\mathrm{1}}}  \refenvsym{,}  \refenvnt{k_{{\mathrm{2}}}}  \refenvsym{)} =  \token{Timeout} $
            \item $\mathcal{V}^{\mathrm{rt} }  \refenvsym{(}   \refenvnt{e} ^{ 0  }   \refenvsym{,}  E  \refenvsym{,}  R  \refenvsym{,}   \token{Dom}_{V}( \Psi_{{\mathrm{1}}} )   \refenvsym{,}  S_{{\mathrm{1}}}  \refenvsym{,}  \refenvnt{k_{{\mathrm{2}}}}  \refenvsym{)} = \token{Done} \, \refenvsym{(}  \refenvnt{v}  \refenvsym{,}   \token{Dom}_{V}( \Psi_{{\mathrm{2}}} )   \refenvsym{,}  S_{{\mathrm{2}}}  \refenvsym{)}$, $ \Psi_{{\mathrm{2}}} \mid \Theta_{{\mathrm{2}}} \vdash^{ \gamma_{{\mathrm{1}}} } \refenvnt{v} : \refenvnt{A} \ \token{val} $ and\/ $\Psi_{{\mathrm{2}}}  \vdash  S_{{\mathrm{2}}}  \colon  \Theta_{{\mathrm{2}}}$ for some $\refenvnt{v}$, $\Psi_{{\mathrm{2}}}$ and\/ $\Theta_{{\mathrm{2}}}$ such that $ \Psi_{{\mathrm{1}}} \subseteq \Psi_{{\mathrm{2}}} $ and\/ $ \Theta_{{\mathrm{1}}} \subseteq \Theta_{{\mathrm{2}}} $.
          \end{itemize}

    \item If\/ $ \Psi_{{\mathrm{1}}} \mid^{   \gamma _{ 0 }  ,\dotsc,  \gamma _{  \refenvnt{k_{{\mathrm{1}}}} ^{+}  }   }  \varepsilon \vdash^{   \gamma _{ 0 }  ,\dotsc,  \gamma _{  \refenvnt{k_{{\mathrm{1}}}} ^{+}  }   }_{ R }   \refenvnt{e_{{\mathrm{1}}}} ^{  \refenvnt{k_{{\mathrm{1}}}} ^{+}   }  : \refenvnt{A} $, $ \Psi_{{\mathrm{1}}} \mid \Theta_{{\mathrm{1}}} \vdash^{  \gamma _{ 0 }  } E \ \token{wf} $ and\/ $\Psi_{{\mathrm{1}}}  \vdash  S_{{\mathrm{1}}}  \colon  \Theta_{{\mathrm{1}}}$, then either of the following holds.
          \begin{itemize}
            \item $\mathcal{V}^{\mathrm{fut} }  \refenvsym{(}  \refenvnt{k_{{\mathrm{1}}}}  \refenvsym{,}   \refenvnt{e_{{\mathrm{1}}}} ^{  \refenvnt{k_{{\mathrm{1}}}} ^{+}   }   \refenvsym{,}  E  \refenvsym{,}  R  \refenvsym{,}   \token{Dom}_{V}( \Psi_{{\mathrm{1}}} )   \refenvsym{,}  S_{{\mathrm{1}}}  \refenvsym{,}  \refenvnt{k_{{\mathrm{2}}}}  \refenvsym{)} =  \token{Timeout} $
            \item $\mathcal{V}^{\mathrm{fut} }  \refenvsym{(}  \refenvnt{k_{{\mathrm{1}}}}  \refenvsym{,}   \refenvnt{e_{{\mathrm{1}}}} ^{  \refenvnt{k_{{\mathrm{1}}}} ^{+}   }   \refenvsym{,}  E  \refenvsym{,}  R  \refenvsym{,}   \token{Dom}_{V}( \Psi_{{\mathrm{1}}} )   \refenvsym{,}  S_{{\mathrm{1}}}  \refenvsym{,}  \refenvnt{k_{{\mathrm{2}}}}  \refenvsym{)} = \token{Done} \, \refenvsym{(}   \refenvnt{e_{{\mathrm{2}}}} ^{ \refenvnt{k_{{\mathrm{1}}}}  }   \refenvsym{,}   \token{Dom}_{V}( \Psi_{{\mathrm{2}}} )   \refenvsym{,}  S_{{\mathrm{2}}}  \refenvsym{)}$,\quad $ \Psi_{{\mathrm{2}}} \mid^{   \gamma _{ 1 }  ,\dotsc,  \gamma _{  \refenvnt{k_{{\mathrm{1}}}} ^{+}  }   }  \varepsilon \vdash^{   \gamma _{ 1 }  ,\dotsc,  \gamma _{  \refenvnt{k_{{\mathrm{1}}}} ^{+}  }   }_{  \token{id}  }   \refenvnt{e_{{\mathrm{2}}}} ^{ \refenvnt{k_{{\mathrm{1}}}}  }  : \refenvnt{A} $ and\/ $\Psi_{{\mathrm{2}}}  \vdash  S_{{\mathrm{2}}}  \colon  \Theta_{{\mathrm{2}}}$ for some $\refenvnt{e_{{\mathrm{2}}}}$, $\Psi_{{\mathrm{2}}}$ and\/ $\Theta_{{\mathrm{2}}}$ such that $ \Psi_{{\mathrm{1}}} \subseteq \Psi_{{\mathrm{2}}} $ and\/ $ \Theta_{{\mathrm{1}}} \subseteq \Theta_{{\mathrm{2}}} $.
          \end{itemize}
  \end{enumerate}
\end{theorem}
Type soundness of the surface type system then follows as a special case of the theorem above, by instantiating $\Psi_{{\mathrm{1}}}$ to $ \varepsilon $, $\gamma_{{\mathrm{1}}}$ to $ \exclam $, and $\Theta_{{\mathrm{1}}}$ to $ \varepsilon $.
\begin{corollary}[Soundness of Surface Typing]
  If\/ $ \varepsilon \vdash^{  \exclam  }  \refenvnt{e} ^{ 0  }  : \refenvnt{A} $, then either of the following holds.
  \begin{itemize}
    \item $\mathcal{V}^{\mathrm{rt} }  \refenvsym{(}   \refenvnt{e} ^{ 0  }   \refenvsym{,}  \varepsilon  \refenvsym{,}   \token{id}   \refenvsym{,}  \varepsilon  \refenvsym{,}  \varepsilon  \refenvsym{,}  \refenvnt{k_{{\mathrm{2}}}}  \refenvsym{)} =  \token{Timeout} $
    \item $\mathcal{V}^{\mathrm{rt} }  \refenvsym{(}   \refenvnt{e} ^{ 0  }   \refenvsym{,}  \varepsilon  \refenvsym{,}   \token{id}   \refenvsym{,}  \varepsilon  \refenvsym{,}  \varepsilon  \refenvsym{,}  \refenvnt{k_{{\mathrm{2}}}}  \refenvsym{)} = \token{Done} \, \refenvsym{(}  \refenvnt{v}  \refenvsym{,}  N  \refenvsym{,}  S  \refenvsym{)}$ where $ \Psi \mid \Theta \vdash^{  \exclam  } \refenvnt{v} : \refenvnt{A} \ \token{val} $, $ \token{Dom}_{V}( \Psi )  = N$ and $\Psi  \vdash  S  \colon  \Theta$ for some $\Psi$ and $\Theta$.
  \end{itemize}
\end{corollary}

\subsection{Safety of Offline Code Generation}
As another important property, we prove the safety of offline code generation.
This property states that the content of generated code remains well-typed.
In particular, we expect that generated code does not depend on runtime objects such as store or reserved names, which is also discussed in earlier staged calculi with mutable state~\cite{conf/aplas/KiselyovKS16,journals/pacmpl/XieWNY23,journals/pacmpl/LiKXY26}.
It is clear that we can discard the store, since expressions do not directly depend on locations.
It is less trivial if we can discard reserved names, since they can be used in generated code.
However, we can show that generated code will not use reserved names if it is well-typed at the initial classifier, as stated in the following lemma.
\begin{lemma}\label{claim:discard-reserved-names}
  If\/ $ \Psi \mid^{  \exclam  }  \varepsilon \vdash^{  \exclam  }_{  \token{id}  }   \refenvnt{e} ^{ 0  }  : \refenvnt{A} $, then $ \varepsilon \mid^{  \exclam  }  \varepsilon \vdash^{  \exclam  }_{  \token{id}  }   \refenvnt{e} ^{ 0  }  : \refenvnt{A} $.
\end{lemma}
We can then prove safety of offline code generation, stated as follows.
\begin{theorem}[Safety of Offline Code Generation]
  \label{claim:safety-offline-code-gen}
  If\/ $ \varepsilon \vdash^{  \exclam  }  \refenvnt{e_{{\mathrm{1}}}} ^{ 0  }  :  \langle \refenvnt{A} \rangle^{  \exclam  }  $ and\/ $\mathcal{V}^{\mathrm{rt} }  \refenvsym{(}   \refenvnt{e_{{\mathrm{1}}}} ^{ 0  }   \refenvsym{,}  \varepsilon  \refenvsym{,}   \token{id}   \refenvsym{,}  \varepsilon  \refenvsym{,}  \varepsilon  \refenvsym{,}  \refenvnt{k_{{\mathrm{2}}}}  \refenvsym{)} = \token{Done} \, \refenvsym{(}  \refenvnt{v}  \refenvsym{,}  N  \refenvsym{,}  S  \refenvsym{)}$, then $\refenvnt{v} =  \langle   \refenvnt{e_{{\mathrm{2}}}} ^{ 0  }   \rangle $ and\/ $ \varepsilon \vdash^{  \exclam  }  \refenvnt{e_{{\mathrm{2}}}} ^{ 0  }  : \refenvnt{A} $ for some $ \refenvnt{e_{{\mathrm{2}}}} ^{ 0  } $.
\end{theorem}

\subsection{Mechanization in Rocq}
We mechanized in Rocq the definitions and metatheory of our calculus presented in \Zcref{sec:runtime-typing,sec:metatheory}. In particular, the mechanization establishes the two main results of the paper:
\Zcref{claim:type-soundness,claim:safety-offline-code-gen}.
The formalization comprises 66,201 lines of Rocq source, much of which was generated from the definitions and theorem statements in the paper and its supplementary material using OpenAI Codex with GPT-5.5. The authors reviewed and corrected the generated code, checked the complete development with Rocq, and verified its correspondence with the source material. They take full responsibility for its correctness.

We use the locally nameless representation with cofinite quantification over
fresh names~\cite{conf/popl/AydemirCPPW08} for binders introduced by
polymorphic classifiers. We do not use such a representation for
object-language variables: they are represented directly as names, and their
binding structure is managed by the operational semantics through fresh-name
generation and renaming environments. As a result, the mechanization does not
require reasoning about alpha-equivalence or substitution for object-language
variables; locally nameless reasoning is needed only for classifier binders.

\section{Related Work}\label{sec:relatedwork}
\subsection{Earlier proposals with Refined Environment Classifiers}
Earlier REC-based systems are restricted to two-level staging and support neither runtime execution nor CSP~\cite{conf/aplas/KiselyovKS16,conf/gpce/OishiK17,conf/gpce/IsodaYK24,conf/popl/LeeXKY26}. In that setting, classifiers need only be tracked at a single stage, simplifying the type systems. In contrast, supporting multi-level staging, \lstinline|run|, and scoped CSP requires typing contexts and judgments that track scoping dependencies across multiple stages. Classifier polymorphism poses an additional challenge: although the need for it had previously been noted informally, integrating it into earlier systems was nontrivial.

To address these challenges, we redesign typing contexts and judgments to manage scoping information uniformly across multiple stages. The resulting system supports multi-level code generation, \lstinline|run|, and scoped CSP, using the visibility relation to capture the classifier scoping required for safe nested code generation. The same context structure yields natural surface typing rules for polymorphic classifiers simply by adding the corresponding context form, an advantage over earlier REC proposals.

\subsection{Typing Disciplines for MetaML-style Code Generation with Effects}
\label{sec:relatedwork:typing}
Early work on multi-stage programming contrasted two approaches to typing MetaML-style code generation. The type $\bigcirc A$ of \lamcirc represents open code and accommodates code with free variables, but supports neither runtime execution nor effectful code generation~\cite{conf/lics/Davies96,journals/jacm/Davies17}. The type $\Box A$ of \lambox represents only closed code, but safely supports runtime execution and effects because scope extrusion cannot occur~\cite{journals/jacm/DaviesP01,conf/icalp/CalcagnoMT00}. The need for both capabilities motivated several systems combining the two types~\cite{conf/esop/MoggiTBS99,conf/gpce/CalcagnoTHL03,conf/ppdp/YuseI06}. Among them, Calcagno et al.~\cite{conf/gpce/CalcagnoTHL03} proposed a sound system with mutable state that permits only closed code to be executed or stored, preventing scoping errors. Rather than combining separate types for open and closed code, \emph{environment-classifier}\footnote{Despite the similarity in terminology, environment classifiers and refined environment classifiers serve distinct purposes: the former track the staging dependencies of generated code, whereas the latter track its scoping dependencies.} systems annotate a single code type with staging information, distinguishing potentially open code valid at a particular stage from closed code portable across stages~\cite{conf/popl/TahaN03,conf/esop/CalcagnoMT04}. This supports safe runtime execution and could likely accommodate mutable state by restricting storage to closed code, though this has not been formally studied.

Xie et al.~\cite{journals/pacmpl/XieWNY23} and Li et al.~\cite{journals/pacmpl/LiKXY26} formalized MetaML-style typed calculi with mutable state and proved the safety of offline code generation; Li et al. additionally support runtime execution. Both focus on the interaction between mutable state and offline code generation, showing that generated code is independent of generation-time state, such as store locations. Consistent with this focus, they restrict mutable references to ground types, simplifying their formalizations but excluding references containing code and hence scope extrusion through mutable state.

Contextual types are another approach to typing open code, though usually in calculi unlike MetaML; see \Zcref{sec:relatedwork:contextualtypes}. One exception is \lambra, a MetaML-style calculus with mutable state whose contextual code types explicitly list the free variables available to a code fragment~\cite{conf/esop/Rhiger12}. However, they do not account for visibility and thus admit well-typed programs exhibiting scope extrusion, as explained in \Zcref{sec:informaldescription:r2ecs:subtlety} and demonstrated in the supplementary material. This limitation appears difficult to address in a contextual type system.

\subsection{Reasoning about Dynamic Aspects of Effectful Code Generation}
Practical multi-stage languages tend to adopt simpler type systems based on \lamcirc, allowing dynamic errors~\cite{thesis/Rompf12,thesis/Stucki23,conf/flops/Kiselyov14}.
This motivates studying the dynamic behavior of effectful code generation, including erroneous behavior such as scope extrusion.

Moggi and Fagorzi's combinator-based calculus~\cite{conf/fossacs/MoggiF03} makes scope extrusion through mutable state observable by treating fresh-name generation as an effect and generated names as part of the computation state. Combinator-based REC calculi adopt a similar view~\cite{conf/aplas/KiselyovKS16,conf/gpce/OishiK17,conf/gpce/IsodaYK24}, but further distinguish harmful scope extrusion, where the final result of code generation contains a free generated name.

Lee et al.~\cite{conf/popl/LeeXKY26} study dynamic detection of scope extrusion in code generation with effect handlers, exploring the trade-off between precision and performance. Such mechanisms could complement our static discipline in practical extensions, particularly gradual typing~\cite{conf/gpce/YaguchiK25,conf/tyde/ChenSSCMVL25}, and could be incorporated into our operational semantics.

Yin et al.~\cite{conf/lics/Yin26} proposed a typed multi-stage calculus
with mutable state and strong type-safety guarantees. They also developed
a fully abstract trace model based on operational game semantics, providing
a semantic foundation for effectful code generation.
We note, however, that while they present their language as MetaML-style,
we classify it as a contextual-type-based calculus; see \Zcref{sec:relatedwork:contextualtypes}.

Lightweight Modular Staging (LMS)~\cite{thesis/Rompf12} is a practical Scala framework that addresses scope extrusion through language design: it encapsulates effects such as deduplication and code motion within its staging primitives, allowing programmers to write staged programs in a pure style while benefiting from effectful code generation. This design makes scope extrusion less likely, though not impossible.

\subsection{Cross-Stage Persistence}
Cross-stage persistence (CSP) allows variables or values from one stage to be used at later stages, but admits several implementations. Xie et al.~\cite{journals/pacmpl/XieWNY23} classify them as \emph{Value-based CSP}, \emph{Path-based CSP}, and \emph{Heap-based CSP}. Heap-based CSP stores values in the heap and embeds their locations in generated code for resolution at runtime; it corresponds to the original notion of CSP in MetaML~\cite{journals/tcs/TahaS00,conf/popl/TahaN03} and to what we call traditional CSP in this paper.

Our \emph{scoped CSP} allows generated code to contain variable names resolved at runtime against the surrounding lexical environment. This mechanism is not new: a similar one appears in the staged C variant \textsc{`C}~\cite{journals/toplas/PolettoHEK99}, showing that scoped CSP arises naturally from runtime execution under lexical scoping.
Although more restrictive than heap-based CSP, scoped CSP supports similar use cases and fits naturally with our R2EC-based typing discipline~(\Zcref{sec:surface-typing}).
By contrast, heap-based CSP requires non-trivial type-system extensions to ensure the safety of offline code generation~\cite{conf/flops/HanadaI14,journals/pacmpl/LiKXY26}. Comparing their use cases and type systems more closely remains future work.

\subsection{MetaML-style MSP and Contextual-Types-based MSP}\label{sec:relatedwork:contextualtypes}
Besides MetaML-style MSP, another approach to MSP is based on contextual types~\cite{conf/popl/KimYC06,journals/jfp/NanevskiP05,journals/pacmpl/ParreauxVSK18,conf/esop/MuraseNI23}. From a syntactic perspective, the main difference between the two approaches lies in how they treat free variables in code fragments. In MetaML-style MSP, free variables in code fragments can be bound by the surrounding context. In contrast, in contextual-types-based calculi, free variables in code fragments are bound by the context annotation on a quotation or contextual type, rather than directly by the surrounding context.
An advantage of the contextual-types-based approach is that it is largely orthogonal to other features, since open code fragments are not lexically bound by surrounding contexts. For example, several contextual-types-based calculi have been proposed in combination with effectful features such as mutable state~\cite{conf/popl/KimYC06,conf/lics/Yin26} and session-typed communication~\cite{journal/pacmpl/SanoGKPT25, conf/esop/AngeloIMV26}. By contrast, subtle interactions between MetaML-style MSP and effectful computation have long been a challenge. From this perspective, the main result of this paper is significant because it provides a typing discipline for MetaML-style MSP that composes cleanly with mutable state while ensuring safety.

Since the two approaches offer different strengths, they can be hybridized to combine their advantages; see Chiang and Xie~\cite{journals/pacmpl/ChiangX25}. Thus, they are better viewed as complementary rather than competing approaches to MSP.

\subsection{Logical Foundations of Multi-Stage Programming}
Type systems for multi-stage programming are known to have logical foundations in modal logic~\cite{journals/jacm/Davies17,journals/jacm/DaviesP01,journals/tocl/NanevskiPP08,journals/corr/abs-1010-3806}.
In a recent preprint, Murase and Maniwa~\cite{journals/corr/abs-2602-09462}
proposed \emph{bounded modal logic} as a logical foundation for MetaML-style metaprogramming,
generalizing the S4 modal lambda calculus to make the scope dependencies
of code fragments explicit.
Their formalization is similar to ours in that it uses classifiers to annotate modal types and supports polymorphism over classifiers. However, it does not account for scope visibility and would therefore admit programs that are undesirable in our setting. Elaborating their logic to provide a logical foundation for our type system is an interesting direction for future work.

\section{Conclusion and Future Work}\label{sec:conclusion}
In this paper, we proposed a novel type system based on refined environment classifiers (RECs) for MetaML-style multi-stage programming with mutable state.
We showed that extending RECs beyond earlier settings is nontrivial: typing multi-level programs requires tracking not only variable scopes but also the scopes of classifiers themselves.
We formalized this refinement through reachability and visibility and called the resulting typing discipline \emph{refined\textsuperscript{2} environment classifiers} (R2ECs).
We further incorporated polymorphic classifiers to support more general and reusable staged code generators.

For the full calculus, we proved type soundness and the safety of offline code generation, demonstrating that R2ECs support expressive, effectful multi-stage programming while preventing harmful scope extrusion. We also presented a definitional-interpreter semantics that incorporates scoped CSP and renaming environments, providing an operational account of lexical scoping and a basis for future semantic studies. Taken together, these contributions establish a theoretical foundation for effectful MetaML-style MSP supporting multi-level code generation, runtime execution, scoped CSP, and classifier polymorphism.

As future work, we plan to investigate practical extensions, including subtyping~\cite{conf/esop/Rhiger12,conf/aplas/KiselyovKS16}, type inference~\cite{conf/esop/CalcagnoMT04,conf/aplas/KiselyovKS16}, analytic metaprogramming~\cite{journals/pacmpl/ParreauxVSK18,thesis/Stucki23,journals/pacmpl/ChiangX25}, effect handlers~\cite{conf/popl/LeeXKY26,conf/gpce/IsodaYK24}, and module systems~\cite{journals/pacmpl/XieWNY23,journals/pacmpl/ChiangYWX24,journals/scp/SuwaI26}.

\section*{Data-Availability Statement}
We provide a supplementary document and an artifact package.
The supplementary document contains material omitted from the main text, including full definitions, paper proofs of the metatheoretic properties, and counterexamples showing that $\lambda^{[]}$, a MetaML-style calculus with mutable state~\cite{conf/esop/Rhiger12}, is not type sound, as discussed in~\Zcref{sec:relatedwork}. The supplementary document has been uploaded separately and is also included in the artifact package described below.

The artifact package contains a browser-based implementation of our calculus and a Rocq mechanization of the formal definitions and proofs presented in the supplementary document. The implementation supports interactive exploration of the surface type system and the definitional interpreter. We provide Docker images for both components to facilitate reproducibility. The artifact package is available on Zenodo at \url{https://doi.org/10.5281/zenodo.21468038}. For convenience, the implementation can also be tried directly through an interactive web demo at \url{https://www.fos.kuis.kyoto-u.ac.jp/~murase/r2ecs-demo/}.

\bibliographystyle{ACM-Reference-Format}
\bibliography{mybib}

\end{document}